\pdfoutput=1
\documentclass[11pt]{article}
\usepackage{cite}
\usepackage{graphicx}
\usepackage{latexsym}   
\usepackage{mathrsfs}
\usepackage[scr=rsfs,cal=boondox]{mathalfa}
\usepackage[overload]{textcase}

\font\grande=cmr9.5 scaled \magstep4
\font\medio=cmr9.5 scaled \magstep2
\outer\def\beginsection#1\par{\medbreak\bigskip
      \message{#1}\leftline{\bf#1}\nobreak\medskip
\vskip-\parskip
      \noindent}

\begin{document}
\bibliographystyle{unsrt}

\titlepage

\vspace{1cm}
\begin{center}
{\grande Relic gravitons and high-frequency detectors}\\
\vspace{0.5cm}
\vspace{1.5 cm}
Massimo Giovannini \footnote{e-mail address: massimo.giovannini@cern.ch}\\
\vspace{0.5cm}
{{\sl Department of Physics, CERN, 1211 Geneva 23, Switzerland }}\\
\vspace{0.5cm}
{{\sl INFN, Section of Milan-Bicocca, 20126 Milan, Italy}}\\
\vspace*{1cm}
\end{center}
\vskip 0.3cm
\centerline{\medio  Abstract}
\vskip 0.5cm
Cosmic gravitons are expected  in the MHz--GHz regions that are currently unreachable by the operating wide-band interferometers and where  various classes of electromechanical detectors have been proposed through the years. The minimal chirp amplitude detectable by these instruments is often set on the basis of the sensitivities reachable by the detectors currently operating in the audio band. By combining the observations of the pulsar timing arrays, the limits from wide-band detectors and the other phenomenological bounds we show that this requirement is far too generous and even misleading since the actual detection of relic gravitons well above the kHz would demand chirp and spectral amplitudes that are ten or even fifteen orders of magnitude smaller than the ones currently achievable in the audio band, for the same classes of stochastic sources. We then examine more closely the potential high-frequency signals and show that the sensitivity in the chirp and spectral amplitudes must be even smaller than the ones suggested by the direct and indirect constraints on the cosmic gravitons. We finally analyze the high-frequency detectors in the framework of Hanbury-Brown Twiss interferometry and argue that they are actually more essential than the ones operating in the audio band (i.e. between few Hz and few kHz) if we want to investigate the quantumness of the relic gravitons and their associated second-order correlation effects. We suggest, in particular, how the statistical properties of thermal and non-thermal gravitons can be distinguished by studying the corresponding second-order interference effects.
\noindent
\vspace{5mm}
\vfill
\newpage

\renewcommand{\theequation}{1.\arabic{equation}}
\setcounter{equation}{0}
\section{Introduction}
\label{sec1}
Relic gravitons are produced by the pumping action of the space-time curvature prior to matter-radiation equality \cite{AA,AB,AC} and their spectrum extends, in principle, between\footnote{The standard prefixes of the international system of units are used so that, for instance, $1\, \mathrm{aHz} = 10^{-18} \mathrm{Hz}$, $1\, \mathrm{GHz} = 10^{9}\, \mathrm{Hz}$ and so on. The present value of the scale factor is normalized as $a_{0} =1$ and this means that at $\tau_{0}$ the comoving and the physical frequencies coincide. The spectral energy density in critical units is specifically defined later on but it is customary to introduce directly $h_{0}^2 \, \Omega_{gw}(\nu, \tau_{0})$ since this quantity does not depend on the indetermination of the present Hubble rate.} few aHz and $100$ GHz. In the concordance paradigm their spectral energy density in critical units ($h_{0}^2 \Omega_{gw}(\nu, \tau_{0})$ in what follows) is quasi-flat for comoving frequencies $\nu$ larger than $100$ aHz \cite{BA} while below this frequency it scales as $\nu^{-2}$ \cite{BB}. The flatness of $h_{0}^2 \Omega_{gw}(\nu,\tau_{0})$ (for all the wavelengths exiting during a conventional stage of inflationary expansion \cite{CA,CB,CC,CD}  and reentering the Hubble radius when the plasma is dominated by radiation) imposes a low-frequency normalization determined by the tensor to scalar ratio $r_{T}$ evaluated at a conventional frequency $\nu_{p}= 3.092$ aHz that corresponds to a pivot wavenumber $k_{p} = 0.002\,\, \mathrm{Mpc}^{-1}$.  The current analyses suggest $r_{T}(\nu_{p}) = r_{T} < 0.06$ \cite{DA,DB,DC} or even $r_{T} < 0.03$ and while the differences between the determinations of $r_{T}$ are immaterial for the present purposes, it is relevant to stress that the tensor to scalar ratio is not the only source of suppression since, for frequencies larger than the nHz and smaller than the Hz, $h_{0}^2 \Omega_{gw}(\nu,\tau_{0})$ is further damped by the neutrino free-streaming \cite{DD,DE}. 
If we put together the flatness of the spectrum \cite{BA}, the low-frequency normalization \cite{DA,DB,DC}, and the suppression due to neutrino-free streaming (and to other sources \cite{DF}) we obtain that, in the concordance paradigm, $h_{0}^2 \Omega_{gw}(\nu, \tau_{0})$ cannot (optimistically) exceed ${\mathcal O}(10^{-17})$ for comoving frequencies falling between few kHz and $100$ MHz. Even if the smallness of this result only depends on the assumption that radiation suddenly dominates after inflation, 
the wide-band interferometers currently operating cannot probe comoving frequencies larger than the kHz so that it is tempting to consider the possibility of detecting cosmic gravitons with electromechanical detectors. In this case the high-frequency wave may interact both with the electromagnetic field and with the field of elastic deformations of the detector (see \cite{DF} for a recent review including a discussion of these detectors).

 One of the first detectors proposed at high frequencies is the so-called Bragisnky-Menskii (toroidal) wave-guide \cite{EA1,EA2} where an electromagnetic wave-packet propagates and the presence of a gravitational wave eventually shifts the electromagnetic frequency.  Through the years it was realized that not only dynamical electromagnetic fields can be used to detect gravitational radiation but also the static ones. Microwave cavities with superconducting walls have then been proposed in the 1970s and 1980s \cite{EB, cav1a, EC,EC2} for the detection of small harmonic displacements and a number of prototypes have been studied \cite{ED,EE,EF,EG}. While the first prototypes in the mid 1980s could resolve chirp amplitudes ${\mathcal O}(10^{-17})$ the potential sensitivities reached the level of $10^{-20}$ twenty years later \cite{EF,EG} and they might be today 
 comparable with the typical chirp amplitudes probed by wide-band 
 interferometers in a much lower frequency range. Microwave cavities operate 
 in fact as electromagnetic resonators with two levels and they could detect, in principle, relic gravitons between few GHz and $0.1$ THz  \cite{MG1,MG2,MG3}. The analysis of electromagnetic cavities has been complemented by the use of dynamical electromagnetic fields, such as for instance, waveguides \cite{EH,EI}. A proposal for the observation of relic gravitons at $100$ MHz has been illustrated in Refs. \cite{EL,EM}. Other interesting  detectors have been described and partially built \cite{EN,EO,EP} with frequency of operation of the order of 100 MHz and possibly even higher. The detection of relic gravitons in the MHz region has also been seriously considered by using small (i.e. 75 cm) interferometers \cite{EQ}.  High-frequency detectors may be the sole chance of resolving single-gravitons \cite{FA} and this may happen by conversion to photons in a strong magnetic field \cite{FB} with experimental techniques very similar to the ones employed for the scrutiny of axion-like particles\cite{FC} (see also \cite{FA1,FA2,FA3,FA4,FD} for some other papers with similar inspiration). 

Even if this paper does not pretend to suggest new types of high-frequency instruments, it is amusing that most of the reported attempts consider a success the detection of a chirp amplitude as small as the one currently assessed by wide-band detectors\footnote{ 
The sensitivities of these instruments can be expressed in terms of the minimal 
detectable chirp and spectral amplitudes denoted, respectively, by $h_{c}(\nu,\tau_{0})$ and $S_{h}(\nu,\tau_{0})$. Th accurate definition of these variables is one of the themes of section \ref{sec2}. We note that various classes of high-frequency detectors currently 
suggested as novel are in fact reprises of ideas of the 1980s and 1990s.}. In the 1980s the coupled cavities in the MHz range could detect typical chirp amplitudes ${\mathcal O}(10^{-17})$. These sensitivities improved by $4$ or even $5$ orders of magnitude so that, today, the minimal detectable $h_{c}(\nu,\tau_{0})$ is comparable with the on 
probed by wide-band interferometers but in a much higher frequency range.
There are therefore a number of suggestions on how to improve these sensitivities but it is 
difficult to gauge the feasibility of these suggestions that are often purely theoretical. For this reason  we intend to clarify here how small should be the minimal chirp amplitude to be relevant for the detection of relic graviton backgrounds 
at high-frequencies. Indeed the relic graviton backgrounds exhibiting a large signals in the MHz region \cite{MG1,MG2,MG3} have been originally taken as the main motivation for the analysis of high-frequency detectors and we are today witnessing a similar trend that also includes the detectors of axion-like particles. 

The goals of more recent studies, by admission of the authors, generically target signals from the early Universe (i.e. relic gravitons) but nonetheless the range of the minimal detectable chirp amplitude is ${\mathcal O}(10^{-20})$ or marginally smaller in contrast 
with what could be deduced from more accurate theoretical analyses. If we take at face value the current amplitude of the relic graviton background coming from the concordance paradigm we would have, rather optimistically, that the minimal 
detectable $h_{c}(\nu,\tau_{0})$ should be between $15$ or $20$ orders of magnitude smaller in the MHz range. Moreover, according 
to the current constraints in the audio band  $h_{c}(\nu,\tau_{0}) \leq {\mathcal O}(10^{-24})$. As we shall see, this constraint cannot be naively rescaled at higher frequencies and, for this reason, a sound strategy suggests not only to enforce the relevant constraints but also to examine the broad classes of high-frequency signals. 

We find necessary to spell out unambiguously the sensitivity goals that must be required if we want to target the signals coming from the early evolution of the plasma prior to matter-radiation equality. In the current literature this aspect is quite confusing also because the relevant bounds of the problem involve not only the determination of the tensor to scalar ratio but also the pulsar timing arrays, the limits from the interferometric detectors in the kHz region and the constraints coming from big-bang nucleosynthesis. Besides the current bounds it is equally essential to examine the high-frequency signals that can be either thermal or non-thermal. We also point out that the second-order interference effects (associated with the interferometric techniques developed by Hanbury-Brown and Twiss \cite{HBT0a,HBT0b}) can be used to distinguish between thermal and non-thermal sources in the high-frequency domain.  This is, in our opinion, one of the novel possibilities 
associated with the high-frequency instruments.

Before concluding this introductory considerations it is useful to stress, as already mentioned above, that the current analyses \cite{DA,DB,DC} suggest at upper limit on the tensor to scalar ratio $r_{T}$. The considerations developed here deal 
with high-frequency gravitons while $r_{T}$ sets the low-frequency normalization. In this sense the value of $r_{T}$ 
is not directly relevant to illustrate the interplay between relic gravitons and high-frequency detectors. 
At the same time different values of $r_{T}$ may modify the allowed regions of the parameter space. For all the numerical estimates
discussed hereunder we shall be assuming that $r_{T} = {\mathcal O}(0.06)$ even if lower values of $r_{T}$ can be discussed 
with a similar approach.

The layout of this paper is, in short, the following. In section \ref{sec2} we set the basic notations and introduce the mutual connections between the observables that are employed throughout the investigation. Section \ref{sec3} is devoted to the direct and indirect constraints on the diffuse backgrounds of relic gravitons. Particular attention is paid to the current limits from the interferometers in the audio band, to the measurements of the pulsar timing arrays and to the big-bang nucleosynthesis bounds. At the end of section \ref{sec3} we preliminarily assess the required sensitivity in the high-frequency domain. The concrete realizations of high-frequency signals are considered in section \ref{sec4} by distinguishing the two broad categories of thermal and non-thermal gravitons. In section \ref{sec5} we finally argue that if high-frequency detectors will ever be able to resolve bunches of relic gravitons, then it will possible to distinguish the origin and the correlation properties of the signals by analyzing the second-order interference effects associated with the intensities (rather than with the amplitudes).

\renewcommand{\theequation}{2.\arabic{equation}}
\setcounter{equation}{0}
\section{Chirp amplitude, spectral energy density and spectral amplitude}
\label{sec2}
The connection between the spectral amplitude, the chirp amplitude 
and the spectral energy density ultimately depends on the way 
the energy-momentum pseudo-tensor of the relic gravitons is assigned. 
As discussed in Ref. \cite{MG4}, different prescriptions lead 
to expressions of the energy density that do not generally agree 
for typical wavelengths larger than the Hubble radius. If the energy momentum pseudo-tensor is defined from the variation of the second-order action with respect 
to the background metric the corresponding energy density 
is consistently defined in all the kinematical regions. This 
approach corresponds ultimately to the one pioneered 
in Ref. \cite{AC}. We then start from the action of the gravitons in a Friedmann-Robertson-Walker background\footnote{As usual $M_{P} = G^{-1/2}$ is the Planck mass 
while $\overline{M}_{P}$ is the reduced Planck mass defined in Eq. (\ref{FUND1}). 
The Greek (lowercase) indices run over the four space-time dimensions while the Latin (lowercase) are purely spatial. 
The signature of the metric is mostly minus [i.e. $(+,\, -,\,-,\, -)$]. } \cite{AC,MG4}
\begin{equation}
S_{g} = \frac{\overline{M}_{P}^2}{8} \int d^{4} x\, \, \sqrt{-\overline{g}} \, \overline{g}^{\mu\nu} \partial_{\mu} h_{i\, j} \, \partial_{\nu} h^{i\, j},\qquad \overline{M}_{P} = \frac{M_{P}}{\sqrt{8 \pi}},
\label{FUND1}
\end{equation}
where $\overline{g}_{\mu\nu}$ is the background metric, $g$ its determinant and $h_{i\,j}$ is the tensor amplitude. By definition the tensor amplitude is both {\em solenoidal
and traceless}. The energy-momentum pseudo-tensor can be 
then derived by functional variation of Eq. (\ref{FUND1}) 
with respect to the background metric and the result is:
\begin{equation}
{\mathcal T}_{\mu}^{\,\,\,\nu} = \frac{\overline{M}_{P}^2}{4}\biggl[ \partial_{\mu} \, h_{i\, j} 
\partial^{\nu} \, h^{i\, j} - \frac{1}{2} \biggl(\overline{g}^{\alpha\beta} \partial_{\alpha} \, h_{i\, j} \,\partial_{\beta} \, h^{i\, j}\biggr) \, \delta_{\mu}^{\,\,\,\nu} \biggr].
\label{FUND2}
\end{equation}
\subsection{The spectral energy density}
We now specialize to the case of conformally flat background geometries that are 
observationally preferred \cite{DA,DB,DC} and set $\overline{g}_{\mu\nu} = a^2(\tau) \eta_{\mu\nu}$ where $a(\tau)$ is the scale factor and $\tau$ is the conformal time 
coordinate; in this case the energy density from the $(0\,0)$ 
component of Eq. (\ref{FUND2}) is given by:
\begin{equation}
\rho_{gw} = \frac{\overline{M}_{P}^2}{8\, a^2} \biggl(\partial_{\tau} h_{i\,j} \,\partial_{\tau} h^{i\,j} +  \partial_{k} h_{i\,j} \,\partial^{k} h^{i\,j}\biggr).
\label{FUND3}
\end{equation}
Equations (\ref{FUND2})--(\ref{FUND3}) are not sufficient to define the spectral energy density since we need 
to introduce an averaging scheme as originally suggested in Refs. \cite{HA,HB}. In what follows we shall 
assume a stochastic average that does not necessarily imply an underlying quantum mechanical interpretation 
even if, as we are going to argue, this is probably the most interesting physical case. Within this approach the gravitational radiation is characterized by two power spectra. In Fourier space the tensor amplitude is given by 
\begin{equation}
h_{i\, j}(\vec{k}, \tau) = \frac{1}{( 2 \pi)^{3/2}} \int d^{3} \, x e^{i \vec{k}\cdot \vec{x}} \, \, h_{i\, j}(\vec{x}, \tau), \qquad \qquad 
h_{i\, j}^{\ast}(\vec{k}, \tau) = h_{i\, j}(- \vec{k}, \tau).
\label{FUND4}
\end{equation}
The expectation values of the Fourier amplitude and of its time derivative are therefore 
defined as\footnote{As usual the two tensor polarizations are defined as $e^{\oplus}_{i\, j}(\hat{k}) = \hat{m}_{i}\, \hat{m}_{j} + \hat{n}_{i}\, \hat{n}_{j}$ and  $e^{\otimes}_{i\, j}(\hat{k}) = \hat{m}_{i}\, \hat{n}_{j} - \hat{n}_{i}\, \hat{m}_{j}$, where $\hat{m}$, $\hat{n}$ and $\hat{k}$ are a triplet of mutually orthogonal unit 
vectors. Note that the sum over the polarizations can be written as $\sum_{\alpha} \, e^{(\alpha)}_{i\, j}(\hat{k}) 
\,\,e^{(\alpha)}_{m\, n}(\hat{k}) = 4 {\mathcal S}_{i\,j\,m\,n}(\hat{k})$ where ${\mathcal S}_{i\,j\,m\,n}(\hat{k})$ is defined in Eq. (\ref{FUND7}).}
\begin{eqnarray}
\langle h_{i\, j}(\vec{k}, \tau) \, h_{m\, n}(\vec{p}, \tau) \rangle = \frac{2\pi^2}{k^3} \,P_{T}(k, \tau) \, {\mathcal S}_{i\,j\,m\,n}(\hat{k}) \, \delta^{(3)}(\vec{k}+ \vec{p}), 
\label{FUND5}\\
\langle \partial_{\tau}h_{i\, j}(\vec{k}, \tau) \, \partial_{\tau} h_{m\, n}(\vec{p}, \tau) \rangle = \frac{2\pi^2}{k^3} \,Q_{T}(k, \tau) \, {\mathcal S}_{i\,j\,m\,n}(\hat{k}) \, \delta^{(3)}(\vec{k}+ \vec{p}), 
\label{FUND6}
\end{eqnarray}
where ${\mathcal S}_{i\,j\,m\, n}(\hat{k})$ is transverse, traceless and can be defined in terms of the projectors $p_{i\,j}(\hat{k})= (\delta_{i\, j} - \hat{k}_{i}\, \hat{k}_{j})$:
\begin{equation}
{\mathcal S}_{i\,j\,m\, n}(\hat{k}) = \frac{1}{4} \bigl[p_{i\,m}(\hat{k}) \, p_{j\, n}(\hat{k}) + p_{i\,n}(\hat{k}) \, p_{j\, m}(\hat{k}) - p_{i\,j}(\hat{k}) \, p_{m\, n}(\hat{k})].
\label{FUND7}
\end{equation}
In terms of the two tensor power spectra $P_{T}(k,\tau)$ and $Q_{T}(k,\tau)$ 
we can obtain the average energy density $\overline{\rho}_{gw} = \langle \rho_{gw}(\vec{x},\tau) \rangle$ and the result follows thanks to Eqs. (\ref{FUND5})--(\ref{FUND6}) after inserting Eq. (\ref{FUND4}) into Eq. (\ref{FUND3}):
\begin{equation}
\overline{\rho}_{gw} = \frac{\overline{M}_{P}^2}{8 \, a^2 }\int \frac{d\, k}{k} \biggl[ Q_{T}(k,\tau) + k^2 P_{T}(k,\tau)\biggr].
\label{FUND8}
\end{equation}
Finally, from Eq. (\ref{FUND8}) we can deduce the spectral energy density in critical units namely
\begin{equation}
\Omega_{gw}(k,\tau) = \frac{1}{\rho_{crit}} \,\, \frac{d \overline{\rho}_{gw}}{d \ln{k}} = \frac{k^2 P_{T}(k,\tau)}{24 \, H^2 \, a^2}\biggl[ 1 + \frac{Q_{T}(k,\tau)}{k^2 \, P_{T}(k,\tau)}\biggr],
\label{FUND9}
\end{equation}
where $\rho_{crit} = 3 \, H^2 \, \overline{M}_{P}^2$.
Equation (\ref{FUND9}) has been purposely written by factoring the contribution of $P_{T}(k,\tau)$ since when all the frequencies are larger than the expansion rate at the corresponding epoch the second term inside the squared bracket is at most of order $1$:
\begin{equation}
\frac{Q_{T}(k,\tau)}{k^2 \, P_{T}(k,\tau)} = 1 + {\mathcal O}\biggl(\frac{a^2\, H^2}{k^2}\biggr), \qquad k \gg a\, H.
\label{FUND10}
\end{equation}
This means that in the high-frequency limit (which is the one discussed here) 
$k^2 \, P_{T}(k,\tau)$ and $Q_{T}(k,\tau)$ have the same weight in Eq. (\ref{FUND9}) and equally 
contribute to the spectral energy density in critical units.
\subsection{The chirp and the spectral amplitude}
The chirp amplitude $h_{c}(k,\tau)$ is defined, from the expectation value 
of the tensor amplitudes; more specifically we write the two-point function as:
\begin{equation}
\langle h_{i\, j}(\vec{x}, \tau) \,\, h^{i\,j}(\vec{x}+ \vec{r}, \tau) \rangle = 2 \int \frac{d\, k}{k} \, h_{c}^2(k,\tau)
 \, j_{0}(k \, r),
\label{FUND11}
\end{equation}
where $j_{0}(k,\, r) = \sin{k\,r}/(k\, r)$. If we now recall Eqs. (\ref{FUND4})--(\ref{FUND5}) 
we can easily deduce that the tensor power spectrum $P_{T}(k,\tau)$ 
is twice the square of the chirp amplitude, i.e. $P_{T}(k,\tau) = 2 \, h_{c}^2(k,\tau)$ which 
means, in particular that Eq. (\ref{FUND5}) can also be written as 
\begin{equation}
\langle h_{i\, j}(\vec{k}, \tau) \, h_{m\, n}(\vec{p}, \tau) \rangle = \frac{4\pi^2}{k^3} \,h_{c}^2(k, \tau) \, {\mathcal S}_{i\,j\,m\,n}(\hat{k}) \, \delta^{(3)}(\vec{k}+ \vec{p}).
\label{FUND12}
\end{equation}
Both the tensor power spectrum and the chirp amplitude are dimensionless. It is 
also possible to introduce another quantity, namely the spectral amplitude $S_{h}(k,\tau)$ 
that can be defined in terms of the chirp amplitude and of the power spectrum:
\begin{equation}
2 \nu\, S_{h}(k,\tau) =  2 h_{c}^2(k,\tau) =  P_{T}(\nu,\tau),\qquad k = 2\pi\nu,
\label{FUND13}
\end{equation}
implying that $S_{h}(k,\tau)$ has dimensions of an inverse frequency (or of  a time). Both the chirp and the spectral amplitudes 
are defined solely in terms of the tensor amplitude. This means, in particular, 
that they apply {\em when the relevant wavelengths are shorter than the Hubble radius}.
Only in this regime the spectral energy density can be explicitly related 
both to $h_{c}^2(k,\tau)$ and to $S_{h}(k,\tau)$. Recalling Eqs. (\ref{FUND9})--(\ref{FUND10}) we can therefore obtain that the chirp amplitude and $\Omega_{gw}(k,\tau)$ are related as:
\begin{equation}
\Omega_{gw}(k,\tau) = \frac{k^2}{12 \, H^2\, a^2} P_{T}(k,\tau) = \frac{k^2}{6 \, H^2\, a^2} \,h_{c}^2(k,\tau).
\label{FUND14}
\end{equation}
If we now use the comoving frequency instead of the comoving wavenumber 
Eq. (\ref{FUND14}) reads: 
\begin{equation}
\Omega_{gw}(\nu,\tau) = \frac{\pi^2 \nu^2}{3 \, H^2\, a^2} P_{T}(\nu,\tau) = \frac{2\,\pi^2\, \nu^2}{3 \, H^2\, a^2} \,h_{c}^2(\nu,\tau),
\label{FUND15}
\end{equation}
where we recall that, in natural units $ k = \omega = 2\pi \,\nu$ (see also Eq. (\ref{FUND13})). The same strategy leads to 
the relation between $S_{h}(\nu,\tau)$ and $\Omega_{gw}(\nu,\tau)$
\begin{equation}
\Omega_{gw}(\nu,\tau) =  \frac{2\,\pi^2\, \nu^3}{3 \, H^2\, a^2} \,S_{h}(\nu,\tau),\qquad \nu\, S_{h}(\nu,\tau) = h_{c}^2(\nu, \tau).
\label{FUND16}
\end{equation}
Equations (\ref{FUND15})--(\ref{FUND16}) are valid in the case of a generic conformal time 
$\tau$. At the present time $\tau_{0}$ we shall normalize the scale factor to $1$ (i.e. $a_{0} \to 1$) 
so that physical and comoving frequencies coincide today but not in the past.  It is already clear 
from Eqs. (\ref{FUND15})--(\ref{FUND16}) that for a given spectral energy density 
the values of $h_{c}(\nu,\tau_{0})$ and $S_{h}(\nu,\tau_{0})$ decrease at high frequencies. This means, for instance, that for a nearly scale-invariant $\Omega_{gw}(\nu,\tau_{0})$ the minimal detectable $h_{c}(\nu,\tau_{0})$ must be, comparatively, much smaller at higher frequencies. The question we ought to address concerns exactly the smallness of the chirp and spectral amplitudes of the potential cosmological signals. As we shall see $h_{c}(\nu,\tau_{0})$ will have to be typically much smaller than the values currently measured in the audio band by the operating detectors.

\subsection{Gravitons and the high-frequency limit}
In the high-frequency limit relic gravitons behave effectively like gas of relativistic (massless) species
whose barotropic index is $1/3$, exactly as in the case of photons.
Indeed, from  Eq. (\ref{FUND1}) the energy momentum pseudo-tensor can also be written as:
\begin{equation}
{\mathcal T}_{\mu}^{\nu} =  \overline{u}_{\mu} \, \overline{u}^{\nu} \rho_{gw} - \overline{{\mathcal P}}_{\mu}^{\,\,\,\,\nu} p_{gw},
\label{FUND17}
\end{equation}
where $\overline{g}_{\mu\nu} \, \overline{u}^{\mu} \, \overline{u}^{\nu} =1$ and 
$\overline{{\mathcal P}}_{\mu\nu} = (\overline{g}_{\mu\nu} - \, \overline{u}_{\mu} \, \overline{u}_{\nu})$.
In Eq. (\ref{FUND13}) $\rho_{gw}$ has the form already given in Eq. (\ref{FUND1}) 
while $p_{gw}$ is 
\begin{equation} 
p_{gw} = - \frac{1}{3} \overline{{\mathcal P}}_{\mu\nu} \, {\mathcal T}^{\mu\nu}
= \frac{\overline{M}_{P}^2}{8 a^2} \biggl[ \partial_{\tau} h_{i\,j} \partial_{\tau} h^{i\, j} - \frac{1}{3} \partial_{\tau} h_{i\,j} \partial^{k} h^{i\, j}\biggr].
\label{FUND18}
\end{equation}
Recalling Eqs. (\ref{FUND4}) and (\ref{FUND5})--(\ref{FUND6}), in the mean pressure
\begin{equation}
\overline{p}_{gw} = \frac{\overline{M}_{P}^2}{8 \, a^2 }\int \frac{d\, k}{k} \biggl[ Q_{T}(k,\tau) - \frac{k^2}{3} P_{T}(k,\tau)\biggr],
\label{FUND19}
\end{equation}
the first term inside the squared bracket of the integrand dominates against the second
in the high-frequency limit $k \gg a\, H$. Thanks to the expression of 
$\overline{\rho}_{gw}$ of Eq. (\ref{FUND8}) we have that 
\begin{equation}
\overline{p}_{gw} = \frac{\overline{\rho}_{gw}}{3} + \frac{\overline{M}_{P}^2}{12\, a^2} 
\int \frac{d\,k}{k} {\mathcal O}\biggl(\frac{a^2\, H^2}{k^2}\biggr),\qquad \qquad k \gg \, a\, H,
\label{FUND20}
\end{equation}
where Eq. (\ref{FUND10}) has been used. Neglecting the subleading contributions 
in the limit $k \gg a\, H$ the barotropic index associated with the high-frequency 
gravitons is $1/3$ as in the case of a relativistic gas of massless species.
In analogy with the spectral energy density 
in critical units introduced in Eq. (\ref{FUND9}) we can also define the 
spectral pressure in critical units, namely 
\begin{equation}
\Sigma_{gw}(k,\tau) = \frac{1}{\rho_{crit}} \,\, \frac{d \overline{p}_{gw}}{d \ln{k}} = \frac{k^2 P_{T}(k,\tau)}{36\, H^2 \, a^2}\biggl[ 1 + {\mathcal O}\biggl(\frac{a^2\, H^2}{k^2}\biggr)\biggr],
\label{FUND21}
\end{equation}
which also implies, in the limit $k \gg a\, H$ that $\Sigma_{gw}(k,\tau) = \Omega_{gw}(k,\tau)/3$. Recalling finally Eqs. (\ref{FUND15})--(\ref{FUND16}) 
the relations of $\Sigma_{gw}(\nu,\tau)$ with the chirp and with the spectral amplitudes 
follow from the same class of considerations. It is interesting to appreciate that 
the results of Eqs. (\ref{FUND20})--(\ref{FUND21}) are not generic since 
in the low-frequency regime $k \ll a\, H$ the barotropic index switches from $1/3$ 
to $-1/3$ \cite{MG4} (see also \cite{AC}) but this result is modified if the background 
contracts instead of expanding. We finally note that the low-frequency limit is 
comparatively more sensitive to the specific form 
of the energy-momentum pseudo-tensor \cite{MG4,HA,HB} but this observation 
will not have any impact in the present case.

\renewcommand{\theequation}{3.\arabic{equation}}
\setcounter{equation}{0}
\section{Direct and indirect bounds at high-frequency}
\label{sec3}
The current bounds on the relic graviton backgrounds at low frequencies 
imply a series of limits on the chirp and spectral amplitudes 
in the MHz and GHz regions. We are now going to examine 
three qualitatively different sets of direct and indirect constraints
on the diffuse backgrounds of relic gravitons. In particular these requirements 
include, in various combinations, the bounds 
from the operating interferometers in the audio band, the limits 
from the pulsar timing arrays in the nHz region and the BBN constraints 
that apply to the whole spectrum of relic gravitons.

\subsection{Limits from wide-band detectors}
Starting from 2004 and 2005 \cite{WB1,WB2} the wide-band detectors provided a series of limits on the relic graviton backgrounds. These limits are customarily phrased in terms of the spectral energy density and for a selected bunch of typical slopes. The parametrization employed by the Ligo, Virgo and Kagra collaborations \cite{WB3} is, in short, the following:
\begin{equation}
\Omega_{gw}(\nu,\tau_{0}) = \overline{\Omega}_{\sigma} \, (\nu/\nu_{ref})^{\sigma}, \qquad \nu_{ref} = {\mathcal O}(60) \mathrm{Hz},\qquad \sigma \geq 0.
\label{BB1}
\end{equation}
The reference frequency appearing in Eq. (\ref{BB1}) is of the order 
of $60$ Hz even if the various sets of bounds involved, through the years, slightly different 
ranges. The constant amplitude $\Omega_{\sigma}$ of Eq. (\ref{BB1}) depends upon the value of $\sigma$. While in Ref. \cite{WB1} the bound on the scale invariant spectrum was 
quite generous (i.e. $\Omega_{0} < 23$), it became more stringent already in Ref. \cite{WB2} 
(i.e. $\overline{\Omega}_{0} < 8.4 \times 10^{-4}$). 
Today the most recent bounds on the relic graviton backgrounds have been reported\footnote{Since they are superseded by the latest constraints, we do not mention, for the sake of conciseness, the subsequent bounds that have been reported between $2005$ and $2019$ by the wide-band detectors (see e.g. \cite{WB4,WB5,WB6,WB7}). All these subsequent bounds have been reviewed, for instance, in Ref. \cite{DF} and they all are less stringent than the ones of Ref. \cite{WB3}.} by Ref. \cite{WB3} and they can be summarized as follows:
\begin{eqnarray}
\overline{\Omega}_{0} &<&  5.8\times 10^{-9}, \qquad 20 \, \mathrm{Hz} < \nu_{ref}< 76.6 \, \mathrm{Hz},
\label{BB20}\\
\overline{\Omega}_{2/3} &<&  3.4\times 10^{-9}, \qquad 20 \, \mathrm{Hz} < \nu_{ref}< 90.6 \, \mathrm{Hz},
\label{BB2a}\\
\overline{\Omega}_{3} &<&  3.9\times 10^{-10}, \qquad 20 \, \mathrm{Hz} < \nu_{ref}< 291.6 \, \mathrm{Hz}.
\label{BB2b}
\end{eqnarray}
Strictly speaking the bounds of Eqs. (\ref{BB20}) and (\ref{BB2a})--(\ref{BB2b}) only apply for a handful 
of spectral indices but, in what follows, we assume that they
also hold in all the intermediate cases and, in particular, when the 
spectrum is nearly scale-invariant. When the value of $\sigma$ increases the bound becomes more restrictive once $\nu_{ref}$ is kept fixed. The three results of  Eqs. (\ref{BB20}) and (\ref{BB2a})--(\ref{BB2b}) can be unified in a single interpolating formula, for  $\log{\overline{\Omega}}_{\sigma}$.
From this result and from Eqs. (\ref{FUND15})--(\ref{FUND16}) we can deduce an interpolating formula for the spectral and for the chirp amplitudes. For instance, in the case of $h_{c}(\nu,\tau_{0})$ we obtain
\begin{equation}
\log{h_{c}(\nu,\tau_{0})} \leq -\,23.794 -0.167\, \sigma -0.335\, \sigma^2 + \log{h_{0}} + (\sigma -1/2) \log{(\nu/\nu_{ref})},
\label{NOT2}
\end{equation}
where we used that $\log{\overline{\Omega}_{\sigma}} \leq -8.236 -0.335 \, \sigma -\, 0.018 \, \sigma^2$.  To avoid potential confusions we note that the bounds coming from the wide-band 
detectors constrain, for technical reasons related with the form of the signal to noise ratio,
only $\Omega_{gw}(\nu,\tau_{0})$ which does depend on the Hubble rate.
We deal instead with $h_{0}^2 \, \Omega_{gw}(\nu,\tau_{0})$ that is independent\footnote{As usual $h_{0}$ is the Hubble rate expressed in units of $100\,\mathrm{Hz}\, \mathrm{km}/\mathrm{Mpc}$ and since $\Omega_{gw}(\nu,\tau_{0})$ denotes the spectral energy density in critical units, $h_{0}^2$ appears 
in its denominator (see, for instance, Eq. (\ref{FUND9}) and comment thereafter). For this reason it is common practice to phrase the discussions directly in terms of  $h_{0}^2 \Omega_{gw}(\nu)$ that is independent of the specific value of $h_{0}$.} of the actual value $h_{0}$. When comparing the limits of wide-band detectors with the other bounds we must therefore specify the range of $h_{0}$ that we broadly take between $0.6$ and $0.7$. This is the reason why $h_{0}$ explicitly appears in Eq. (\ref{NOT2}). 

We conclude this discussion by noting that Eq. (\ref{NOT2}) applies, strictly speaking, for $\nu = {\mathcal O}(\nu_{ref})$ and in this range of frequencies we would have $h_{c}(\nu, \tau_{0}) = {\mathcal O}(10^{-24})$.
For instance if $\sigma =0$, $\nu=100\, \mathrm{Hz}$ and $\nu_{ref}= 60\,\mathrm{Hz}$
Eq. (\ref{NOT2}) implies $h_{c}(\nu,\tau_{0}) \leq 6.75\times 10^{-25}$.  If $\sigma>1$ 
the results are similar provided $\nu = {\mathcal O}(\nu_{ref})$ however, as the frequency 
increases the bound of Eq. (\ref{NOT2}) becomes apparently less restrictive. If we suppose, for instance, that $\nu = {\mathcal O}(\mathrm{MHz})$ and $\sigma =3$, Eq. (\ref{NOT2}) 
would imply that $h_{c}(\nu, \tau_{0}) \leq 3.8 \times 10^{-23}$; this would seem a less restrictive 
upper limit but this way of reasoning is actually misleading. In fact, we would have that for $\nu = {\mathcal O}(\mathrm{MHz})$ a plausible value of $\overline{\Omega}_{3}$  (e.g. 
$\overline{\Omega}_{3} =10^{-10}$, see Eq. (\ref{BB1})) would produce, according to Eq.  (\ref{BB2b}),$\Omega_{gw}(\nu,\tau_{0}) = 4.62 \times 10^{2}$ which is grossly incompatible with few other bounds, including the BBN limit discussed later in this section. 

\subsection{Limits from the pulsar timing arrays}
The pulsar timing arrays (PTA) recently reported an evidence potentially attributed to the relic gravitons.  Using the spectral energy density in critical units as a pivotal variable, this purported signal should tentatively fall in the interval: 
\begin{equation}
10^{-8.86}  \,\,q_{0}^2\,\, <  h_{0}^2\,\Omega_{gw}(\nu)< \,\,q_{0}^2 \,\,10^{-9.88} , \qquad 3\,\,\mathrm{nHz} \, < \nu< \,100 \,\, \mathrm{nHz},
\label{BBP1}
\end{equation}
where the values of $q_{0}$ depend on the specific experimental determination; for instance the Parkes Pulsar Timing Array (PPTA) collaboration \cite{PPTA} suggests $q_{0}= 2.2$; 
the  International Pulsar Timing Array (IPTA) estimates $q_{0}= 2.8$ \cite{IPTA} while the EPTA (European Pulsar Timing Array) \cite{EPTA} gives $q_{0} = 2.95$.  The results of PPTA, IPTA and EPTA seem, at the moment, broadly compatible with the NANOgrav 12.5 yrs data \cite{NANO1} implying $q_{0} =1.92$. If we take the average of the four measurements presented so far we obtain $\overline{q}_{0} = 2.467$ which implies\footnote{If  $q_{0} \to 1$, Eq. (\ref{BBP1}) would imply that the energy density in the nHz domain is comparatively smaller than the Ligo-Virgo-Kagra constraint. However $q_{0}$ {\em is not} $1$.}
\begin{equation} 
10^{-9.09} \biggl(\frac{\overline{q}_{0}}{2.467}\biggr)^2 \leq h_{0}^2 \, \Omega_{gw}(\nu) \leq 10^{-8.07} \biggl(\frac{\overline{q}_{0}}{2.467}\biggr)^2.
\label{BBP2}
\end{equation}
This means that Eq. (\ref{BBP1}) is always {\em more} constraining than Eq. (\ref{BBP2}) 
even if we choose the smallest value of $q_{0}$ which is the one 
associated with the NANOgrav estimate\cite{NANO1}: if $q_{0} =1.92$ we get from Eq. (\ref{BBP2}) that $h_{0}^2 \Omega_{gw}(\nu) \leq 10^{-8.29}$ which is always larger than the value of Eq. (\ref{BBP1}). Even if these bounds are less relevant at higher frequencies it is wise to bear 
them in mind since they may affect indirectly the low-frequency part of a potential signal. In all the cases discussed here the bounds of the PTA are always satisfied.

\subsection{Limits from big-bang nucleosynthesis}
Since the additional relativistic species increase the expansion rate at the nucleosynthesis time by  affecting directly the abundances of the light elements (and in particular of the $^{4}$He), it is possible to set a bound on the possible presence of relic gravitons and this constraint is customarily phrased as\cite{BBN1,BBN2,BBN3}:
\begin{equation}
h_{0}^2  \int_{\nu_{bbn}}^{\nu_{max}}
  \Omega_{gw}(\nu,\tau_{0}) d\ln{\nu} = 5.61 \times 10^{-6} \Delta N_{\nu} 
  \biggl(\frac{h_{0}^2 \Omega_{\gamma0}}{2.47 \times 10^{-5}}\biggr),
\label{BBN1}
\end{equation}
where $\Omega_{\gamma0}$ is the critical fraction of photons 
in the concordance paradigm. In Eq. (\ref{BBN1})
$\nu_{bbn}= {\mathcal O}(10^{-2})$ nHz is the big-bang nucleosynthesis frequency 
and $\nu_{max}$ corresponds instead to the maximal frequency of the spectrum. In the case of the relic gravitons produced within the concordance scenario
$\nu_{max} = {\mathcal O}(100)$ MHz. As we are going to see in section \ref{sec4} $\nu_{max}$ depends on the post-inflationary expansion rate and, for this reason, 
we use the notation $\overline{\nu}_{max}$ to indicate the maximal frequency 
in the context of the concordance paradigm:
\begin{equation}
\overline{\nu}_{max}= 269.33 \,\biggl(\frac{r_{T}}{0.06}\biggr)^{1/4} \,\biggl(\frac{{\mathcal A}_{{\mathcal R}}}{2.41\times 10^{-9}}\biggr)^{1/4} \, \biggl(\frac{h_{0}^2 \, \Omega_{R0}}{4.15 \times 10^{-5}}\biggr)^{1/4} \,\,\,\mathrm{MHz},
\label{BBN1a}
\end{equation}
where ${\mathcal A}_{{\mathcal R}}$ is the amplitude of the power spectrum 
of curvature inhomogeneities at the pivot scale $k_{p}= 0.002 \, \mathrm{Mpc}^{-1}$
and $\Omega_{R0}$ denotes the critical fraction of the massless species 
at the present time. As we are going to see later on $\nu_{max} > \overline{\nu}_{max}$ 
if the post-inflationary expansion rate is slower than radiation 
while $\nu_{max} < \overline{\nu}_{max}$ if the post-inflationary expansion rate is 
faster than radiation. 
Since $\Delta N_{\nu}$ ranges from $\Delta N_{\nu} \leq 0.2$ 
to $\Delta N_{\nu} \leq 1$, Eq. (\ref{BBN1}) can be interpreted as un upper bound on $h_{0}^2 \Omega_{gw}(\nu,\tau_{0})$ 
\begin{equation}
h_{0}^2  \int_{\nu_{bbn}}^{\nu_{max}}
  \Omega_{gw}(\nu,\tau_{0}) d\ln{\nu} < 5.61\times 10^{-6} \biggl(\frac{h_{0}^2 \Omega_{\gamma0}}{2.47 \times 10^{-5}}\biggr). 
\label{BBN2}
\end{equation}
If we consider, for the sake of simplicity, the case of an exactly 
scale-invariant spectral slope and use the notation of Eq. (\ref{BB1})
we have that Eq. (\ref{BBN2}) is logarithmically sensitive 
to the (huge) frequency range:
\begin{equation}
h_{0}^2 \overline{\Omega}_{0} < \frac{5.61\times 10^{-6} \Delta N_{\nu}}{\ln{(\overline{\nu}_{max}/\nu_{bbn}})} \biggl(\frac{h_{0}^2 \Omega_{\gamma0}}{2.47 \times 10^{-5}}\biggr) = 1.22\times 10^{-7},
\label{BBN3}
\end{equation}
where $\overline{\nu}_{max}$ has been already given in Eq. (\ref{BBN1a})
and $\nu_{bbn}$ is:
\begin{equation}
\nu_{bbn}= 2.3\times 10^{-2} \biggl(\frac{g_{\rho}}{10.75}\biggr)^{1/4} \biggl(\frac{T_{bbn}}{\,\,\mathrm{MeV}}\biggr) 
\biggl(\frac{h_{0}^2 \Omega_{R0}}{4.15 \times 10^{-5}}\biggr)^{1/4}\,\,\mathrm{nHz}.
\label{BBN4}
\end{equation}
Note that in Eq. (\ref{BBN4}) $g_{\rho}$ is the effective number of relativistic species associated with the energy density.
If we now compare the result of Eqs. (\ref{BB20}) and (\ref{BBN3}) 
we can see that the current limits from interferometers are $100$ times 
more constraining than the nucleosynthesis bounds even if Eq. (\ref{BB20}) applies in a much narrower frequency range within 
the audio band. If $\sigma \neq 0$ the limit on $\overline{\Omega}_{\sigma}$ 
may become more constraining especially when $\sigma>0$; the 
limit can be phrased as 
\begin{equation}
h_{0}^2 \overline{\Omega}_{\sigma} < \frac{5.61\times 10^{-6}\, \Delta N_{\nu}}{c(\nu_{bbn}, \overline{\nu}_{max})} \,\biggl(\frac{h_{0}^2 \Omega_{\gamma0}}{2.47 \times 10^{-5}}\biggr),
\label{BBN5}
\end{equation}
where $c(\nu_{bbn}, \overline{\nu}_{max})= [(\overline{\nu}_{max}/\nu_{ref})^{\sigma}
- (\overline{\nu}_{bbn}/\nu_{ref})^{\sigma}]/\sigma$. As long as $\sigma >0$ 
the upper limit of integration is far more relevant that the lower 
one and for $\nu_{ref} = {\mathcal O}(60)\mathrm{Hz}$ 
the bound of Eq. (\ref{BBN5}) is generally more constraining than Eqs. (\ref{BB2a})--(\ref{BB2b}).
We could for instance consider, for the sake of illustration, the cases 
$\sigma =3/2$ and $\sigma=3$; in these cases Eq. (\ref{BBN5})
suggests: 
\begin{eqnarray} 
&& \overline{\Omega}_{3/2} \leq 2.56 \times 10^{-15},\qquad \nu_{ref} = 60 \, \, \mathrm{Hz},
\label{BBN6a}\\
&& \overline{\Omega}_{3} \leq 5.61 \times 10^{-25},\qquad \nu_{ref} = 60 \, \, \mathrm{Hz},
\label{BBN6b}
\end{eqnarray}
where we took, for simplicity, $\Delta N_{\nu} =1$ and $h_{0} =0.6$. For lower values 
of $\Delta N_{\nu}$ and larger $h_{0}$ the results of Eqs. (\ref{BBN6a})--(\ref{BBN6b}) 
are marginally more constraining. What made the conditions Eqs. (\ref{BBN6a})--(\ref{BBN6b})
more constraining than Eqs. (\ref{BB2a})--(\ref{BB2b}) is the requirement that 
$\sigma$ remains the same between $\nu_{bbn}$ and $\overline{\nu}_{max}$.
Moreover $\overline{\Omega}_{\sigma}$ just denotes the amplitude of the spectral energy density 
at $\nu_{ref}$. If we would deal instead with the parametrization $\Omega_{gw}(\nu,\tau_{0}) =
\overline{\Omega}_{\sigma}^{(max)} (\nu/\overline{\nu}_{max})^{\sigma}$ we would conclude that the limits 
(\ref{BBN6a})--(\ref{BBN6b})  become, respectively, $\overline{\Omega}_{3/2}^{(max)} \leq 2.33 \times 10^{-5}$
and $\overline{\Omega}_{3}^{(max)} \leq 4.67 \times 10^{-5}$, as qualitatively expected from Eq. (\ref{BBN3}). Indeed, from a qualitative 
viewpoint, when the spectral slope increases (i.e. $\sigma>0$), the constraint of Eq. (\ref{BBN3}) 
mainly comes from the upper limit of integration, i.e. for all the 
frequencies close to $\nu_{max}$ (or $\overline{\nu}_{max}$, in the case of the concordance 
paradigm).

\subsection{The required sensitivity in the high-frequency domain}
The three classes of constraints discussed in the previous subsections 
limit the physical region for the chirp and for the spectral amplitudes so 
that the minimal detectable $h_{c}(\nu,\tau_{0})$ and $S_{h}(\nu,\tau_{0})$ 
 (i.e. $h_{c}^{(min)}$ and $S_{h}^{(min)}$ in what follows) can be already estimated in a 
 preliminary perspective. This estimate is however reduced even further 
 in the context of specific signals, as we are going to see in the 
 following section.  
\begin{figure}[!ht]
\centering
\includegraphics[height=6.5cm]{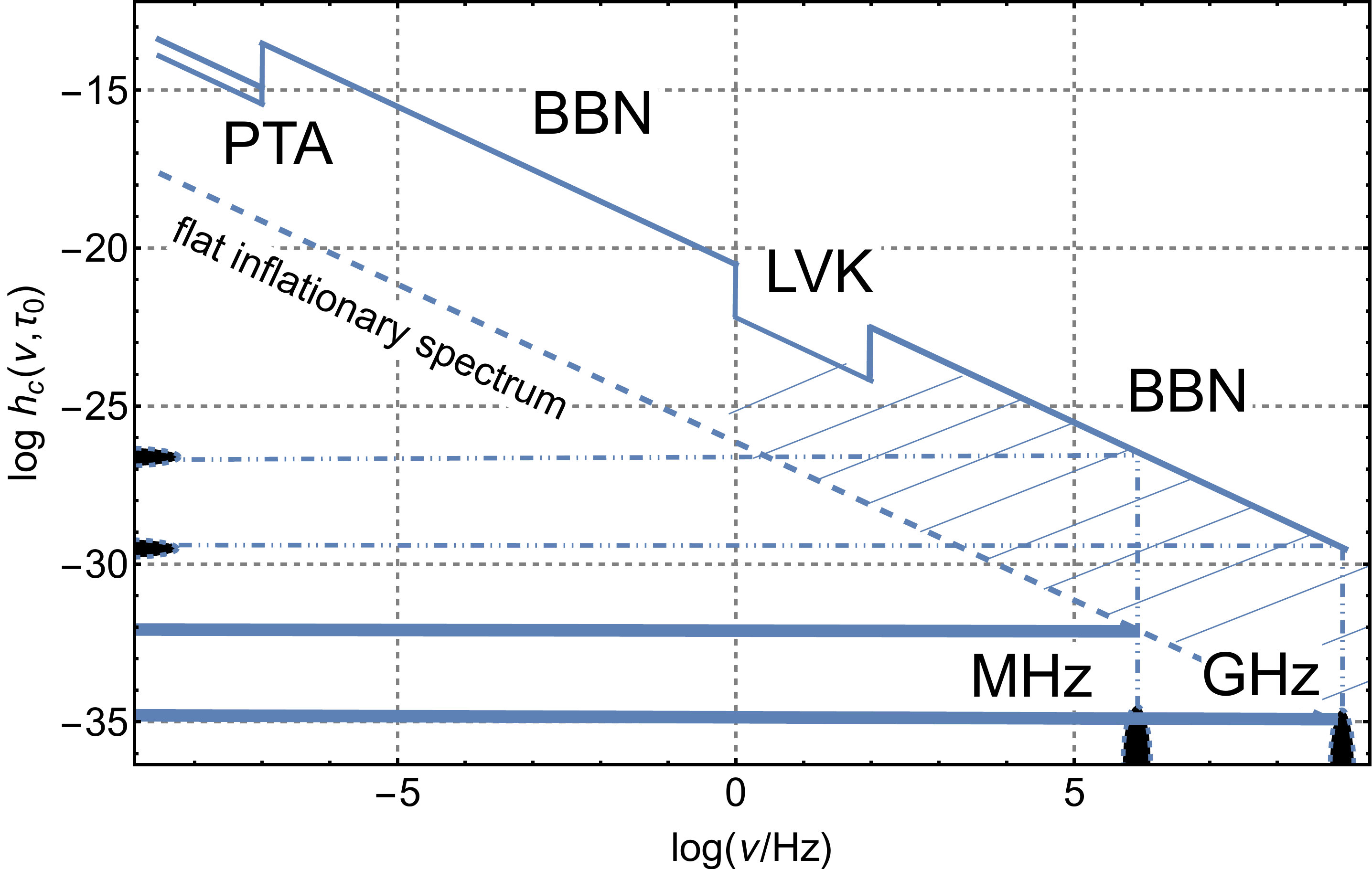}
\caption[a]{The allowed phenomenological region for the chirp 
amplitude is illustrated. Common logarithms are employed on both 
axes. The various acronyms refer to the corresponding constraints already mentioned 
in the text. In particular we report the regions explored by the pulsar timing arrays (PTA), 
the current bounds of the Ligo-Virgo-Kagra collaboration (LVK) and the big-bang nucleosynthesis (BBN) limit. For the present discussion the relevant frequency range starts around $100$ kHz and extends above the GHz; in this range the interesting values of $h_{c}(\nu,\tau_{0})$ 
fall within the diagonal stripe bounded by the phenomenological constraints and by 
the conventional inflationary signal.}
\label{FIGURE1}      
\end{figure}
For this purpose, both in Figs.  \ref{FIGURE1} and \ref{FIGURE2} we consider the case of a flat spectrum 
with arbitrary amplitude with the aim of determining the values of $h_{c}(\nu,\tau_{0})$ and $S_{h}(\nu,\tau_{0})$ that are 
generally compatible with the current bounds. In particular 
the double line in the leftmost region of both plots corresponds to the PTA requirement in the case $\overline{q}_{0}= 2.47$ while the dashed line illustrates the inflationary signal. The assumption here is that 
all the modes reenter during radiation and this means, combining the various 
sources of damping, that, at most $h_{0}^2 \Omega_{gw}(\nu, \tau_{0}) = {\mathcal O}(10^{-17})$ for $\nu > {\mathcal O}(\mathrm{Hz})$. If there are devices operating in the MHz or GHz regions we may now 
ask what the sensitivity goals should be if the aim is the detection 
of a potential cosmological signal. According to Fig. \ref{FIGURE1} we have that the 
typical sensitivity in $h_{c}(\nu,\tau_{0})$ should be 
\begin{equation}
{\mathcal O}(10^{-32}) \leq h_{c}(\nu,\tau_{0}) \leq {\mathcal O}(10^{-27}), \qquad \nu = {\mathcal O}(\mathrm{MHz}).
\label{FB1}
\end{equation}
Since for flat spectral energy density $h_{c}(\nu,\tau_{0})$ 
scales as the inverse frequency, in the GHz region 
we should have instead that 
\begin{equation}
{\mathcal O}(10^{-35}) \leq h_{c}(\nu,\tau_{0}) \leq {\mathcal O}(10^{-30}), \qquad \nu = {\mathcal O}(\mathrm{GHz}).
\label{FB2}
\end{equation}
The figures of Eqs. (\ref{FB1})--(\ref{FB2}) must be compared with the trend of the 
sensitivities of the various instruments reported in the literature. For instance 
coupled microwave cavities with superconducting walls
 \cite{EB, cav1a, EC,EC2} could reach $h_{c}(\nu, \tau_{0}) = {\mathcal O}(10^{-17})$ 
 in the mid 1980s \cite{ED,EE,EF,EG}. The potential improvements in the quality 
 factors of the cavities suggested the possibility of reaching $h_{c}(\nu, \tau_{0}) = {\mathcal O}(10^{-21})$ \cite{EF,EG} (see also \cite{MG2}). Further improvements along the same directions  might suggest that today we could reach, with some luck, the region  $h_{c}(\nu, \tau_{0}) = {\mathcal O}(10^{-24})$ for typical frequencies $\nu \geq \mathrm{MHz}$. Figure \ref{FIGURE1} and the results of 
 Eqs. (\ref{FB1})--(\ref{FB2}) already clarify that even reaching (in the MHz region) 
 the current sensitivity of the wide-band detectors (operating in the audio band)
 is insufficient to cut through the region of a potential signal associated 
 with the relic gravitons. This last statement follows directly from the 
 discussion of Ligo-Virgo-Kagra bound  \cite{WB3} that implies 
 $h_{c}(\nu, \tau_{0}) = {\mathcal O}(10^{-24})$ for $\nu = {\mathcal O}(60)$ Hz.
\begin{figure}[!ht]
\centering
\includegraphics[height=6.5cm]{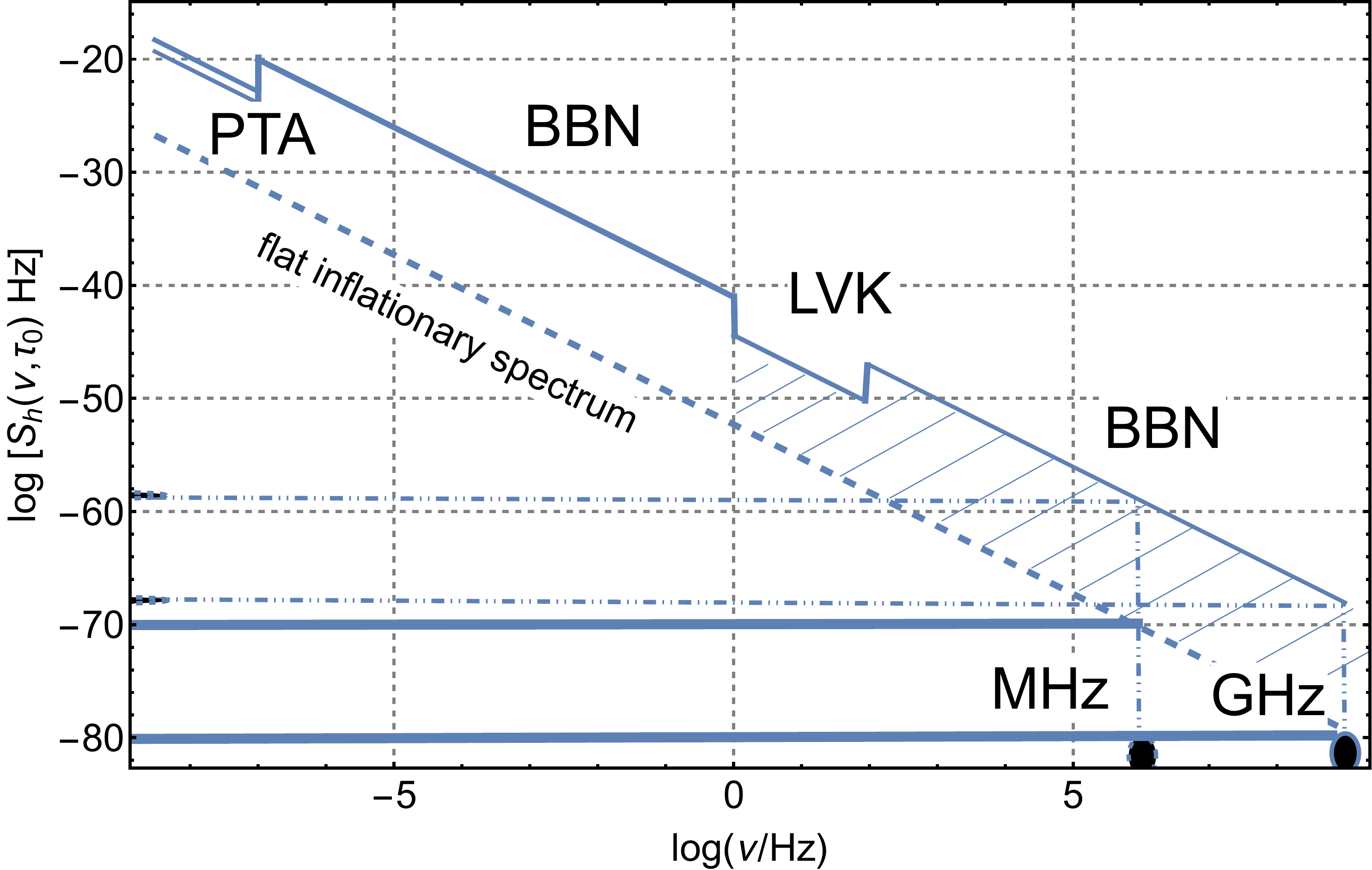}
\caption[a]{We illustrate the spectral 
amplitude $S_{h}(\nu, \tau_{0})$ and its allowed phenomenological region. Common logarithms are employed on both 
axes and all the other notations reproduce exactly the ones of Fig. \ref{FIGURE1}.
Note, however, that the frequency scaling of the spectral and of the chirp amplitudes are markedly different.}
\label{FIGURE2}      
\end{figure}
The same results of Fig. \ref{FIGURE1} can be rephrased in terms of the spectral amplitude.
While $h_{c}(\nu, \tau_{0})$ the spectral amplitude is measured in inverse Hz (or seconds).
According to Fig. \ref{FIGURE2} we have that the 
typical sensitivity in $S_{h}(\nu,\tau_{0})$ should be
\begin{equation}
{\mathcal O}(10^{-69})\,\,\, \mathrm{Hz}^{-1}  \leq S_{h}(\nu,\tau_{0}) \leq {\mathcal O}(10^{-60})\,\,\, \mathrm{Hz}^{-1}, \qquad \nu = {\mathcal O}(\mathrm{MHz}),
\label{FB3}
\end{equation}
for a cosmological signal in the MHz region. If we move from the MHz to the GHz 
the minimal detectable $S_{h}(\nu,\tau_{0})$ gets even smaller 
and it falls between $10^{-68}\,\,\, \mathrm{Hz}^{-1}$ and $10^{-80}\,\,\, \mathrm{Hz}^{-1}$:
\begin{equation}
{\mathcal O}(10^{-79})\,\,\, \mathrm{Hz}^{-1}  \leq S_{h}(\nu,\tau_{0}) \leq {\mathcal O}(10^{-69})\,\,\, \mathrm{Hz}^{-1}, \qquad \nu = {\mathcal O}(\mathrm{GHz}).
\label{FB4}
\end{equation}
Instead of using the spectral amplitude  there are some who prefer to use $\sqrt{S_{h}(\nu,\tau_{0})}$ (measured in units $\mathrm{Hz}^{-1/2}$); we shall use indifferently either $S_{h}(\nu,\tau_{0})$ or its square root depending on the convenience.
In summary we can say that between the MHz and the GHz the minimal detectable chirp amplitude must be, respectively, $h_{c}^{(min)} = {\mathcal O}(10^{-27})$ 
(or smaller) and $h_{c}^{(min)}  = {\mathcal O}(10^{-30})$ (or smaller). Similarly for the spectral amplitude we should have $S_{h}^{(min)} \leq {\mathcal O}(10^{-58})\, \mathrm{Hz}^{-1}$ n the MHz domain and   $S_{h}^{(min)} \leq {\mathcal O}(10^{-67})\, \mathrm{Hz}^{-1}$ in the GHz range. The general requirements stemming from the current phenomenological bounds are complemented by more concrete physical considerations in section \ref{sec4}. 

\renewcommand{\theequation}{4.\arabic{equation}}
\setcounter{equation}{0}
\section{Thermal and non-thermal gravitons}
\label{sec4}
The dashed lines in Figs. \ref{FIGURE1} and \ref{FIGURE2} correspond 
to the spectral energy density of the concordance scenario where $h_{0}^2 \Omega_{gw}(\nu,\tau_{0})$ is approximately scale-invariant. This means that the averaged multiplicity $\overline{n}(\nu,\tau_{0})$ of the produced gravitons is strongly non-thermal and it approximately scales as $\nu^{-4}$ in the MHz--GHz domain. The frequency dependence of the average multiplicity follows from the flatness of the spectral energy density; indeed we can always write that $d \overline{\rho}_{gw} = 2 \, k \, \overline{n}(k,\tau_{0})\, d^{3}k/(2\pi^3)$ where the factor $2$ counts the two polarizations of the graviton. The spectral energy density in critical units depends on the averaged multiplicity and the result is:
\begin{equation}
h_{0}^2 \Omega_{gw}(\nu, \tau_{0}) = \frac{h_{0}^2}{\rho_{crit}} \frac{d \overline{\rho}_{gw}}{d\ln{\nu}} = \frac{128\, \pi^3}{3} \frac{\nu^4 }{H_{0}^2 \, M_{P}^2} \, \overline{n}(\nu,\tau_{0}),
\label{NP1}
\end{equation}
where we used that $k = 2\,\pi\,\nu$; moreover we traded $\overline{M}_{P}$ for $M_{P}$ according to the relation $\overline{M}_{P} = M_{P}/\sqrt{8 \pi}$ already mentioned in Eq. (\ref{FUND1}). In Eq. (\ref{NP1}) the present value of the scale factor is normalized as $a_{0} =1$; this means that, at the present time, comoving and physical temperatures coincide; the same observation holds also for the frequencies and for the wavenumbers. We now recall that in the concordance paradigm the spectral energy density appearing in Eq. (\ref{NP1}) can also be expressed, with compact notations, as \cite{DF}: 
\begin{eqnarray}
h_{0}^2 \Omega_{gw}(\nu, \tau_{0}) &=& {\mathcal N}_{\rho} \, r_{T}(\nu_{p}) \biggl(\frac{\nu}{\nu_{p}}\biggr)^{n_{T}} {\mathcal U}_{low}^2(\nu/\nu_{eq}),
\nonumber\\
{\mathcal N}_{\rho} &=& 4.165\times10^{-15} \biggl(\frac{h_{0}^2 \Omega_{R0}}{4.15\times 10^{-15}}\biggr),
\label{NP2}
\end{eqnarray}
where $\nu_{p} = k_{p}/(2\pi) = 3.092\, \mathrm{aHz}$ and the spectral index $n_{T}$ can be estimated from the consistency relation as $n_{T} = - r_{T}/8 \ll 1$; $\Omega_{R0}$ is the total critical fraction associated with the relativistic species (as implied by the minimal version of the concordance paradigm). In Eq. (\ref{NP2}) ${\mathcal U}_{low}(\nu/\nu_{eq})$ is the low-frequency transfer function that goes to $1$ for typical frequencies larger than the equality frequency $\nu_{eq} = 113.182 \,\, (h_{0}^2 \, \Omega_{M\,0}) \, \mathrm{aHz}$ while it scales as $\nu^{-2}$ in the opposite limit
\footnote{The late-time effects associated with the free-streaming of the neutrinos \cite{STRNU1,STRNU2,STRNU3} are formally included in the expression of ${\mathcal U}_{low}(\nu/\nu_{eq})$ but they are not essential for the present ends since we are interested in the high-frequency range (see also Ref. \cite{DF} and discussion therein).}. For  $\nu > \mathrm{Hz}$ we can therefore estimate  $h_{0}^2 \Omega_{gw}(\nu, \tau_{0})= {\mathcal O}(10^{-16.5}) (\nu/\nu_{p})^{n_{T}}$ and then Eq. (\ref{NP1}) suggests that the average multiplicity of the gravitons is given by:
\begin{equation}
\overline{n}(\nu,\tau_{0}) = {\mathcal O}(10^{19.78}) \,
 \biggl(\frac{\nu}{\mathrm{kHz}}\biggr)^{n_{T} - 4}.
\label{NP3}
\end{equation}
It then follows that since $n_{T} = -\,r_{T}/8$ and $r_{T} \leq {\mathcal O}(0.06)$ \cite{DA,DB,DC} the signal of the concordance 
paradigm in the kHz range consists of roughly $10^{20}$ pairs of gravitons so the averaged multiplicity scales as $\nu^{-4}$ with the comoving frequency. But this means that as $\nu$ increases we will necessarily hit a typical frequency 
where $\overline{n}(\nu, \tau_{0}) \to 1$; this is the maximal frequency of the spectrum corresponding to the production of a single 
pair of gravitons with opposite comoving three-momenta. From Eqs. (\ref{NP1})--(\ref{NP2}) we then heuristically obtain an estimate of the averaged multiplicity:
\begin{equation}
\overline{n}(\nu,\tau_{0}) \simeq \biggl(\frac{\nu}{\overline{\nu}_{max}}\biggr)^{-4}, \qquad \qquad \overline{\nu}_{max} = {\mathcal O}(200) \, \, \mathrm{MHz}.
\label{NP4}
\end{equation}
The frequency $\overline{\nu}_{max}$ has been already introduced
in Eq. (\ref{BBN1a}) and it is ultimately associated with the curvature scale at the end of inflation;
the heuristic estimate of Eq. (\ref{NP4}) can be corroborated by a more direct derivation (see below Eq. (\ref{NT18}) and discussion therein).  
If Eq. (\ref{NP4}) is evaluated for $\nu = {\mathcal O}(\mathrm{kHz})$ we obtain that $\overline{n}(\nu,\tau_{0}) = {\mathcal O}(10^{20})$ which is the figure appearing Eq. (\ref{NP3}). In the non-thermal case gravitons are typically 
produced from the vacuum and this is what happens when the total number of $e$-folds is larger than ${\mathcal O}(60)$
as we shall assume throughout for the illustrative purposes of the present discussion. The smallness of the high-frequency signal associated with the standard inflationary spectrum (illustrated 
with the dashed line in Figs. \ref{FIGURE1} and \ref{FIGURE2}) ultimately follows from Eqs. (\ref{NP3})--(\ref{NP4}).  

There are however situations where $\overline{n}(\nu,\tau_{0})$ at high-frequencies  is less suppressed than in the concordance paradigm. The simplest example is the case of a thermal  background where the averaged multiplicity corresponds 
to the Bose-Einstein occupation number i. e. $\overline{n}(k,\tau_{0})=(e^{k/T_{g\,0}} -1)^{-1}$,
where $T_{g\,0}$ is the present temperature of the gravitons. Since 
$\overline{n}(\nu,\tau_{0})$ scales as $(T_{g\,0}/\nu)$ at low-frequencies,   
Eq. (\ref{NP1}) suggests that, in the same limit, $h_{0}^2 \Omega_{gw}(\nu, \tau_{0})$ scales as $\nu^3$. A thermal spectrum of gravitons might also have a geometric origin as argued long ago by Parker \cite{park1} (see also \cite{park2,park3}). This idea can now be realized, in a more recent  perspective, in some classes of bouncing scenarios 
where the averaged multiplicity grows at low-frequencies as $\nu^{3}$ it is exponentially suppressed above the maximal frequency (see, for instance, \cite{DF} and discussion therein). It can also happen that the averaged multiplicity of non-thermal gravitons at high-frequency is much less suppressed than in the case of Eqs. (\ref{NP3})--(\ref{NP4}) even if the 
underlying signal is fully compatible with the patterns of the concordance 
paradigm at low-frequencies. Instead of simply estimating $h_{c}(\nu,\tau_{0})$ and $S_{h}(\nu,\tau_{0})$ from the current constraints (as discussed in section \ref{sec3}) it is now interesting to analyze the spectral and the chirp amplitudes when the averaged multiplicity deviates substantially from Eqs. (\ref{NP3})--(\ref{NP4}).

\subsection{Thermal gravitons}
\subsubsection{Graviton decoupling}
The evolution of the plasma may produce a thermal spectrum 
when relic gravitons decouple. Since the cross section for the interaction of two gravitons $\sigma_{g}$ and the reaction rate $\Gamma_{g}$ are given, respectively, by: 
\begin{equation}
\sigma_{g} = \ell_{P}^2 \biggl(\frac{T}{\overline{M}_{P}}\biggr)^2, \qquad  \Gamma_{int} \simeq \sigma_{g} T^3,\qquad \ell_{P}= 1/\overline{M}_{P},
\end{equation}
we have that $\Gamma_{int} < H$ provided $T < \overline{M}_{P}$. The decoupling temperature of the gravitons is smaller than the current temperature of the photons $T_{\gamma 0}$ and a similar hierarchy of temperatures also arises in the case of (massless) neutrinos\footnote{It is well established  
that $T_{\gamma 0}$ (i.e. the CMB temperature) is given by $T_{\gamma 0} = (2.72548 \pm 0.00057) \, \mathrm{K}. $. \cite{CMB2,CMB4,CMB5} and this is the value assume throughout 
the discussion for actual estimates.}. Denoting, respectively,  with $T_{\gamma}(\tau_{b})$ and $T_{\gamma}(\tau_{a})$ the photon temperatures before and after graviton decoupling we have, from the evolution of the entropy density, that
\begin{equation}
g_{s}(t_{b}) \,a^3(\tau_{b}) \, T_{\gamma}^{3}(\tau_{b}) = g_{s}(\tau_{a}) \,a^3(\tau_{a}) \, T_{\gamma}^{3}(\tau_{a}),
\label{DF1}
\end{equation}
where $g_{s}(\tau)$ is the effective number of relativistic species associated with the entropy density; this 
number does not necessarily coincide with the effective number of relativistic species appearing in the 
energy density and already introduced in Eq. (\ref{BBN4}). Before graviton decoupling 
the total number of relativistic degrees of freedom is given by $g_{s}(\tau_{b}) = 2 + g_{s}(\tau_{i})$
(where, as usual, the $2$ counts the two polarizations of the graviton and $g_{s}(\tau_{i})$ corresponds 
to the total number of relativistic degrees of freedom of the particle physics model). On a general 
ground we shall be assuming that  $g_{s}(\tau_{i}) \geq 106.75$ since $106.75$ is obtained in the context of the standard description of strong and electroweak interactions\footnote{When all 
the species of the plasma are in local thermal equilibrium  $g_{s}(\tau_{i})=g_{\rho}(\tau_{i})$. }. The explicit value of $g_{s}(\tau_{a})$ can be instead written as:
\begin{equation}
g_{s}(\tau_{a}) = g_{s}(\tau_{0}) + 2 \biggl[\frac{T_{g}(\tau_{a})}{T_{\gamma}(\tau_{a})}\biggr]^3,
\label{DF3}
\end{equation}
where we took into account that, in general, the temperatures of the gravitons and of 
the photons {\em after} graviton decoupling are different; $g_{s}(\tau_{0})$ 
measures the number of degrees of freedom associated with the entropy density at the present time.
A simple counting that includes three species of massless neutrinos\footnote{In the minimal version of the concordance 
scenario the neutrinos are massless. Even if they are not massless the difference is immaterial for the present purposes where 
the relevant point is the ratio between the neutrino mass and the MeV scale; such a quantity is anyway ${\mathcal O}(10^{-6})$.} suggests that
$g_{s}(\tau_{0}) = 2 + (7/8) \times 3\times (4/11) =3.91$
where now the $2$ counts the two polarzations of the photon while the second term counts 
the two helicities of the neutrinos (whose associated temperature is $(4/11)^{1/3}$ times 
smaller than the one of the photons). We can finally 
recall that the temperatures of the gravitons before and after decoupling 
are related as:
\begin{equation}
a^{3}(\tau_{b}) \,\, T_{g}^3(\tau_{b}) = a^{3}(\tau_{a}) \,\, T_{g}^3(\tau_{a}).
\label{DF4}
\end{equation}
If we now divide Eq. (\ref{DF1}) by Eq. (\ref{DF4}) and take into account that 
before graviton decoupling the temperatures of photons and gravitons coincide, i.e. 
$T_{\gamma}(\tau_{b}) = T_{g}(\tau_{b})$. We then arrive at the following expression
\begin{equation}
g_{s}(\tau_{i}) + 2 = \biggl[g_{s}(\tau_{0}) + 2 \frac{T^3_{g}(\tau_{a})}{T^3_{\gamma}(\tau_{a})}\biggr]
\frac{T^3_{\gamma}(\tau_{a})}{T^3_{g}(\tau_{a})},
\label{DF5}
\end{equation}
implying that after graviton decoupling (and in particular at the present time) 
the temperature of the gravitons is always smaller than the one of the photons 
\begin{equation}
T_{g\, 0} = \epsilon_{g} \, T_{\gamma\, 0}, \qquad \epsilon_{g} = \biggl[\frac{g_{s}(\tau_{0})}{g_{s}(\tau_{i})}\biggr]^{1/3} < 1,
\label{DF6}
\end{equation} 
where, by definition, $T_{g\, 0}= T_{g}(\tau_{0})$. Since $g_{s}(\tau_{i}) \geq 106.75$, $\epsilon_{g}$ is always smaller than $1$ and, in particular,
\begin{equation}
\epsilon_{g} \leq 0.3321 \,\biggl[\frac{g_{s}(t_{0})}{3.91}\biggr]^{1/3} \,\,\,\biggl[\frac{g_{s}(t_{i})}{106.75}\biggr]^{-1/3},
\label{DF6a}
\end{equation}
which also implies that 
\begin{equation}
T_{g\, 0} = \epsilon_{g} \,\, T_{\gamma\,0} \leq 0.9051\,\, \biggl(\frac{T_{\gamma\, 0}}{2.72548\, \mathrm{K}}\biggr)\,\,\, \mathrm{K}.
\label{DF6b}
\end{equation}

\subsubsection{The graviton spectrum and its temperature}
Since  $T_{g\,0} \leq 0.9051$ K and  $T_{g\, 0} < T_{\gamma\, 0}$  the graviton black-body is always suppressed in comparison with the photon black-body by a factor  $\epsilon_{g}^4$ and the spectral 
energy density in critical units becomes therefore:
\begin{equation}
h_{0}^2 \Omega_{gw} (\nu, \tau_{0}) = \frac{15}{\pi^4} \,\,h_{0}^2 \Omega_{\gamma 0}\,\,\epsilon_{g}^4 \,  F(x_{g}),\qquad F(x_{g}) = \frac{x_{g}^4}{e^{x_{g}} -1},
\label{gravBB6}
\end{equation}
where $x_{g}= k/T_{g\, 0}= 2 \pi \nu/T_{g\,0}$ ultimately depends on the frequency so that we can write
\begin{equation}
x_{g}(\nu) = \frac{0.0176}{\epsilon_{g}} \, \biggl(\frac{\nu}{\mathrm{GHz}}\biggr) \biggl(\frac{T_{\gamma0}}{2.7254\,\, \mathrm{K}}\biggr)^{-1}.
\label{gravBB3ab}
\end{equation}
The maximal frequency of the spectral energy density coincides with the maximum of $F[x_{g}(\nu)]$
and since $x_{g}(\nu_{max,\,g}) =3.9206$ we also have that\footnote{In this section we denoted 
the maximal frequency of the thermal gravitons by $\nu_{max,\, g}$ just to distinguish it 
from the photon case indicated by $\nu_{max,\, \gamma}$.}
\begin{equation}
\nu_{max,\,\, g} = 73.943 \biggl(\frac{T_{g\, 0}}{0.9051 \, \mathrm{K}}\biggr) \, \mathrm{GHz}.
\label{gravBB7}
\end{equation}
while in the case of the photons the typical frequency that maximizes 
the corresponding spectral energy density is given by $\nu_{max,\,\,\gamma} = 226.643\,\, \mathrm{GHz}$.
\begin{figure}[!ht]
\centering
\includegraphics[height=6.5cm]{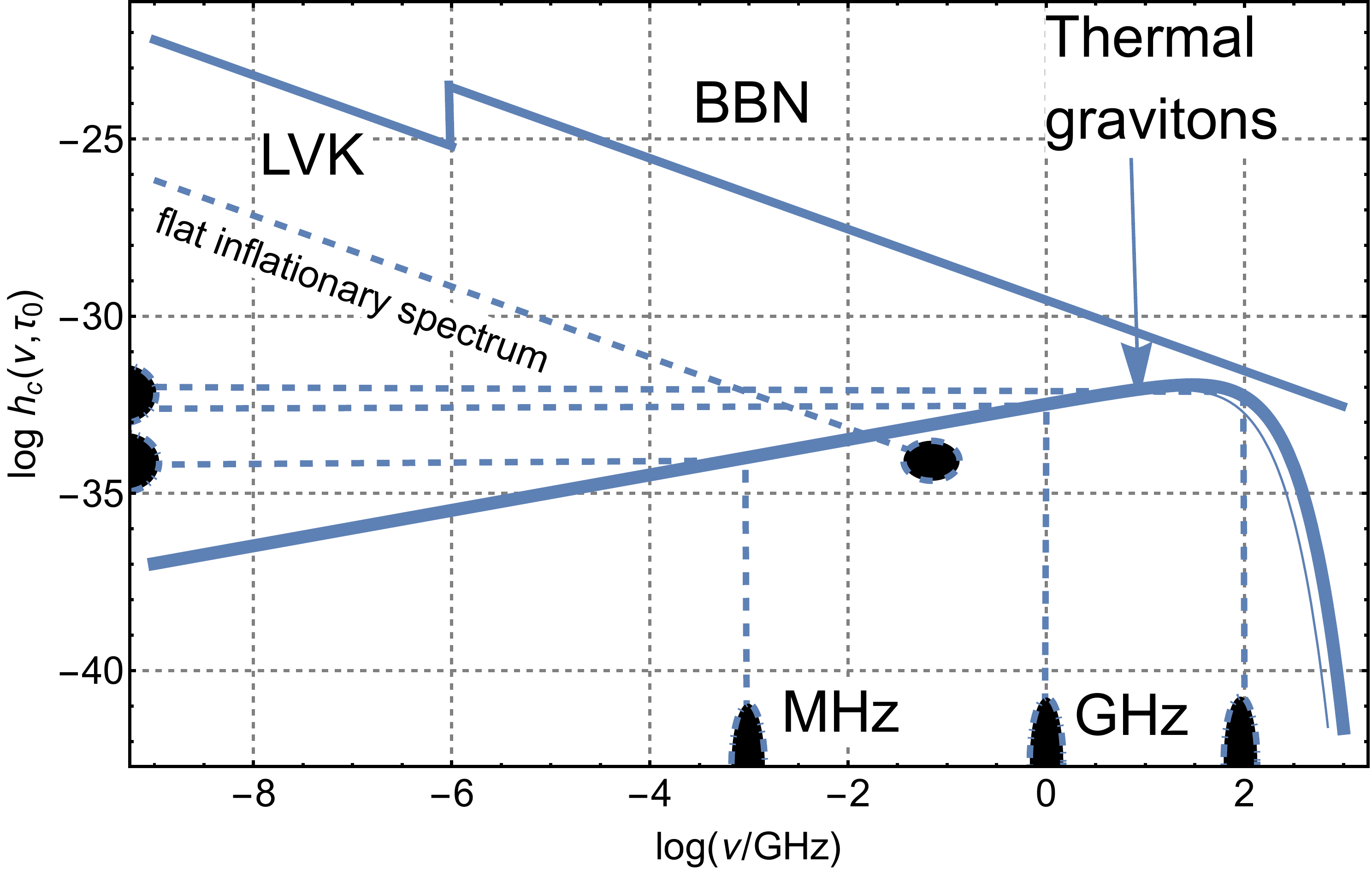}
\caption[a]{The allowed phenomenological region for the chirp 
amplitude is illustrated. Common logarithms are employed on both 
axes. The various acronyms refer to the corresponding constraints already mentioned 
in the text as well as in Figs. \ref{FIGURE2} and \ref{FIGURE3}. In particular we report the regions explored by the pulsar timing arrays (PTA), the current bounds of the Ligo-Virgo-Kagra collaboration (LVK) and the big-bang nucleosynthesis (BBN) limit. For the present ends the relevant frequency range starts around $100$ kHz and extends above the GHz; in this domain the interesting values of $h_{c}(\nu,\tau_{0})$ fall within the diagonal stripe bounded by the phenomenological constraints and by 
the standard inflationary signal.}
\label{FIGURE3}      
\end{figure}
Evaluated at their respective maxima the energy densities of the cosmic 
photons and gravitons are then: 
\begin{eqnarray}
h_{0}^2 \Omega_{\gamma} (\nu_{max,\,\, \gamma} , \tau_{0}) &=& 1.819 \times 10^{-5}  \biggl(\frac{T_{\gamma\, 0}}{2.7254\, \mathrm{K}}\biggr)^{4},
\nonumber\\
h_{0}^2 \Omega_{gw} (\nu_{max,\,\, g} , \tau_{0}) &\leq& 2.213 \times10^{-7}\,\,  \biggl(\frac{T_{g\, 0}}{0.9051\, \mathrm{K}}\biggr)^{4}.
\label{gravBB8}
\end{eqnarray}
While the frequencies of the maxima are comparable (within one order of magnitude), the 
spectral energy density of the gravitons is always smaller than in the case of the photons 
since $h_{0}^2 \Omega_{gw} (\nu , \tau_{0}) \leq {\mathcal O}(10^{-7})$ as long as  $\epsilon_{g} \leq 0.3321$ and $g_{s}(\tau_{i}) \geq 106.75$.
\begin{figure}[!ht]
\centering
\includegraphics[height=6.5cm]{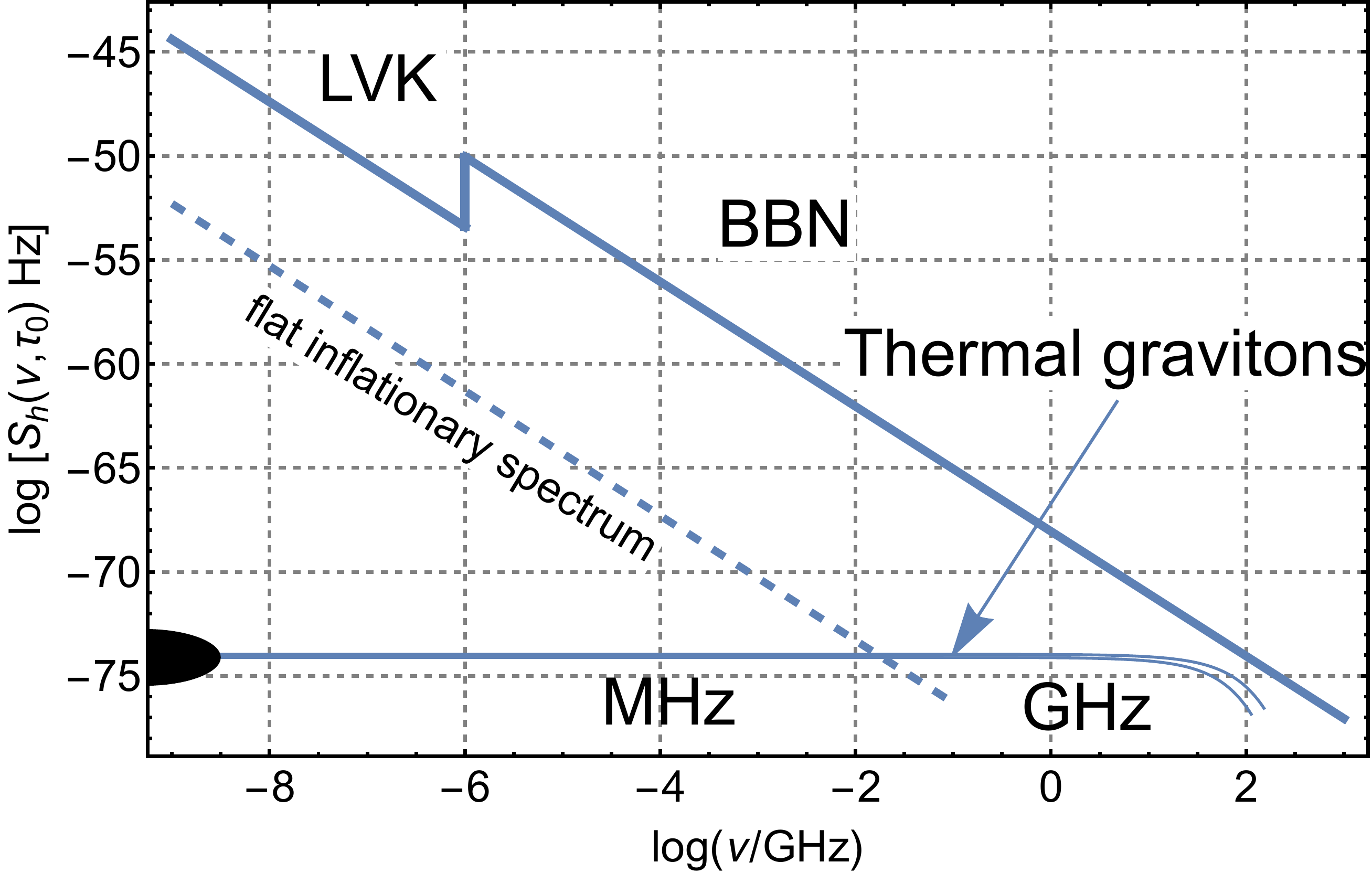}
\caption[a]{We illustrate the spectral amplitude and its allowed phenomenological range with the same notations of Fig. \ref{FIGURE3}; common logarithms are employed on both axes. By comparing 
the two plots we see that while in Fig. \ref{FIGURE3} the chirp amplitude associated with the 
thermal gravitons {\em increases}, the spectral amplitude {\em remains constant} as a function of the comoving frequency. See also, in this respect, the Eqs. (\ref{gravBB11})--(\ref{gravBB12}).}
\label{FIGURE4}
\end{figure}

\subsubsection{The chirp and the spectral amplitudes in the thermal case}
In Fig. \ref{FIGURE3} the allowed phenomenological region for the spectral amplitude is illustrated together with the signal associated with the thermal gravitons. In contrast with the results of the concordance paradigm, as the frequency increases $h_{c}(\nu, \tau_{0})$ gets larger and if 
Eq. (\ref{NP1}) is inserted into Eq.  (\ref{FUND16}), two general expressions for $h_{c}(\nu,\tau_{0})$ and $S_{h}(\nu, \tau_{0})$ can be derived in terms of the averaged multiplicity $\overline{n}(\nu, \tau_{0})$ of produced pairs of gravitons:
\begin{eqnarray}
h_{c}(\nu,\tau_{0}) &=& 7.643 \times 10^{-34} \biggl(\frac{\nu}{\mathrm{GHz}}\biggr) \,\,
\sqrt{\overline{n}(\nu,\tau_{0})}\,,
\label{gravBB9}\\
S_{h}(\nu,\tau_{0}) &=& 5.841 \times 10^{-76} \biggl(\frac{\nu}{\mathrm{GHz}}\biggr) \,\,\overline{n}(\nu,\tau_{0})\,\,\, \mathrm{Hz}^{-1}.
\label{gravBB10}
\end{eqnarray}
Since the Bose-Einstein occupation number scales as $(T_{g\,0}/\nu)$ at low frequencies we see from Eqs. (\ref{gravBB9})--(\ref{gravBB10}) that $h_{c}(\nu,\tau_{0})$ increases as $\sqrt{\nu}$ while $S_{h}(\nu,\tau_{0})$ is constant up to the maximal frequency of the spectrum; both scalings are illustrated, respectively,  in Figs. \ref{FIGURE3} and \ref{FIGURE4}. The 
explicit forms of the chirp and spectral amplitudes 
in the thermal case are therefore:
\begin{eqnarray}
h_{c}(\nu,\tau_{0}) &=& 3.317 \times 10^{-33} \sqrt{\frac{\nu}{\mathrm{GHz}}} \, \sqrt{\frac{h_{0}^2 \, \Omega_{\gamma\, 0}}{2.47 \times 10^{-5}}} \, \sqrt{\frac{\epsilon_{g}}{0.3321}}\,\,\biggl(\frac{T_{\gamma0}}{2.72548\,\, \mathrm{K}}\biggr)^{-3/2},
\label{gravBB11}\\
S_{h}(\nu,\tau_{0}) &=& 1.100 \times 10^{-74} \biggl(\frac{\epsilon_{g}}{0.3321}\biggr) \biggl(\frac{h_{0}^2 \, \Omega_{\gamma\, 0}}{2.47 \times 10^{-5}}\biggr)\,\, \biggl(\frac{T_{\gamma0}}{2.7254\,\, \mathrm{K}}\biggr)^{-3}\,\,\mathrm{Hz}^{-1},
\label{gravBB12}
\end{eqnarray}
where, as already stressed, $\epsilon_{g} \leq 0.3321$ as long as $g_{s}(\tau_{i}) \geq 106.75$.  The figures of Eqs. (\ref{gravBB11})--(\ref{gravBB12}) demonstrate, for instance, that the potential target of 
high-frequency detectors in the MHz and GHz regions cannot be chirp amplitudes ${\mathcal O}(10^{-20})$ or even ${\mathcal O}(10^{-24}$) (i.e. comparable with the current sensitivity of interferometers in the audio band). To put it mildly this requirement would be too generous. Therefore 
if high-frequency detectors could reach sensitivities ${\mathcal O}(10^{-24})$ in the MHz or GHz domains their role could only be marginal for the direct detection of thermal gravitons and, as we shall see, of practically all foreseeable signals from the early Universe. We can therefore conclude that the minimal detectable chirp and spectral amplitudes between the MHz and the GHz should approximately coincide with the largest signals in each 
specific context; in the thermal case this strategy implies:
\begin{equation} 
h_{c}^{(min)} \leq {\mathcal O}(10^{-34}), \qquad S^{(min)}_{h} \leq  {\mathcal O}(10^{-74})\,\, \mathrm{Hz}^{-1}, \qquad \mathrm{MHz} \leq \nu \leq \mathrm{GHz},
\label{gravBB13}
\end{equation}
where we took into account the mild frequency dependence of the chirp amplitude and the constancy of the spectral amplitude below $\nu_{max,\,g}$.

\subsection{Non-thermal gravitons}
Non-thermal gravitons arise from the parametric amplification 
of the quantum fluctuations as in the case 
of the concordance paradigm where the average multiplicity 
has been heuristically introduced in Eq. (\ref{NP3}).
The quantum mechanical description of the process of parametric amplification \cite{QO1} follows from the action of Eq. (\ref{FUND1}) written in the case of a conformally flat background geometry:
\begin{eqnarray}
S_{g} = \frac{1}{8 \ell_{P}^2} \int d^{3} x \int d\tau \biggl[ \partial_{\tau} \mu_{i\,j} \, \partial_{\tau} \mu^{i\,j} 
+ {\mathcal H}^2 \,\mu_{i\,j} \mu^{i\,j}  - \partial_{k} \mu_{i\, j} \partial^{k} \mu^{i\, j} 
- {\mathcal H}\biggl(\mu_{i\,j} \partial \mu^{i\, j} + \mu^{i\, j} \partial_{\tau} \mu_{i\, j}\biggr)  \biggr],
\label{NT1}
\end{eqnarray}
where we introduced the rescaled tensor amplitude $\mu_{i\,j} = a\, h_{i\, j}$; in Eq. (\ref{NT1}) ${\mathcal H} = a^{\prime}/a$ and the prime denotes a derivation with respect to the conformal time coordinate 
$\tau$. 

\subsubsection{The quantum theory of parametric amplification}
The canonical momenta associated with the action (\ref{NT1}) are  $\pi_{i\,j} = (\partial_{\tau} \mu_{i\, j} - {\mathcal H} \mu_{i\, j})/(8 \,\ell_{P}^2)$ and 
the Hamiltonian associated with Eq. (\ref{NT1}) becomes therefore: 
\begin{eqnarray}
H_{g}(\tau) = \int d^{3}x\, \biggl[ 8 \ell_{P}^2 \pi_{i\,j} \, \pi^{i\,j} + \frac{1}{8 \ell_{P}^2} \partial_{k} \mu_{i\, j} \partial^{k} \mu^{i\, j} + {\mathcal H} \biggl(\mu_{i\, j} \, \pi^{i\, j} + \pi_{i\, j} \, \mu^{i\, j}\biggr) \biggr].
\label{NT2}
\end{eqnarray}
From the quantum mechanical viewpoint the process of parametric 
amplification encoded in Eqs. (\ref{NT1})--(\ref{NT2}) is described as the spontaneous or stimulated production of graviton pairs. This problem has many physical and technical analogies with the quantum optical situation \cite{QO1} and the quantum theory of parametric amplification has been originally formulated by Mollow and Glauber \cite{QO1a} (see also \cite{QO2,QO3}
for some early applications of quantum optical concepts to the problem of relic gravitons). Following the same logic \cite{QO1,QO1a} the classical fields appearing in Eq. (\ref{NT2}) are promoted to the status of quantum operators:
\begin{eqnarray}
\widehat{\mu}_{i\, j}(\vec{x}, \tau) &=& \frac{\sqrt{2} \ell_{P}}{(2\pi)^{3/2}} \sum_{\alpha} \int d^{3} k \,\,e^{(\alpha)}_{i\,j}(\hat{k})  \,\, \widehat{\mu}_{\vec{k},\,\alpha}(\tau) \, e^{-i \vec{k}\cdot\vec{x}},
\label{NT3}\\
\widehat{\pi}_{i\, j}(\vec{x}, \tau) &=& \frac{1}{4 \,\sqrt{2}\, \ell_{P}\, (2\pi)^{3/2}} \sum_{\alpha} \int d^{3} k \,\,e^{(\alpha)}_{i\,j}(\hat{k})  \,\, \widehat{\pi}_{\vec{k},\,\alpha}(\tau) \, e^{-i \vec{k}\cdot\vec{x}},
\label{NT4}
\end{eqnarray}
where the index $\alpha$ runs over the two standard tensor polarizations $\oplus$ and $\otimes$. The field operators  $\widehat{\mu}_{\vec{k},\,\alpha}$ and $\widehat{\pi}_{\vec{k},\,\alpha}$ obey the canonical 
commutation relations $[\hat{\mu}_{\vec{k},\, \alpha}, \, \hat{\pi}_{\vec{p},\, \beta}] = i \, \delta_{\alpha\beta} 
\, \delta^{(3)}(\vec{k} + \vec{p})$ and can be expressed in terms of the corresponding creation and annihilation operators: 
\begin{equation}
\widehat{\mu}_{\vec{p} \, \alpha} = \frac{1}{\sqrt{2 p}} \biggl[ \widehat{a}_{\vec{p},\,\alpha} + \widehat{a}^{\dagger}_{-\vec{p},\,\alpha} \biggr], \qquad \widehat{\pi}_{\vec{p} \,\alpha} = - i\,\sqrt{\frac{p}{2}} 
\biggl[ \widehat{a}_{\vec{p},\,\alpha} - \widehat{a}^{\dagger}_{-\vec{p},\,\alpha} \biggr],
\label{NT5}
\end{equation}
where $[\widehat{a}_{\vec{k},\, \alpha}, \, \widehat{a}_{\vec{p},\, \beta}^{\dagger}] = \delta^{(3)}(\vec{k}- \vec{p})$. Inserting Eqs. (\ref{NT3})--(\ref{NT4}) and (\ref{NT5}) into Eq. (\ref{NT2}) the Hamiltonian operator becomes: 
\begin{eqnarray}
\widehat{H}_{g}(\tau) &=& \frac{1}{2} \int d^{3} p \sum_{\alpha=\oplus,\otimes} \biggl\{ \widehat{\pi}_{-\vec{p},\,\alpha}\, \widehat{\pi}_{\vec{p},\,\alpha} + p^2 \widehat{\mu}_{-\vec{p},\,\alpha}\, \widehat{\mu}_{\vec{p},\,\alpha} 
+ {\mathcal H} \biggl[ \widehat{\pi}_{-\vec{p},\,\alpha}\, \widehat{\mu}_{\vec{p},\,\alpha} + 
\widehat{\mu}_{-\vec{p},\,\alpha}\, \widehat{\pi}_{\vec{p},\,\alpha} \biggr] \biggr\}.
\label{NT6}
\end{eqnarray}
In Eqs. (\ref{NT5})--(\ref{NT6}) the presence of both  $\widehat{a}_{\vec{p},\,\alpha}$ and $\widehat{a}_{-\vec{p},\,\alpha}^{\dagger}$ implies that the gravitons are produced in pairs of opposite three-momenta from a state where the total three-momentum vanishes.  This aspect is particularly clear if Eq. (\ref{NT5}) is inserted into Eq. (\ref{NT6}) so that the final result is:
\begin{eqnarray}
\widehat{H}_{g} &=& \frac{1}{2} \int d^{3} p \sum_{\alpha=\oplus,\, \otimes} \biggl\{ p \,\,\biggl[ \widehat{a}^{\dagger}_{\vec{p},\,\alpha} \widehat{a}_{\vec{p},\,\alpha} 
+ \widehat{a}_{- \vec{p},\,\alpha} \widehat{a}^{\dagger}_{-\vec{p},\,\alpha} \biggr] 
+ \lambda \,\,\widehat{a}^{\dagger}_{-\vec{p},\,\alpha} \widehat{a}_{\vec{p},\,\alpha}^{\dagger} 
+ \lambda^{\ast} \,\,\widehat{a}_{\vec{p},\,\alpha} \widehat{a}_{-\vec{p},\,\alpha}
\nonumber\\
&+& \gamma_{-\vec{p},\, \alpha} \widehat{a}_{\vec{p}\,\alpha} + \gamma_{-\vec{p},\, \alpha}^{\ast} \widehat{a}_{\vec{p},\,\alpha}^{\dagger}
+  \gamma_{\vec{p},\, \alpha} \widehat{a}_{- \vec{p},\,\alpha} +  \gamma_{\vec{p},\,\alpha}^{\ast} \widehat{a}_{- \vec{p},\,\alpha}^{\dagger} \biggr\},
\label{NT7}
\end{eqnarray}
where we introduced the notation $ \lambda = i {\mathcal H}$.  The first line of Eq. (\ref{NT7}) 
is responsible for the parametric amplification and it describes the production of pairs
of gravitons with opposite three-momenta. The three 
classes of terms quadratic in the creation and annihilation operators are in fact the generators of the $SU(1,1)$ group and this observation simplifies the calculation of the correlation functions \cite{QO4}. The second line of Eq. (\ref{NT7}) accounts for the presence of a coherent component and it follows from the presence of a coupling proportional to $\mu_{i\, j} \, \Pi^{i\,j}$ where 
$\Pi^{i\,j}$ is the anisotropic stress\footnote{ The coherent component may also be related to the initial conditions but, in this second case, the late time effects 
can only be present if the total number of inflationary $e$-folds is close to the critical one (i.e. $N_{c} = {\mathcal O}(65)$) otherwise the memory of the initial conditions is completely lost. For the sake of generality we shall therefore consider preferentially the cases where $N\gg N_{c}$.}. 
Neglecting, for the moment, the presence of a coherent component, the evolution equations for $\widehat{a}_{\vec{p}}$ and $\widehat{a}_{-\vec{p},\,\alpha}^{\dagger}$
in the Heisenberg description follow from the Hamiltonian  (\ref{NT7}) and they are: 
\begin{eqnarray}
\frac{d \widehat{a}_{\vec{p},\,\alpha}}{d\tau} &=& i \, [ \widehat{H}_{g},\widehat{a}_{\vec{p},\,\alpha}] =  - i\, p \, \widehat{a}_{\vec{p},\,\alpha} -  i \, \lambda \widehat{a}_{- \vec{p},\,\alpha}^{\dagger},
\nonumber\\
\frac{d \widehat{a}_{-\vec{p},\,\alpha}^{\dagger}}{d\tau} &=& i \, [ \widehat{H}_{g},\widehat{a}^{\dagger}_{-\vec{p},\,\alpha}]=  i\, p \, \widehat{a}_{-\vec{p},\,\alpha}^{\dagger} +  i \, \lambda^{\ast} \widehat{a}_{\vec{p},\,\alpha}.
\label{NT8}
\end{eqnarray}
The solution of Eq. (\ref{NT8}) 
can be written in terms of two (complex) time-dependent functions 
$u_{p,\,\alpha}(\tau)$ and $v_{p,\,\alpha}(\tau)$:
 \begin{eqnarray}
\widehat{a}_{\vec{p},\,\alpha}(\tau) &=& u_{p,\,\alpha}(\tau)\,\, \widehat{b}_{\vec{p},\,\alpha}-  
v_{p,\,\alpha}(\tau)\,\, \widehat{b}_{-\vec{p},\,\alpha}^{\dagger},  
\label{NT9}\\
\widehat{a}_{-\vec{p},\,\alpha}^{\dagger}(\tau) &=& u_{p,\,\alpha}^{\ast}(\tau) \,\,\widehat{b}_{-\vec{p},\,\alpha}^{\dagger}  -  v_{p,\,\alpha}^{\ast}(\tau)\,\, \widehat{b}_{\vec{p},\,\alpha}.
\label{NT10}
\end{eqnarray}
If we insert the parametrization of Eqs. (\ref{NT9})--(\ref{NT10}) into Eq. (\ref{NT8}) we obtain the evolution of $u_{p,\,\alpha}(\tau)$ and $v_{p,\,\alpha}(\tau)$, namely
\begin{equation}
u_{p,\,\alpha}^{\prime} = - i p\, u_{p,\,\alpha}  + i \lambda\,\,  v_{p,\,\alpha}^{\ast}, 
\qquad v_{p,\,\alpha}^{\prime} = - i p\, v_{p,\,\alpha} + i \lambda \,\,u_{p,\,\alpha}^{\ast},
\label{NT11}
\end{equation}
where the prime denotes, as usual, a derivation with respect to the cosmic time coordinate $\tau$. The  functions $u_{p,\,\alpha}$ and $v_{p,\,\,\alpha}$ 
are subjected to the conditions $|u_{p,\,\alpha}(\tau)|^2 - |v_{p\,\,\alpha}(\tau)|^2 =1$ 
and they can therefore be parametrized by three real numbers\footnote{One possibility 
is to choose $u_{k,\,\alpha}(\tau) = e^{- i\, \delta_{k,\,\alpha}} \cosh{r_{k,\,\alpha}}$ 
and  $v_{k,\,\alpha}(\tau) =  e^{i (\theta_{k,\,\alpha} + \delta_{k,\,\alpha})} \sinh{r_{k,\,\alpha}}$
as originally suggested, with some slightly different notations, in Ref. \cite{QO2}. The resulting 
evolution equations for the three functions $r_{k,\,\alpha}(\tau)$, $\delta_{k,\,\alpha}(\tau)$ 
and $\theta_{k,\,\alpha}(\tau)$ are however nonlinear and even if their solution 
completely describes the squeezed quantum state of the relic gravitons, we only need 
the average multiplicity of the final state.}.  For the present purposes it is better to avoid the real parts 
of  $u_{p,\,\alpha}$ and $v_{p,\,\alpha}$ and to study the linear combinations 
$(u_{p,\,\alpha} -v^{\ast}_{p,\,\alpha})$ and $(u_{p,\,\alpha} + v^{\ast}_{p,\,\alpha})$ obeying, respectively, 
the following pair of equations: 
\begin{eqnarray}
&& (u_{p,\,\alpha} -v^{\ast}_{p,\,\alpha})^{\prime\prime} + \biggl[ k^2 - \frac{a^{\prime\prime}}{a}\biggr] (u_{p,\,\alpha} -v^{\ast}_{p,\,\alpha}) =0, 
\label{NT12}\\
&&  (u_{p,\,\alpha} +v^{\ast}_{p,\,\alpha})^{\prime\prime} + \biggl[ k^2 - a \biggl(\frac{1}{a}\biggr)^{\prime\prime}\biggr] (u_{p,\,\alpha} +v^{\ast}_{p,\,\alpha}) =0.
\label{NT13}
\end{eqnarray}
Equations (\ref{NT12})--(\ref{NT13}) correspond to the evolution of the mode 
functions of the field and can be solved with the WKB approximation \cite{DF}. 
Since both polarizations obey the same equation we can suppress the polarization index 
and the relevant solutions of Eqs. (\ref{NT12})--(\ref{NT13}) in the regime $k^2 \gg |a^{\prime\prime}/a|$
are:
\begin{eqnarray}
u_{k}(\tau) - v_{k}^{\ast}(\tau) &=& e^{- i k \, \tau_{ex}} \, {\mathcal Q}_{k}(\tau_{ex}, \tau_{re}) \, \biggl(\frac{a_{re}}{a_{ex}}\biggr) \biggl\{ \frac{{\mathcal H}_{re}}{k} \sin{[k ( \tau- \tau_{re})]}
+ \cos[k (\tau - \tau_{re})] \biggr\},
\label{NT13a}\\
u_{k}(\tau) + v_{k}^{\ast}(\tau) &=& i e^{- i k \, \tau_{ex}}\,\,\, {\mathcal Q}_{k}(\tau_{ex}, \tau_{re}) \, \biggl(\frac{a_{re}}{a_{ex}}\biggr) \biggl\{ \frac{{\mathcal H}_{re}}{k} \cos{[k ( \tau- \tau_{re})]}
- \sin[k (\tau - \tau_{re})] \biggr\}.
\label{NT13b}
\end{eqnarray}
where ${\mathcal Q}_{k}(\tau_{ex},\,\tau_{re})$ is given by:
\begin{eqnarray}
{\mathcal Q}_{k}(\tau_{ex}, \tau_{re}) &=& 1 - ( i \, k + {\mathcal H}_{ex}) {\mathcal J}(\tau_{ex}, \tau_{re}),
\nonumber\\
{\mathcal J}(\tau_{ex}, \tau_{re}) &=& \int_{\tau_{ex}}^{\tau_{re}} \frac{a_{ex}^2}{a^2(\tau)} \,\, d\, \tau.
\label{NT16}
\end{eqnarray}
In Eqs. (\ref{NT13a})--(\ref{NT13b}) and (\ref{NT16}) $\tau_{re}$ and $\tau_{ex}$ define 
the turning points where $ k^2 = a^2\, H^2 [ 2 - \epsilon(\tau)] $
where, as usual, $\epsilon = - \dot{H}/H^2$ denotes the slow-roll parameter. If $\epsilon \neq 2$ we have that $k \tau_{re} ={\mathcal O}(1)$ and $k \tau_{ex} = {\mathcal O}(1)$.
Conversely it can happen that $\epsilon\to 2$ in the vicinity of the turning point.
This happens, for instance, when the reentry takes place during the radiation 
epoch; in this case $k \tau_{re} \ll 1$. From Eqs. (\ref{NT13a})--(\ref{NT13b}) the solutions for $u_{k}(\tau)$ and $v_{k}(\tau)$ become:
\begin{eqnarray}
u_{k}(\tau) &=& e^{- i\, k \tau_{ex}}\frac{{\mathcal Q}_{k}(\tau_{ex}, \tau_{re})}{2} \biggl(\frac{a_{re}}{a_{ex}}\biggr) \biggl(1 + i \, \frac{{\mathcal H}_{re}}{k} \biggr) \, e^{- i k (\tau - \tau_{re})},
\label{NT14}\\
v_{k}^{\ast}(\tau) &=& - e^{- i\, k \tau_{ex}}\frac{{\mathcal Q}_{k}(\tau_{ex}, \tau_{re})}{2} \biggl(\frac{a_{re}}{a_{ex}}\biggr) \biggl(1 - i \, \frac{{\mathcal H}_{re}}{k} \biggr) \, e^{i k (\tau - \tau_{re})}.
\label{NT15}
\end{eqnarray}
From the expressions of $u_{k}(\tau)$ and $v_{k}(\tau)$ we can deduce all 
the correlation functions relevant for our discussion starting from the averaged 
multiplicity.
\subsubsection{The averaged multiplicity}
The averaged multiplicity is obtained by computing the mean number of gravitons 
with momentum $\vec{k}$ and $- \vec{k}$, i.e. 
$\langle \hat{N}_{k} \rangle = \langle \widehat{a}_{\vec{k}}^{\dagger} \widehat{a}_{\vec{k}} + \widehat{a}_{-\vec{k}}^{\dagger} \widehat{a}_{-\vec{k}} \rangle$; from Eqs. (\ref{NT9})--(\ref{NT10}) we have, in the unpolarized case, that $\langle \hat{N}_{k} \rangle =
2 |v_{k}(\tau)|^2$. The averaged multiplicity of pairs is then given by 
$\overline{n}(k,\tau) =  |v_{k}(\tau)|^2$ and its explicit expression follows 
from Eq. (\ref{NT15}):
\begin{equation}
\overline{n}(k,\tau) = \frac{1}{4} \biggl(\frac{a_{re}}{a_{ex}}\biggr)^2 \biggl[\biggl(\frac{{\mathcal H}_{re}}{k}\biggr)^2 +1 \biggr] \biggl[ 
1 + (k^2 + {\mathcal H}_{re}) {\mathcal J}^2(\tau_{ex}, \tau_{re}) - 2 {\mathcal H}_{re} {\mathcal J}(\tau_{ex}, \tau_{re})\biggr].
\label{NT17}
\end{equation}
If the reentry takes place close to $\epsilon_{re} \to 2$ 
then $k \tau_{re} \ll 1$ and the term $({\mathcal H}_{re}/k)\gg 1$ dominates
inside the first squared bracket of Eq. (\ref{NT17}).
Conversely if the reentry occurs when $\epsilon_{re}\neq 2$, then $k \tau_{re} \simeq {\mathcal O}(1)$ and $({\mathcal H}_{re}/k) = {\mathcal O}(1)$. We are generally interested in the case when $\tau_{ex}$ falls  
during the inflationary stage (i.e. $ a_{ex} \, H_{ex} = - 1/[(1 - \epsilon) \tau_{ex}]$) and the reentry 
occurs in a decelerated stage of expansion. In a radiation-dominated stage of expansion (i.e. $\epsilon_{re} \to 2$) the average multiplicity of gravitons is then estimated as:
\begin{equation}
\overline{n}(\nu,\tau_{0}) = \frac{1}{4}\biggl(\frac{\nu}{\overline{\nu}_{max}}\biggr)^{- 4 + n_{T}}, \qquad\qquad \nu \gg \nu_{eq}, 
\label{NT18}
\end{equation}
and it holds for typical frequencies larger than the equality frequency; in Eq. (\ref{NT18})
$n_{T} = - 2 \epsilon = - r_{T}/8$ and $\overline{\nu}_{max}$ has been
already introduced in Eq. (\ref{BBN1a}). The terms containing $ {\mathcal J}(\tau_{ex}, \tau_{re})$ 
in Eq. (\ref{NT17}) give subleading contributions that are all negligible.
There is however the possibility that prior to the dominance of radiation (and before the onset of BBN) the background expanded either faster or slower than radiation. If $\tau_{re}$ falls in a stage where the scale factor expands as $a(\tau) \simeq \tau^{\delta}$ (with $\delta \neq 1$) then $\epsilon_{re} \neq 2$ and the averaged multiplicity becomes:
\begin{equation}
\overline{n}(\nu,\tau_{0}) = \frac{1}{2}\biggl(\frac{\nu}{\nu_{max}}\biggr)^{- 4 + m_{T}}, \qquad \nu \geq \nu_{r},
\label{NT18aa}
\end{equation}
where $m_{T}$ and $\nu_{max}$ have now a different meaning in comparison with Eq. (\ref{NT18}). In particular $m_{T}$ s given by:
\begin{eqnarray}
m_{T} = \frac{32 -4 r_{T}}{16 - r_{T}} - 2 \delta \simeq 2 (1 - \delta) + {\mathcal O}(r_{T}),
\label{NT20}
\end{eqnarray}
and the frequency $\overline{\nu}_{max}$ is now replaced by 
\begin{equation}
\nu_{max} = \zeta^{\frac{\delta -1}{2(1+ \delta)}}\,\overline{\nu}_{max}, \qquad \zeta= H_{r}/H_{1},
\label{NT21}
\end{equation}
where $H_{r}$ denotes the curvature scale of radiation dominance 
and $H_{1}$ is the curvature scale at the end of inflation i.e. 
\begin{equation}
\frac{H_{1}}{M_{P}} = \frac{\sqrt{\pi \, r_{T}\, {\mathcal A}_{{\mathcal R}}}}{4} = 5.328 \times 10^{-6} \,\, 
\biggl(\frac{{\mathcal A}_{{\mathcal R}}}{2.41\times 10^{-9}}\biggr)^{1/2} \biggl(\frac{r_{T}}{0.06}\biggr)^{1/2},
\label{NT21a}
\end{equation}
where, as already mentioned after Eq. (\ref{BBN1a}), ${\mathcal A}_{{\mathcal R}}$ denotes the 
amplitude of the power spectrum of curvature inhomogeneities at the pivot scale $k_{p}= 0.002 \, \mathrm{Mpc}^{-1}$.
The average multiplicity of gravitons appearing in Eq. (\ref{NT18aa}) applies for all the frequencies larger than $\nu_{r} = \sqrt{\zeta}\, \overline{\nu}_{max}$, i.e. for all the wavelengths that reenter the Hubble radius prior to radiation dominance. Finally, $\overline{n}(\nu,\tau_{0})$ is again suppressed as $\nu^{-4}$ between $\nu_{eq}$ and $\nu_{r}$ as it happens in the 
 standard case of  Eq. (\ref{NT18}) where the average multiplicity is ${\mathcal O}(10^{20})$ for $\nu = {\mathcal O}(\mathrm{kHz})$ but its value gets suppressed (as $\nu^{-4}$) at higher frequencies. In the 
 case of Eq. (\ref{NT18aa}) the integrals associated with  ${\mathcal J}(\tau_{ex}, \tau_{re})$ may now 
 lead to a logarithmic enhancement that is however effective only in the case $\delta \to 1/2$.

The results of Eqs. (\ref{NT20})--(\ref{NT21}) depend of the post-inflationary expansion rate.
If, after inflation, the Universe expands faster than radiation (i.e. $\delta > 1$ in Eq. (\ref{NT20})) we also have that $m_{T} < 0$ and the average multiplicity is even more suppressed than in the case of Eq. (\ref{NT18}). Conversely, when the post-inflationary expansion rate is slower than radiation (i.e. $\delta < 1$ in Eq. (\ref{NT20})) the average multiplicity is less suppressed than in the case of Eq. (\ref{NT18}) since, in Eq. (\ref{NT18aa}), $m_{T} > 0$. According to Eq. (\ref{NT21}) the post-inflationary rate of expansion also affects the maximal frequency. We repeat that, in Eq. (\ref{NT21}), $\zeta= H_{r}/H_{1} < 1$ defines the ratio between the expansion rates at the beginning of radiation-dominance and at the end of inflation. Consequently $\nu_{max} > \overline{\nu}_{max}$ when the post-inflationary expansion rate is slower than radiation (i.e. $\delta < 1$) while $\nu_{max} < \overline{\nu}_{max}$ when $\delta > 1$ and the rate is faster than radiation.

We finally remark that the expression of Eq. (\ref{NT18aa}) can also be generalized to the situation where the Hamiltonian contains a coherent component and this physical possibility is realized when thee gravitons are produced because of the presence of an anisotropic stress. We consider here, as an example, the case of hybrid inflation \cite{WAT0} where the waterfall fields are amplified with spectral slopes that are even steeper than the ones characterizing the vacuum fluctuations \cite{WAT1,WAT2,WAT3}. The inhomogeneities of the waterfall field induce a secondary graviton spectrum 
between the MHz and the GHz \cite{WAT4} and since the high-frequency slopes are  larger than
$1$ the spectral energy density is practically concentrated in a narrow slice of frequencies around the maximum. Following  the notations of Ref. \cite{WAT4} the waterfall spectrum is parametrized as $P_{\sigma}(k,\tau) = A_{\sigma}^2 (k/k_{max})^{n_{\sigma}-1}$ where $A_{\sigma}$ is expressed in Planck units and the scale-invariant limit corresponds to $n_{\sigma} =1$. The case $n_{\sigma}= 3$ characterizes the slope of quantum (vacuum) fluctuations. If $n_{\sigma} > 3$ the spectral slope is steeper than  in the case of vacuum fluctuations. The amplified spectrum characterizing the waterfall field in hybrid inflation leads to $n_{\sigma} \simeq 4$. This means that the tensor spectrum in Eq. (\ref{NT18aa}) corresponds to $\overline{n}_{T} \simeq 2 (n_{\sigma} -1)$. This means that, overall, the averaged multiplicity of produced gravitons is either scale-invariant (as in the vacuum case) or even increass as $\nu^{2}$.

\subsubsection{Generalizations to multiple phases}
The results discussed in the previous subsection can be generalized to the case of multiple stages of post-inflationary expansion. The scale factor during the $i$-th stage of expansion 
can be parametrized, for instance,  as $a_{i}(\tau) = (\tau/\tau_{i})^{\delta_{i}}$
with $\delta_{i} \neq 1$. If the mode reenter in a stage where $\delta_{i} \to 1$ 
the averaged multiplicity scales as $\nu^{-4}$ and this is why we preferentially 
consider the situation where $\delta_{i} \neq 1$. 
For $\tau > \tau_{i}$ the scale factor during the 
$(i+1)$-th stage of expansion is 
\begin{equation}
a_{i+1}(\tau) \simeq \biggl[\frac{\delta_{i}}{\delta_{i+1}} \biggl(\frac{\tau}{\tau_{i}}-1\biggr)+1 \biggr]^{\delta_{i+1}}, \qquad \qquad \delta_{i+1} \neq 1, \qquad \tau \geq \tau_{i},
\label{NT22}
\end{equation}
where, for the reason given above, $\delta_{i +1} \neq 1$.
Assuming the validity of the consistency relations the value of $m_{T}^{(i)}$ in each of the $i$-th branches now depends 
on $r_{T}$ and $\delta_{i}$:
 \begin{equation} 
m_{T}^{(i)}(r_{T}, \delta_{i}) = \frac{32 - 4 \, r_{T}}{16 -r_{T}} - 2 \delta_{i} = 
2( 1 - \delta_{i}) + {\mathcal O}(r_{T}).
\label{NT23}
\end{equation}
Equation (\ref{NT23}) is consistent with the previous determinations of the spectral index: when 
$\delta_{i} \to 1$ we have that the different $m_{T}^{(i)}$ collapse to $- r_{T}/8$ 
that coincides with the result of Eq. (\ref{NT18}). Similarly, when all the $\delta_{i}$ collapse 
to a single $\delta$ Eq. (\ref{NT23}) reproduces the result of Eq. (\ref{NT20}).
In the case of the multiple post-inflationary stages of expansion the maximal 
frequency $\nu_{max}$ is also affected:
\begin{equation}
 \nu_{max} =  \prod_{i =1}^{N-1} \, \, \zeta_{i}^{\frac{\delta_{i} -1}{2(\delta_{i} +1)}}\, \, \overline{\nu}_{max},
\label{NT24}
\end{equation}
where $\zeta_{i} = H_{i+1}/H_{i} < 1$; as expected, if $\delta_{i} \to 1$ 
in Eq. (\ref{NT24}) for all the $i= 1, \, .\, ., \,. \, N$ we also have that $\nu_{1} = \nu_{max} \to  \overline{\nu}_{max}$. 
The lower frequency of the spectrum is related to the dominance of radiation and it is therefore given by
\begin{equation}
\nu_{r} = \prod_{j = 1 }^{N-1} \, \sqrt{\zeta_{j}} \,\,\, \overline{\nu}_{max} = \sqrt{\zeta} \, \overline{\nu}_{max},
 \label{NT25}
 \end{equation}
where $\nu_{r} = \nu_{N}$ and the second equality follows since, by definition, $ \zeta_{1}\,\zeta_{2} \,.\,.\,. \,\zeta_{N-2}\, \zeta_{N-1} = \zeta$. The notation followed in Eqs. (\ref{NT24})--(\ref{NT25}) 
implies that the maximal frequency coincides with $\nu_{1}$ (i. e. $\nu_{1}\equiv \nu_{max}$) 
while $\nu_{N} \equiv \nu_{r}$. 

For the present ends it is relevant to understand under which circumstances the highest 
frequency of the spectrum is maximized. According to Eqs. (\ref{NT24})--(\ref{NT25}) the largest value of $\nu_{max}$ is realized when all the $\delta_{i}$ are smaller than $1$ (i.e.  $\delta_{i} <1$) and this happens
since, by definition, the $\zeta_{i} < 1$; if some of the $\delta_{i} > 1$ the maximal frequency is comparatively smaller than in the case where all the $\delta_{i}$ are smaller than $1$ (and the 
plasma expands, overall, slower than radiation). For completeness we mention that the intermediate frequencies between $\nu_{max}$ and $\nu_{r}$ can be compactly expressed in a form that interpolates between Eqs. (\ref{NT24}) and (\ref{NT25}):
\begin{equation}
\nu_{m} = \prod_{j=1}^{m-1} \sqrt{\zeta_{j}}\,\, \prod_{i =m}^{N- 1} \, \, \zeta_{i}^{\frac{\delta_{i} -1}{2(\delta_{i} +1)}}\, \, \overline{\nu}_{max}, \qquad\qquad m = 2,\, 3,\, .\,.\,.\, N-2,\, N-1.
\label{NT26}
\end{equation}
We finally mention that depending on the values of  $\delta_{i}$ and $\zeta_{i}$ 
also the maximal number of $e$-folds presently accessible to large-scale observations \cite{MEV0a,MEV0b} gets modified as follows\footnote{For the estimate of Eq. (\ref{NT27})  we 
assume exactly the same late-time parameters mentioned in Eqs. (\ref{BBN1a}) and (\ref{NT21a}).}
\begin{eqnarray}
N_{max}  = 61.88 +\frac{1}{2}\sum_{i}^{N-1} \, \biggl(\frac{\delta_{i} -1}{\delta_{i} + 1}\biggr) \, \ln{\zeta_{i}}.
\label{NT27}
\end{eqnarray}
If we conventionally set $\delta_{i} =1$ the second term in Eq. (\ref{NT27}) disappears and we 
obtain the standard result  implying that ${\mathcal N}_{max} = {\mathcal O}(60)$. 
Moreover, since $\zeta_{i}<1$ we have that ${\mathcal N}_{max} > 60$ when $\delta_{i}<1$ and that ${\mathcal N}_{max} < 60$ in the case $\delta_{i}>1$. When there is a  single phase expanding at a rate that is slower than radiation (as suggested after Eq. (\ref{NT25})), ${\mathcal N}_{max}$ can be as large as $75$. The minimal value of $\zeta$ is estimated by requiring that $H_{r} > 10^{-44} \, M_{P}$ and this result suggests that the plasma is already dominated by radiation for temperatures that are well above the MeV. There are some  possibilities where the MeV-scale reheating temperature could be induced by long-lived massive species with masses close to the weak scale \cite{MEV1,MEV2}. In the present context the condition  $H_{r} \geq 10^{-44} \, M_{P}$ is merely an absolute lower limit on the value of $H_{r}$ and hence on the value of $\zeta$.

\subsubsection{The averaged multiplicity in the non-thermal case}
\begin{figure}[!ht]
\centering
\includegraphics[height=7.5cm]{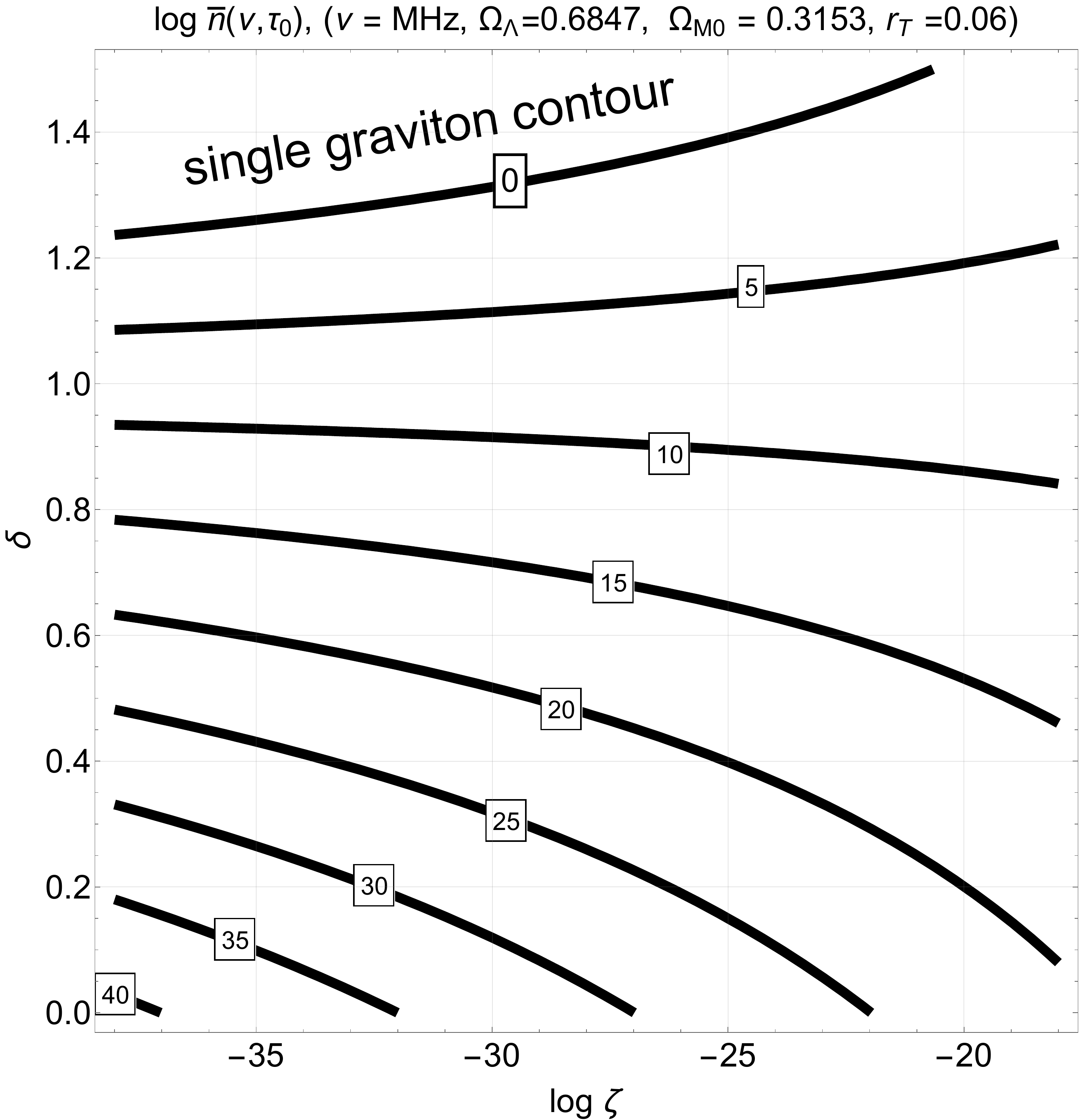}
\includegraphics[height=7.5cm]{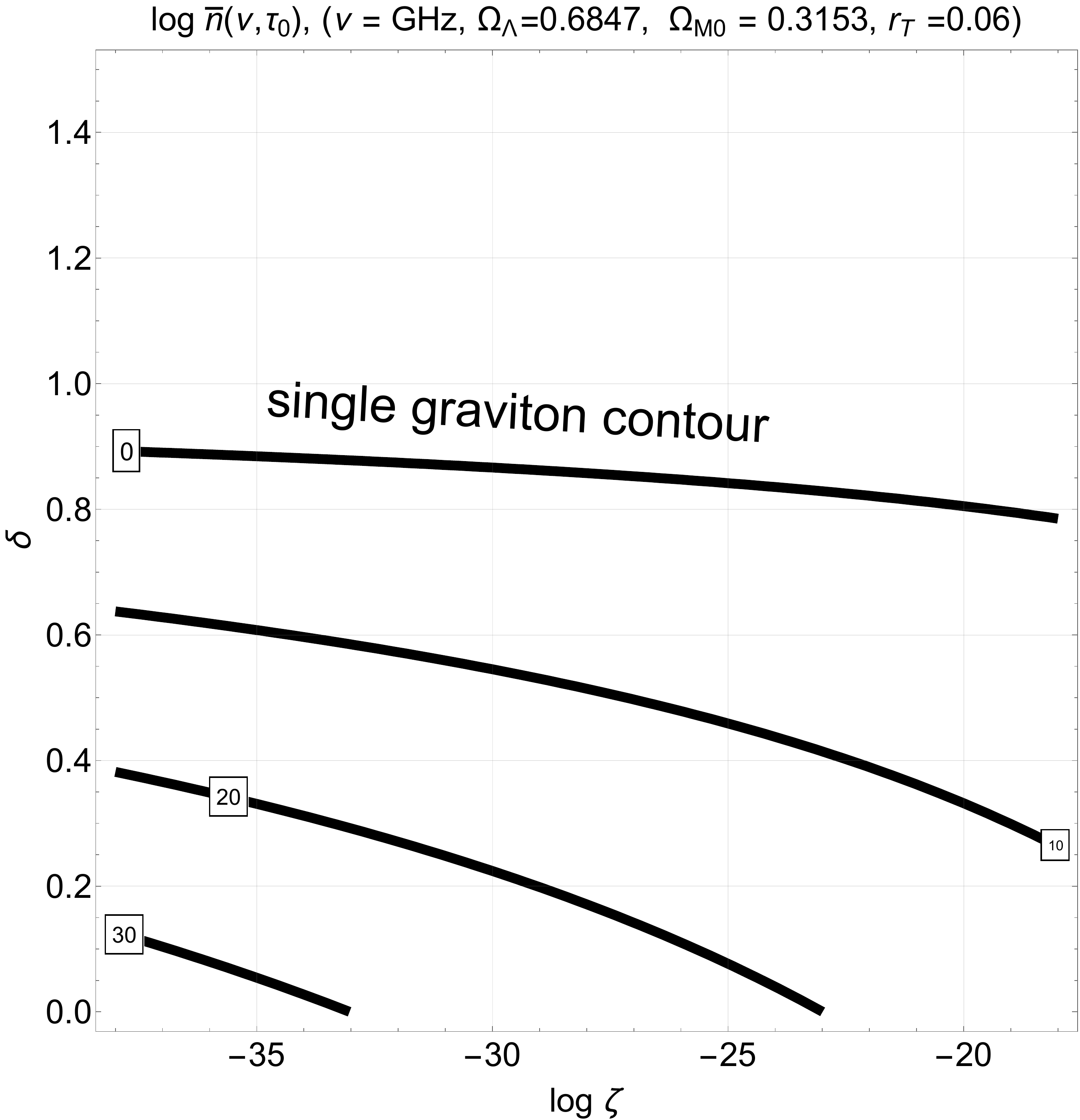}
\caption[a]{The common logarithm of the averaged multiplicity discussed in Eqs. (\ref{NT18aa})
and (\ref{NT20})--(\ref{NT21}) is reported. In the plot at the left we consider the MHz region while in the right plot the typical frequency is the GHz range. The labels appearing on the different curves give the common logarithm of the averaged multiplicities. The contour labeled by $0$ corresponds to the production of a single graviton pair; in this case the common logarithm of the averaged multiplicity vanishes. The production of a single graviton pair pins down the maximal frequency of the spectrum since for higher frequencies gravitons are not produced. A comparison of the two plots implies that gravitons are simultaneously produced in the MHz and GHz regions only when $\delta < 1$. Conversely, when $\delta > 1$ the averaged multiplicity practically vanishes for $\nu = {\mathcal O}(\mathrm{GHz})$.}
\label{FIGURE5}      
\end{figure}
We start this discussion of the averaged multiplicity in the non-thermal case 
by considering Fig. \ref{FIGURE5} where the common logarithms of the averaged multiplicities are illustrated in the $(\log{\zeta},\,\delta)$ plane for typical values of the parameters.
In both plots of Fig. \ref{FIGURE5} the frequencies are fixed in the two ranges that are relevant for 
the present discussion (i.e. $\nu = \mathrm{MHz}$ 
in the plot at the left and $\nu = \mathrm{GHz}$ in the right plot). The labels on the curves denote the common logarithm of $\overline{n}(\nu,\tau_{0})$ for different values of $\delta$ (the expansion rate after inflation) and $\zeta = H_{r}/H_{1}$ where $H_{r}$ is the expansion rate at the onset of the radiation-dominated stage while $H_{1}$ is evaluated at the end of inflation (see Eq. (\ref{NT21a}) and also the discussion after Eq. (\ref{NT27})). This situation is the most relevant for the present ends and the spectrum only consists of two typical frequencies namely $\nu_{r}$ and $\nu_{max}$. The frequencies $\nu_{max} = \zeta^{(\delta -1)/[2 (\delta+1)]} \, \overline{\nu}_{max}$ and $\nu_{r} = \sqrt{\zeta} \,\, \overline{\nu}_{max}$ follow, respectively, from Eqs. (\ref{NT24})--(\ref{NT25}) in the case $N=2$. All the other intermediate frequencies given by Eq. (\ref{NT26}) vanish so that the averaged multiplicity roughly scales as $\nu^{-4}$ for $\nu < \nu_{r}$ while it is less suppressed for $\nu > \nu_{r}$ (see, in this respect, Eqs. (\ref{NT18aa})--(\ref{NT20})). In Fig. \ref{FIGURE5} a particularly interesting contour is the {\em single graviton line}\footnote{For short we are going to refer to this curve as 
the single graviton line even if it corresponds, strictly speaking, to the production 
of a single graviton pair where the gravitons have opposite comoving 
three-momenta.}. Since we use common logarithms 
and since $\overline{n}(\nu,\tau_{0})$ denotes the number of graviton pairs
the single graviton line corresponds, approximately, to the curve labeled by $0$ beyond which no gravitons are produced.
\begin{figure}[!ht]
\centering
\includegraphics[height=7cm]{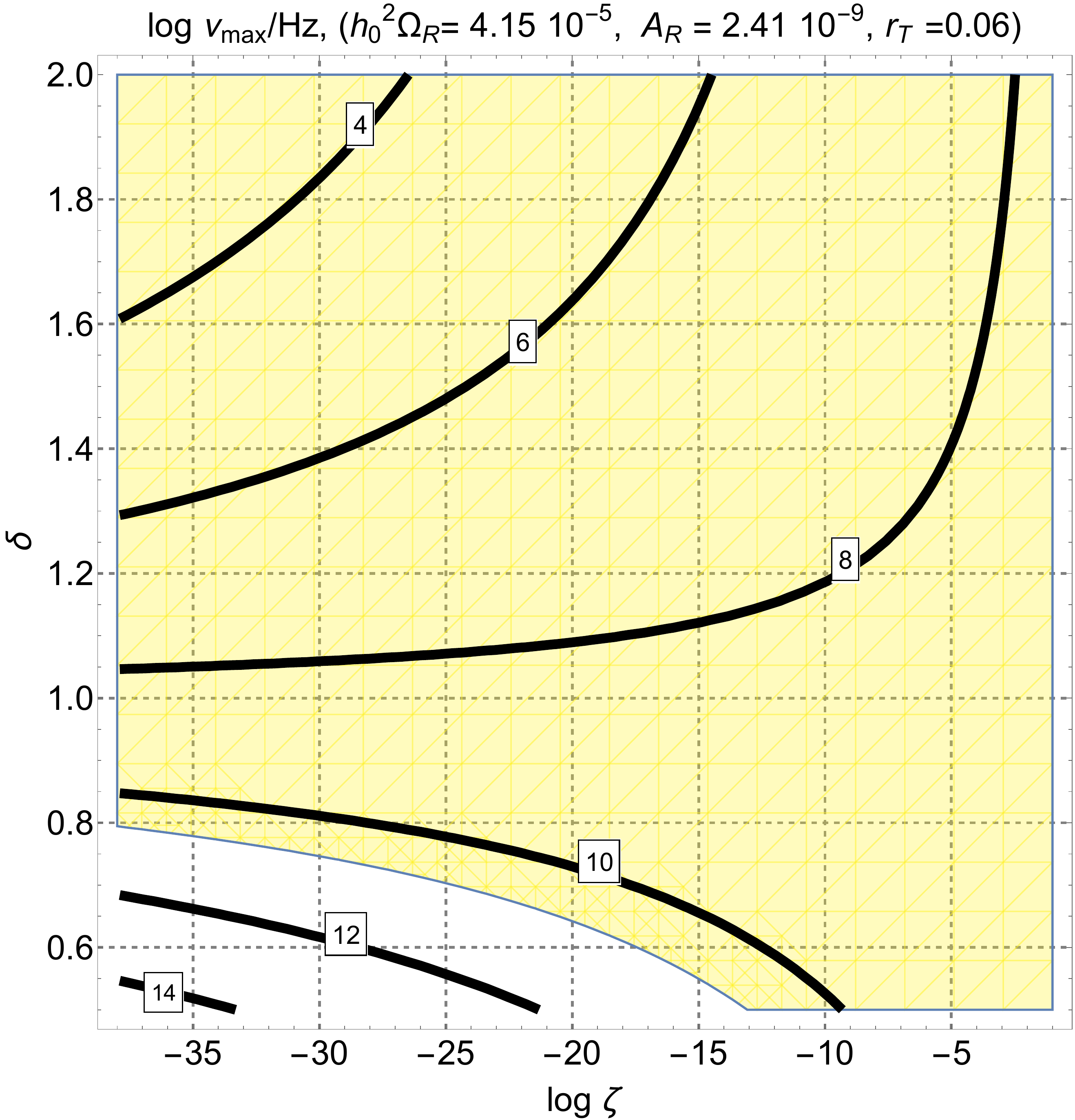}
\includegraphics[height=7cm]{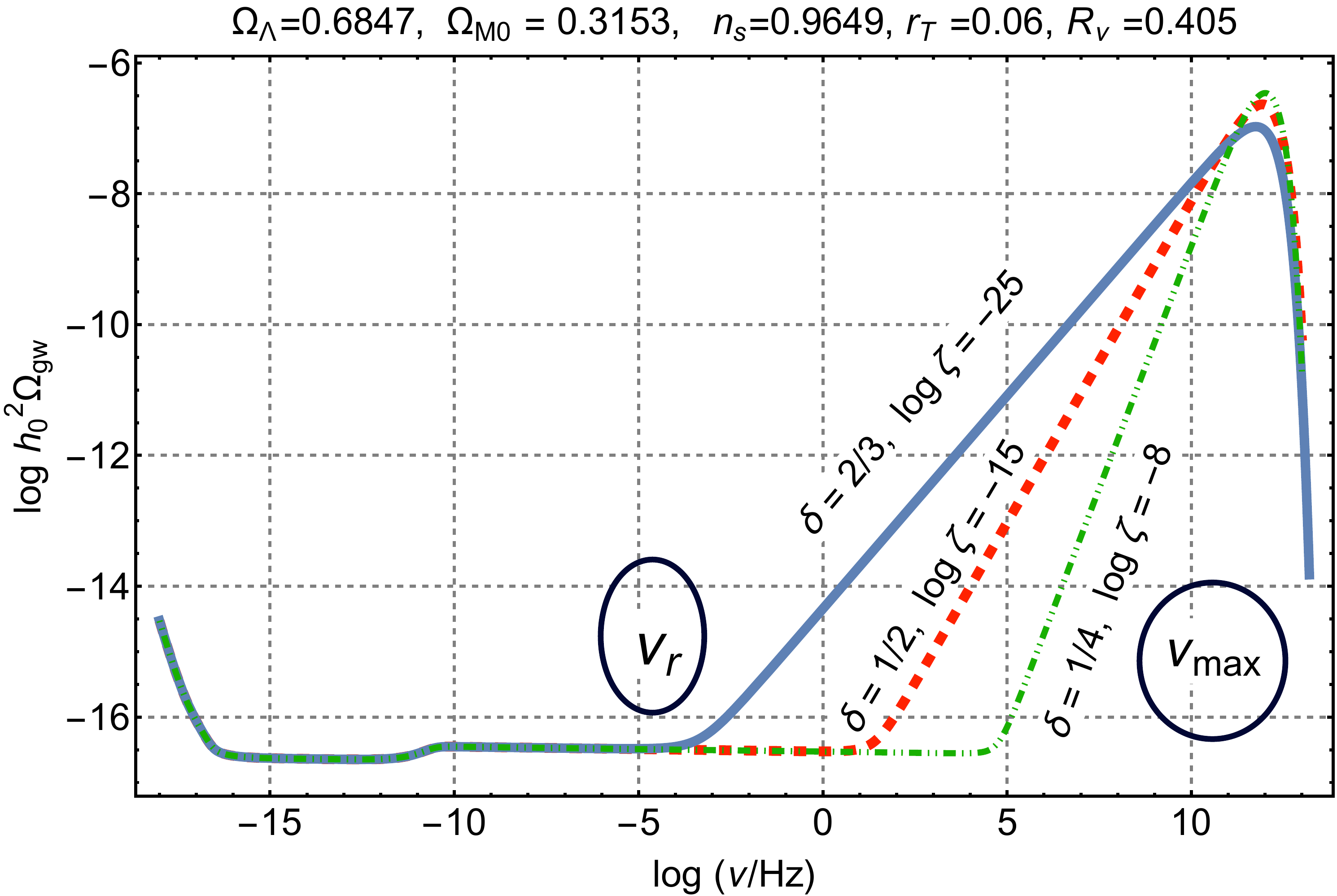}
\caption[a]{ In the plot at the left the 
various labels correspond to the common logarithm of $\nu_{max}$ 
expressed in Hz. In the right plot we report instead the common logarithm 
of the spectral energy density for a handful of different parameters leading 
to a large signal around $\nu_{max}$. What matters, for the present considerations, is the high-frequency region where the average multiplicity is approximately suppresses as 
$\nu^{-2 - 2 \delta}$ for the different values of $\delta$ reported in the right plot. As already illustrated in the previous plots, the single graviton line corresponds to the averaged multiplicity 
when the comoving frequency is exactly ${\mathcal O}(\nu_{max})$. }
\label{FIGURE6}      
\end{figure}

In the left plot of Fig. \ref{FIGURE6} we illustrate the common logarithm of $\nu_{max}$ (expressed in Hz) as a function of the expansion rate and of the length of the post-inflationary phase (i.e. respectively $\delta$ and $\zeta$). The shaded region in the left plot of Fig. \ref{FIGURE6} illustrates the portion of the parameter space  where all the phenomenological constraints discussed of section \ref{sec3} are concurrently satisfied. For the sake of illustration in the right plot of Fig. \ref{FIGURE6} the spectral energy density has been reported as a function of the frequency and for different values of $\delta$ and $\zeta$. 
In the previous sections the complications associated with the free-streaming of the neutrinos and with the other late-time suppressions of the spectral energy density have been neglected; they are instead included in the right plot of Fig. \ref{FIGURE6} that also demonstrates why these complications are not essential in the MHz--GHz domain. From Fig. \ref{FIGURE6} we actually see that the interesting frequency range for the present purposes is much  larger than $\nu_{r}$ and it is worth stressing that the values of $\nu_{max}$ appearing in each contour of Fig. \ref{FIGURE6} {\em decrease} when $\delta >1$ and 
{\em increase} when $\delta <1$. This means that the maximal frequency is comparatively 
larger when the post-inflationary expansion rate is slower than radiation; this is why, ultimately, the region $\delta < 1$ is more constrained. 

In the right plot of Fig. \ref{FIGURE6} 
we also illustrated the mutual positions of 
$\nu_{r}$ and $\nu_{max}$ for $\delta < 1$. As previously discussed 
$(\nu_{max}/ \nu_{r}) = \zeta^{1/(\delta +1)}$ and since 
$H_{1}$ follows from Eq. (\ref{NT21a}), $\zeta = (H_{r}/H_{1}) <1$ estimates the extension of the post-inflationary phase prior to radiation dominance. The radiation must dominate before big-bang nucleosynthesis and for this reason we required $\zeta \geq 10^{-38}$; this is the range of $\zeta$ adopted throughout the present investigation (see also Eq. (\ref{NT27}) and the discussion thereafter).
We see from Fig. \ref{FIGURE5} that, as long as $\delta < 1$, the averaged multiplicity always exceeds the one of the concordance scenario. As a reference value we can compare 
the contours of Fig. \ref{FIGURE5} with $\overline{n}(\nu,\tau_{0}) = {\mathcal O}(10^{3})$ which is 
the averaged multiplicity computed from Eq. (\ref{NT18}) in the case of the concordance paradigm
and for a typical frequency $\nu = {\mathcal O}(10)$ MHz.
\begin{figure}[!ht]
\centering
\includegraphics[height=7.5cm]{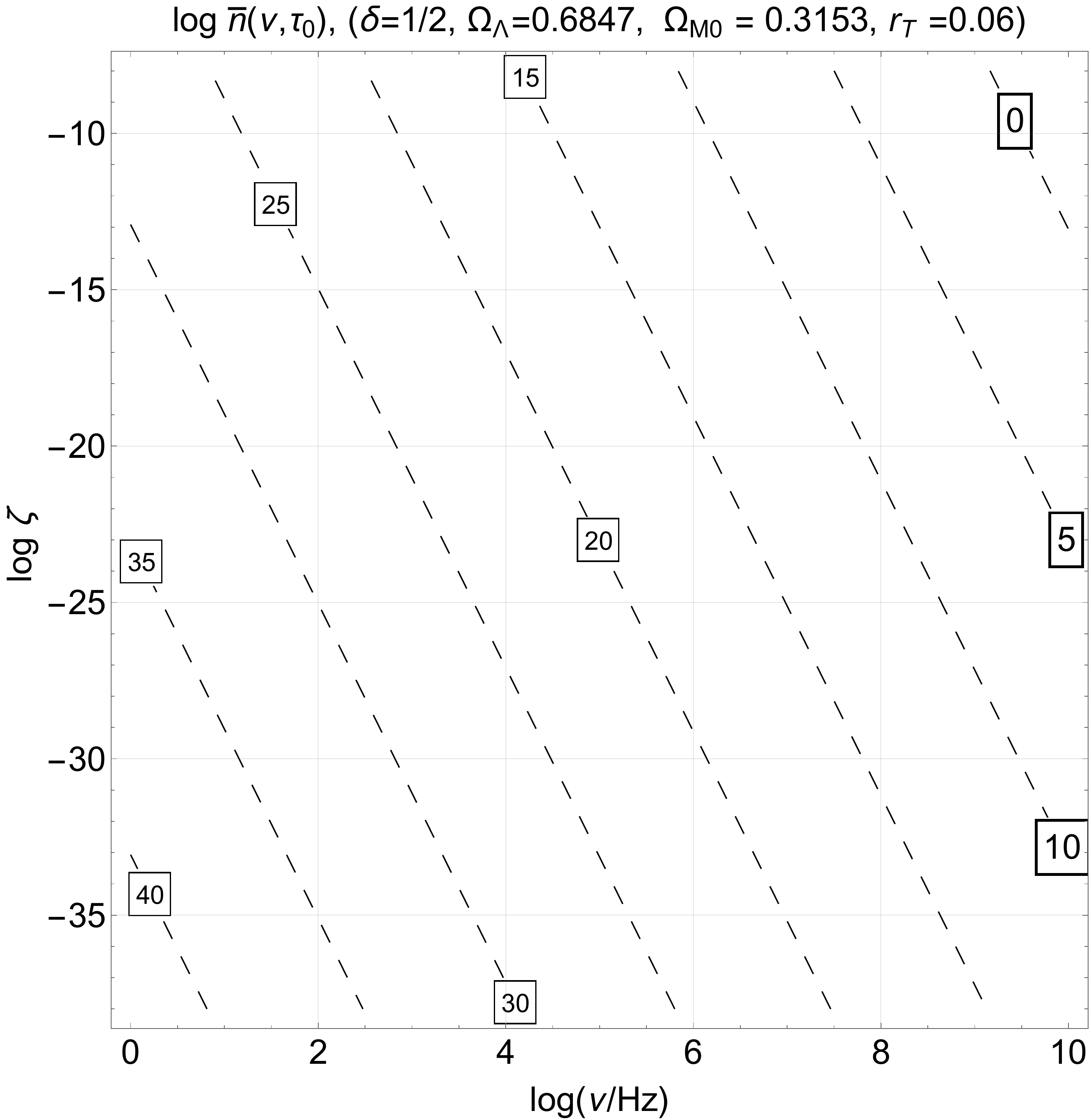}
\includegraphics[height=7.5cm]{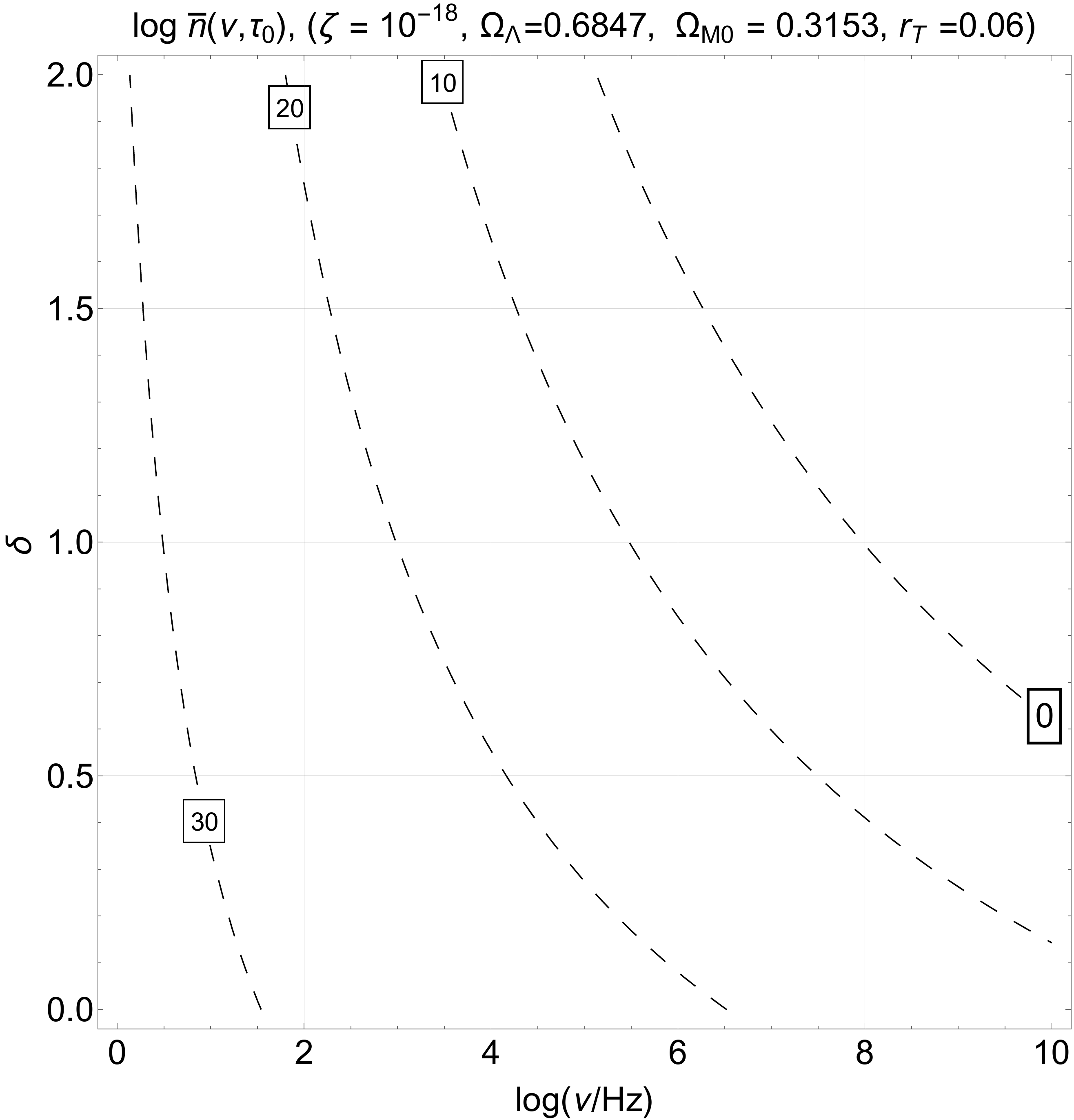}
\caption[a]{While in Fig. \ref{FIGURE5} we fixed the frequency and analyzed the averaged multiplicity in the $(\log{\zeta},\, \delta)$ plane, $\delta$ and $\zeta$ have now been fixed. In particular, in the left plot $\delta \to 1/2$ and the averaged 
multiplicity is examined in the $(\log{\nu},\, \log{\zeta})$ plane. In the right plot $\zeta \to 10^{-18}$ and $\overline{n}(\nu,\tau_{0})$ is illustrated in the $(\log{\nu},\, \delta)$ plane. As in Fig. \ref{FIGURE5} the various labels denote 
the common logarithm of the averaged multiplicity.}
\label{FIGURE7}      
\end{figure}
The empty area in Fig. \ref{FIGURE5} defines the region where gravitons {\em are not} produced but while in Fig. \ref{FIGURE5} {\em the frequency ranges have been fixed}, 
the plots of Fig. \ref{FIGURE7} are obtained, respectively, by fixing {\em the expansion rate}
and {\em the length of the post-inflationary phase}.
In this sense Figs. \ref{FIGURE5} and \ref{FIGURE7} are complementary and demonstrate that the 
largest averaged multiplicities are actually expected for moderate $\zeta$ (i.e. $\zeta \geq 10^{-20}$) and $\delta < 1$. 

In the left plot of Fig. \ref{FIGURE7} the expansion rate $\delta$ is fixed (i.e. 
$\delta \to 1/2$) to a specific value in the range $\delta< 1$ since, according to Fig. \ref{FIGURE5}, this is the situation where the average multiplicity exceeds the value of the concordance paradigm. For example if the post-inflationary 
expansion rate is dominated by a perfect barotropic fluid we have that $\delta = 2/(3 w +1)$ so that 
the case $\delta < 1$ corresponds in fact to stiff equation of state where $ 1/3 < w\leq 1$; in particular, if $w\to 1$ we also have that $\delta \to 1/2$. It can however happen 
that the post-inflationary evolution is dominated by a scalar field; in this case a stiff phase 
with $\delta \to 1/2$ is naturally realized when the potential term vanishes as suggested long ago 
in \cite{MG1,MG2,MG3,MEV3}; the same idea has been discussed later in a number of different 
contexts (see e.g. \cite{MEV4,MEV5,MEV6,MEV7}). The possibility of a long stiff phase is deeply rooted in 
the late-time dominance of the dark energy contribution.
A stiff evolution of the type of the one envisaged in \cite{MG1} can also be realized when a scalar field is coherently oscillating close to its minimum \cite{MEV8} (see also \cite{MEV9}).  
If we assume that the potential can be approximated as $V= V_{1} (\varphi/M)^{2\gamma}$ in the vicinity 
of its minimum then the effective expansion rate under the dominance of the coherent oscillations 
is given by $\delta = 2(\gamma+1)/(4 \gamma -2)$ implying that $\delta < 1$
as long as $\gamma > 2$.
\begin{figure}[!ht]
\centering
\includegraphics[height=5.9cm]{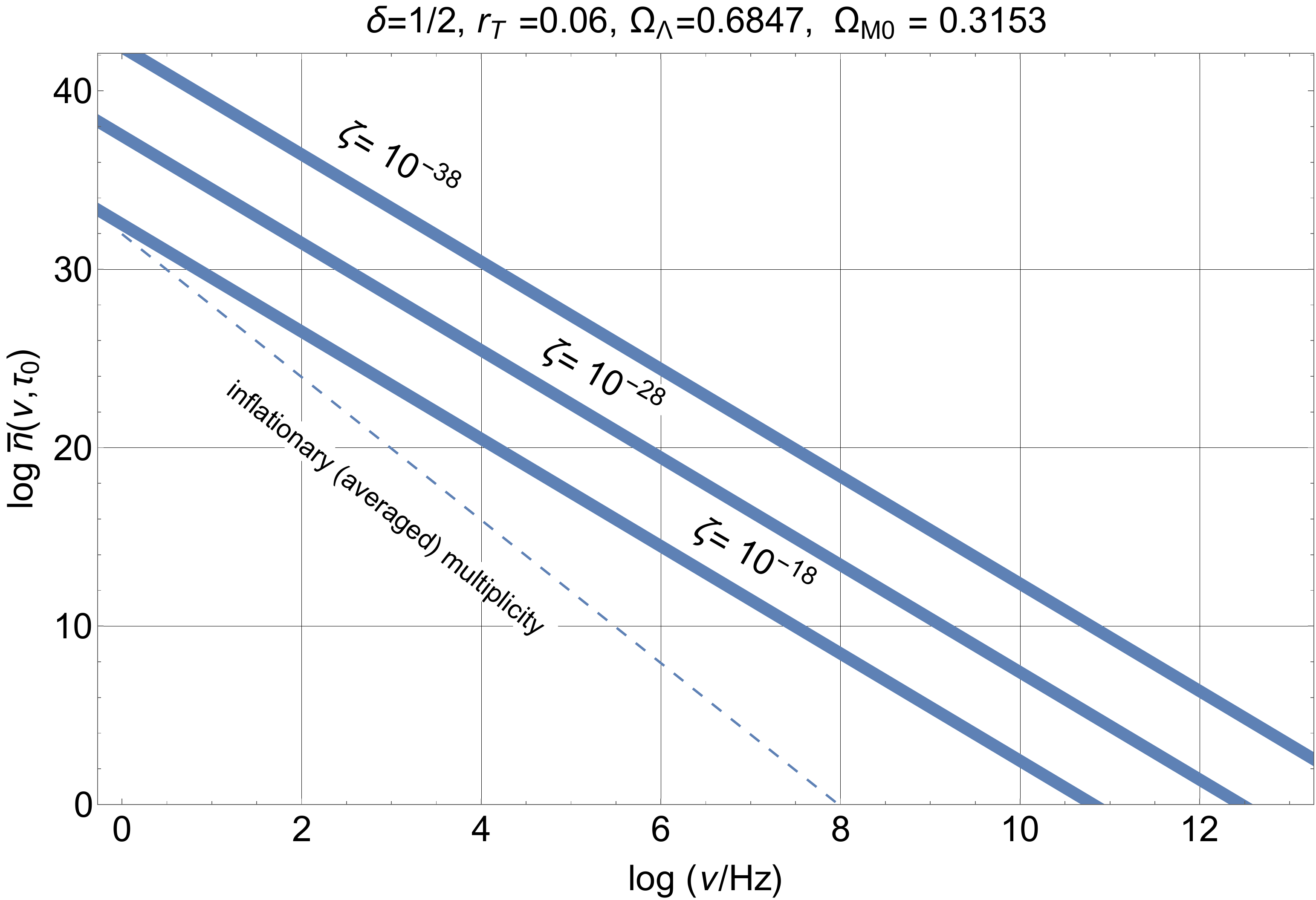}
\includegraphics[height=5.9cm]{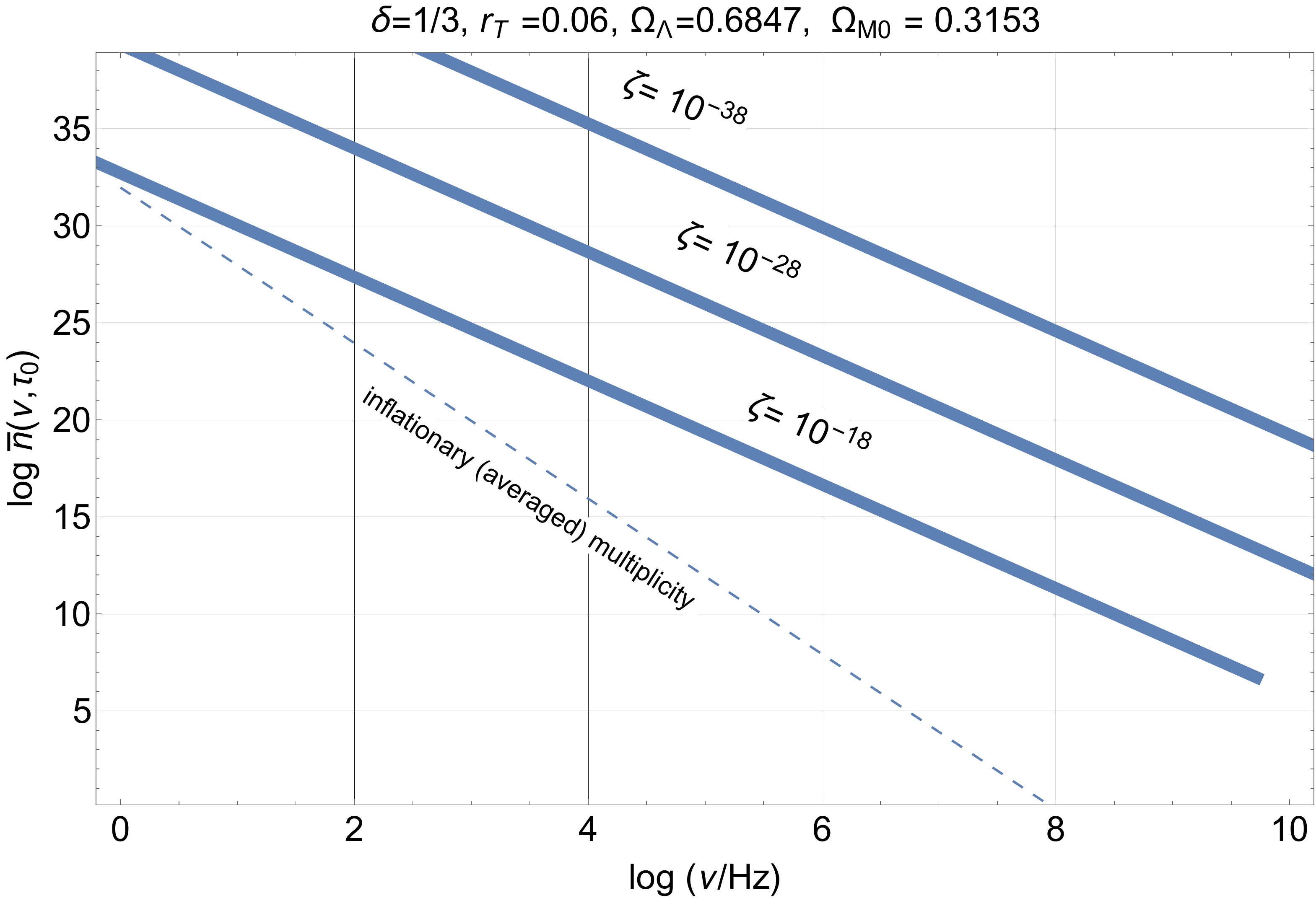}
\includegraphics[height=5.9cm]{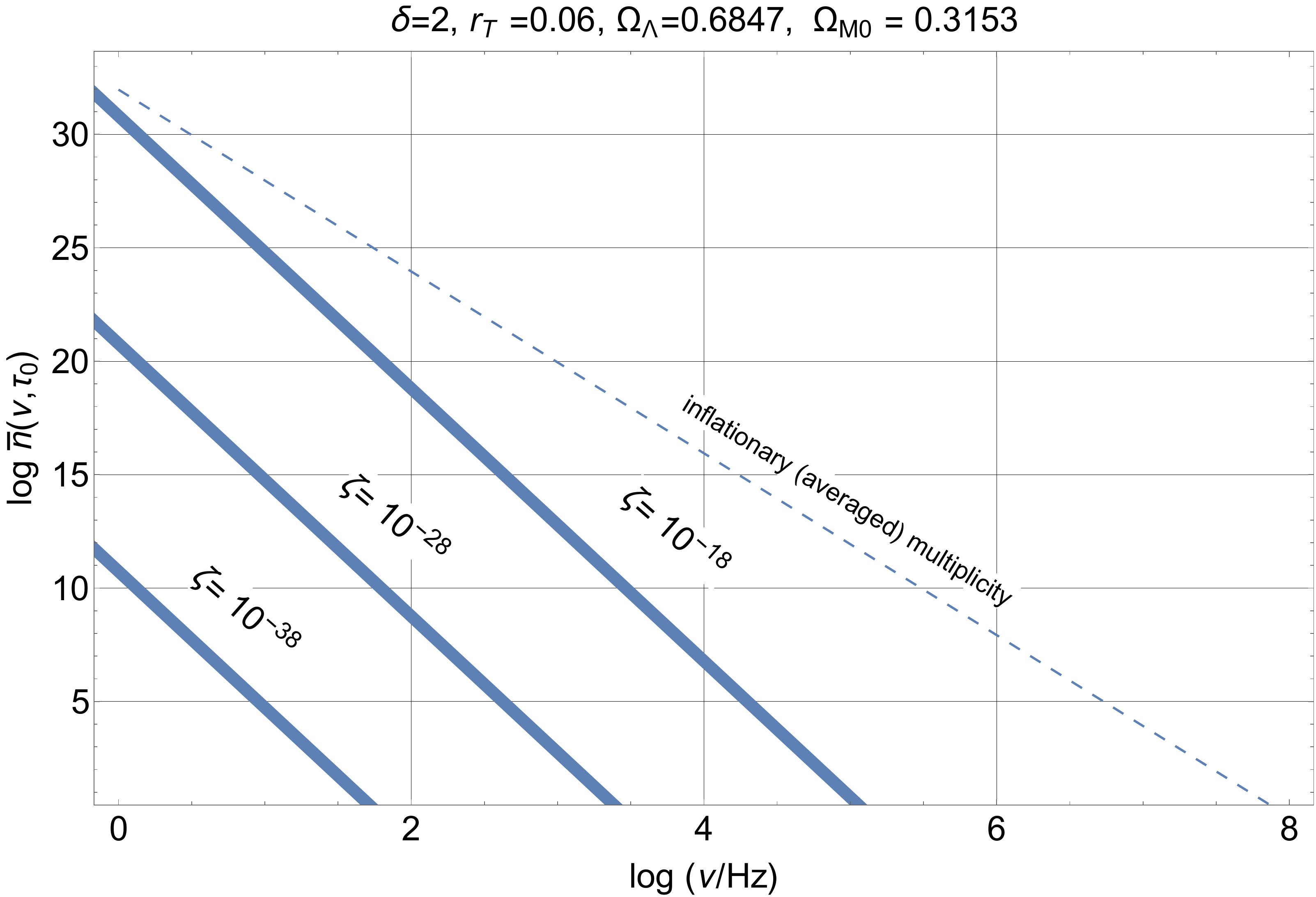}
\includegraphics[height=5.9cm]{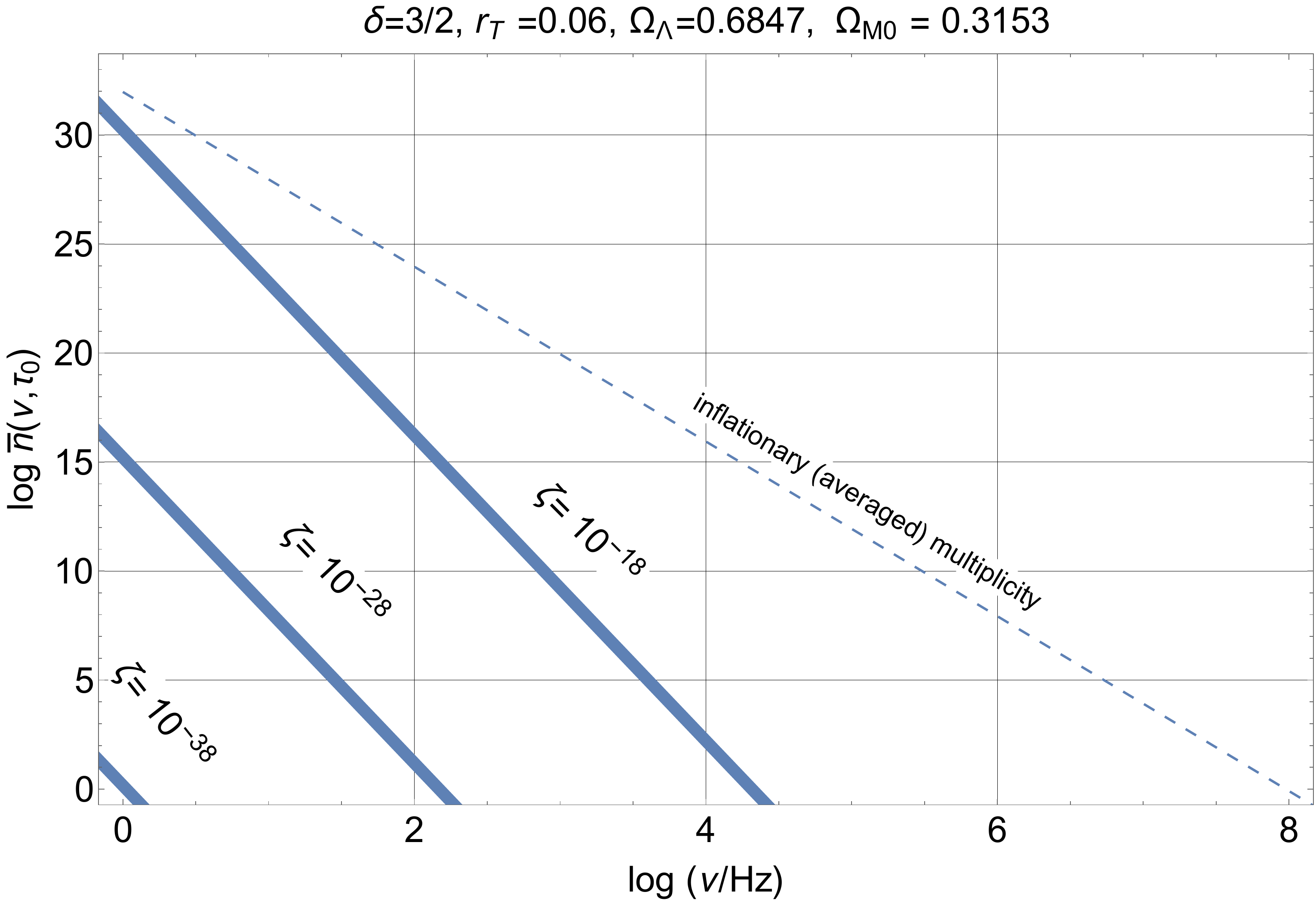}
\caption[a]{The averaged multiplicity is illustrated as a function of the frequency when  
the post-inflationary expansion rate is either slower (two upper plots) or faster (two lower 
plots) than radiation. The dashed line in the four plots represents the common logarithm of the averaged multiplicity computed in the case of the concordance paradigm.}
\label{FIGURE8}      
\end{figure}

All in all we see clearly form Fig. \ref{FIGURE7} that in the MHz range 
the average multiplicity is between $15$ and $20$ orders of magnitude larger 
than in the case of the concordance paradigm and this aspect is also emphasized in Fig. \ref{FIGURE8} 
where we report the common logarithm of the averaged multiplicity 
as a function of the common logarithm of the frequency for different values of $\zeta$ 
and $\delta$. The two upper plots of Fig. \ref{FIGURE8} refer to the case where the post-inflationary expansion rate is slower than radiation: in this situation the averaged multiplicity always exceeds the result of the concordance paradigm represented by the dashed line. The largest multiplicity is obtained for $\zeta_{min} = {\mathcal O}(10^{-38})$ and it corresponds to $H_{r} = {\mathcal O}(10^{-44})\, M_{P}$; in this case the radiation dominance takes place right before BBN (see also Eq. (\ref{NT27}) and discussion thereafter). 

Let us finally consider the possible effects associated with a coherent component since,  
as already mentioned, the inhomogeneities of the waterfall field may induce a secondary 
graviton spectrum between the MHz and the GHz \cite{WAT4} with high-frequency slopes
that can be as steep as the vacuum fluctuations and, in some cases, even steeper.
Since the slopes are larger than in the cases considered 
previously in this section the spectral energy density is practically concentrated in a narrow band of frequency around the maximum (see \cite{WAT4} and discussion therein). This is why the averaged multiplicity of produced gravitons is either scale-invariant (as in the vacuum case) or it may even  increase.
\begin{figure}[!ht]
\centering
\includegraphics[height=5.8cm]{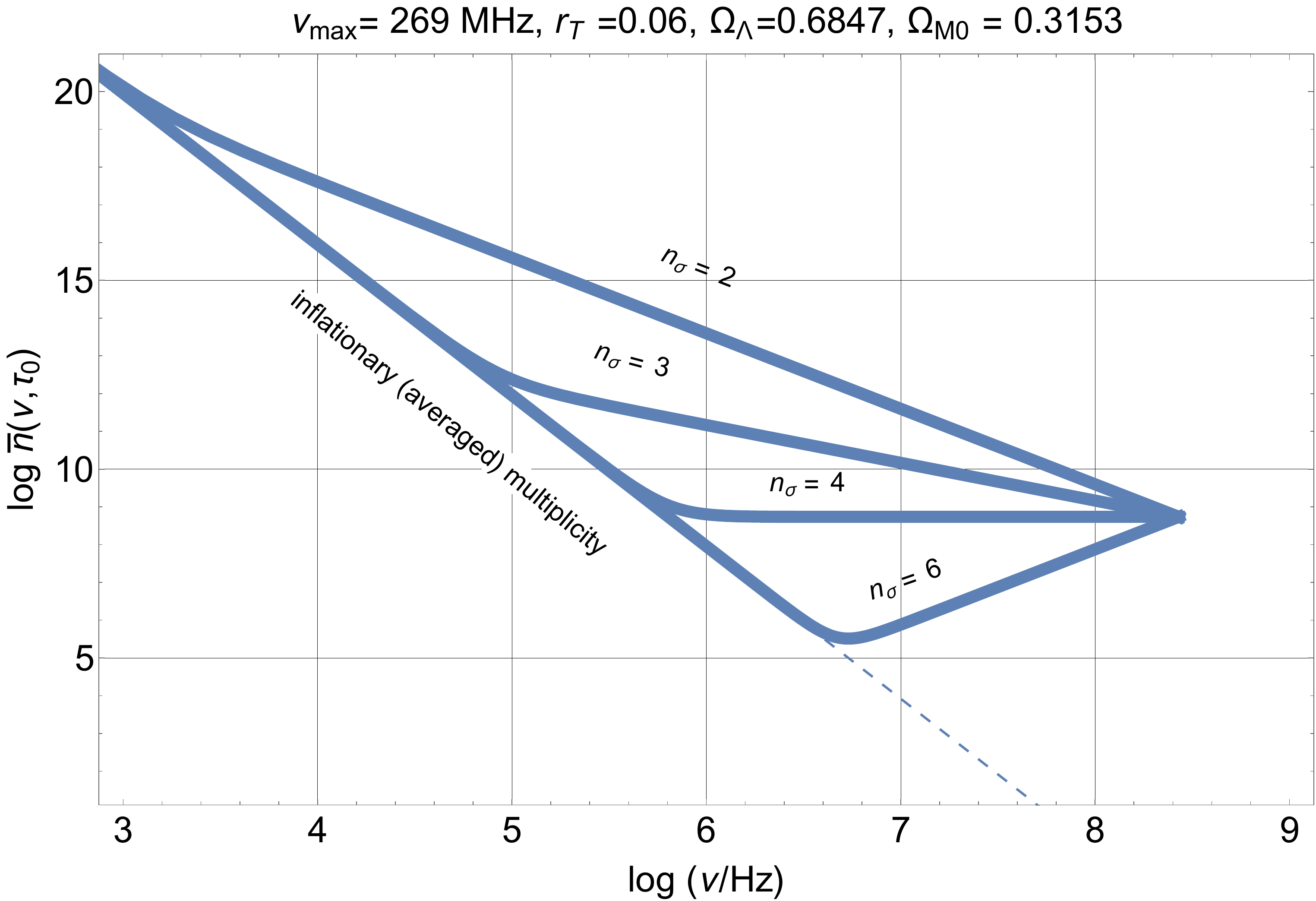}
\includegraphics[height=5.8cm]{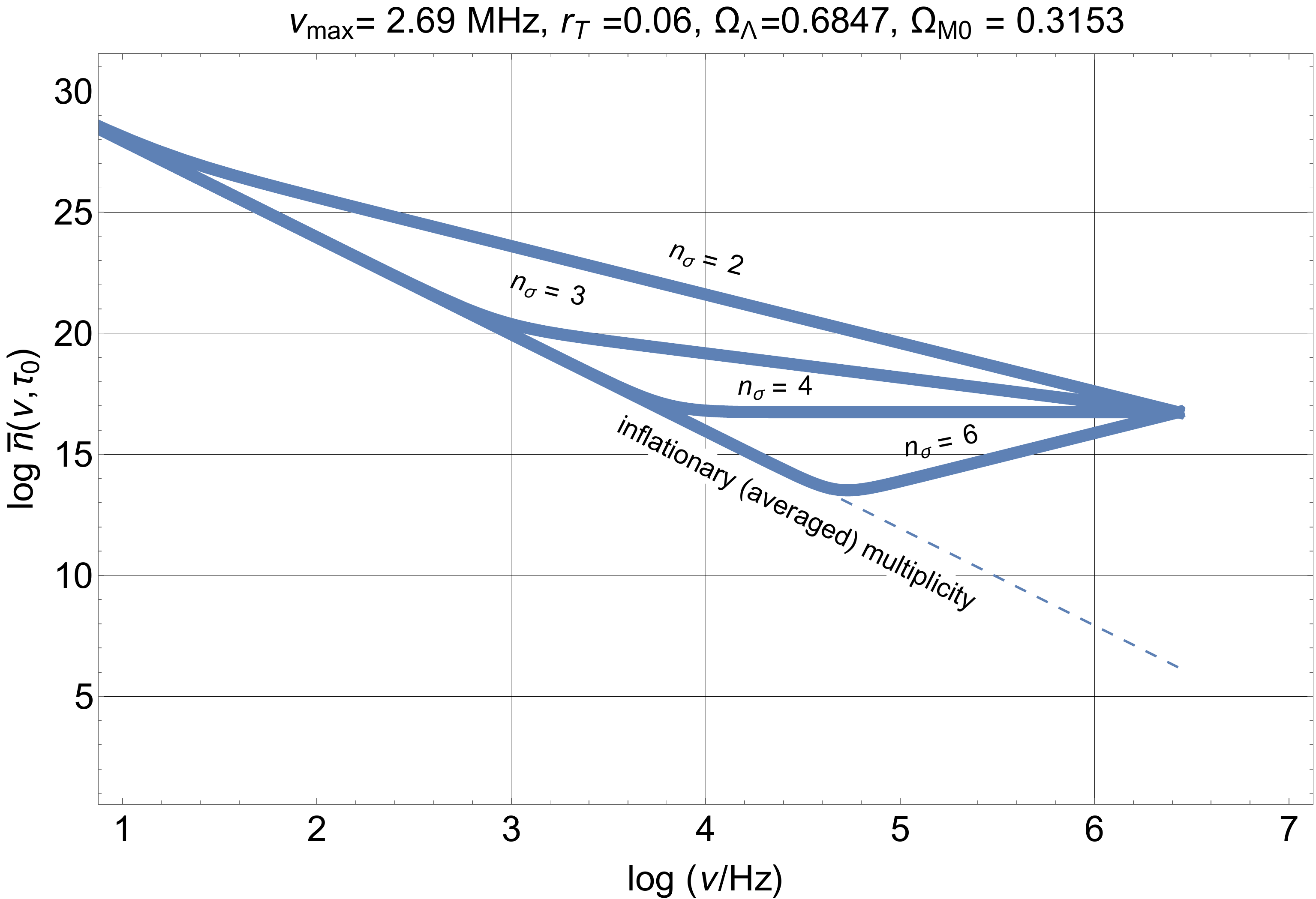}
\caption[a]{The averaged multiplicity in the presence of a large coherent component is illustrated as a function of the frequency. The left and the right plots corresponds, respectively, to $\nu_{max} \simeq \overline{\nu}_{max}$ and  $\nu_{max}= {\mathcal O}(10^{-2})\overline{\nu}_{max}$. The dashed line denote the averaged multiplicity of the concordance paradigm.}
\label{FIGURE9}      
\end{figure}
 In Fig. \ref{FIGURE9} we illustrate various cases for two specific maximal frequencies: in the 
left plot the maximal frequency coincides with $\overline{\nu}_{max}= {\mathcal O}(269) \, \mathrm{MHz}$ 
while in the plot at the right $\nu_{max} = 10^{-2} \, \overline{\nu}_{max}$. 
We remind that the averaged multiplicity scales, in this case, as 
 $(\nu/\nu_{max})^{2 (n_{\sigma} -3)}$ and when $n_{\sigma}\to 3$ $\overline{n}(\nu,\tau_{0})$
 is, approximately, frequency-independent as in the vacuum case.
 We actually remind that. In the other cases $\overline{n}(\nu,\tau_{0})$ 
always grows sharply but the absolute value of the averaged multiplicity remains 
generally smaller than ${\mathcal O}(10^{25})$. This happens because of the 
BBN constraint that forbids arbitrary large amplitudes of the spectral energy density 
in the high-frequency region.

In the previous figures we always assumed $r_{T} = 0.06$ but a similar analysis can be conducted 
for smaller values of $r_{T}$. In particular smaller valued of $r_{T}$ do not affect the slope
$m_{T}$ introduced in Eq. (\ref{NT23}) but they suppress the low-frequency normalization. 
There are other collateral effects associated with a drastic reduction of $r_{T}$ but they 
will only be swiftly mentioned here. In general terms we could however say 
that the theoretical considerations reported before do not assume any specific value of $r_{T}$ 
which is instead relevant for the explicit numerical evaluations.

The first observation is that, in the non-thermal case, the maximal frequency of the spectrum depends on $r_{T}$ 
(see, for instance, Eqs. (\ref{BBN1a}) and (\ref{NT21})). This means that a reduction of $r_{T}$ (while the other 
parameters are kept fixed) entails a reduction of the maximal frequency. The maximal frequency may therefore 
shift from the MHz region to the audio band. This effect may however be compensated by different values 
of $\zeta$ and $\delta$. Since the maximal frequency depends on $r_{T}$ also 
the spectral energy density at the maximum depends on $r_{T}$. In particular 
we have that 
\begin{equation}
 \Omega_{gw}(\nu_{max}, \tau_{0}) = \frac{8}{3 \pi} \, \Omega_{R0} \biggl(\frac{H_{1}}{M_{P}}\biggr)^{4/(\delta+1)} \, \biggr(\frac{H_{r}}{M_{P}}\biggr)^{2(\delta -1)/(\delta+1)}.
\label{RRR1}
\end{equation}
If we now use the consistency relations that have been assumed throughout this discussion, Eq. (\ref{RRR1}) 
can be written as 
\begin{equation}
h_{0}^2 \Omega_{gw}(\nu_{max}, \tau_{0}) = {\mathcal C}(\delta)  h_{0}^2 \Omega_{R0} (r_{T}\,\, {\mathcal A}_{{\mathcal R}})^{2/(\delta+1)} \, \biggr(\frac{H_{r}}{M_{P}}\biggr)^{2(\delta -1)/(\delta+1)},
\label{RRR2}
\end{equation}
where ${\mathcal C}(\delta) = (1/6)(16/\pi)^{(\delta-1)/(\delta+1)}$. This means that, provided the consistency relations are 
enforced, $h_{0}^2 \Omega_{gw}(\nu_{max}, \tau_{0})$ scales as $r_{T}^{2/(\delta+1)}$. However, since $H_{r}$, $\delta$ and $r_{T}$ are all physically independent, a reduction of $r_{T}$ does not necessarily imply a suppression of the maximum of the spectral energy density.

\subsection{The chirp and spectral amplitudes}
\begin{figure}[!ht]
\centering
\includegraphics[height=7.5cm]{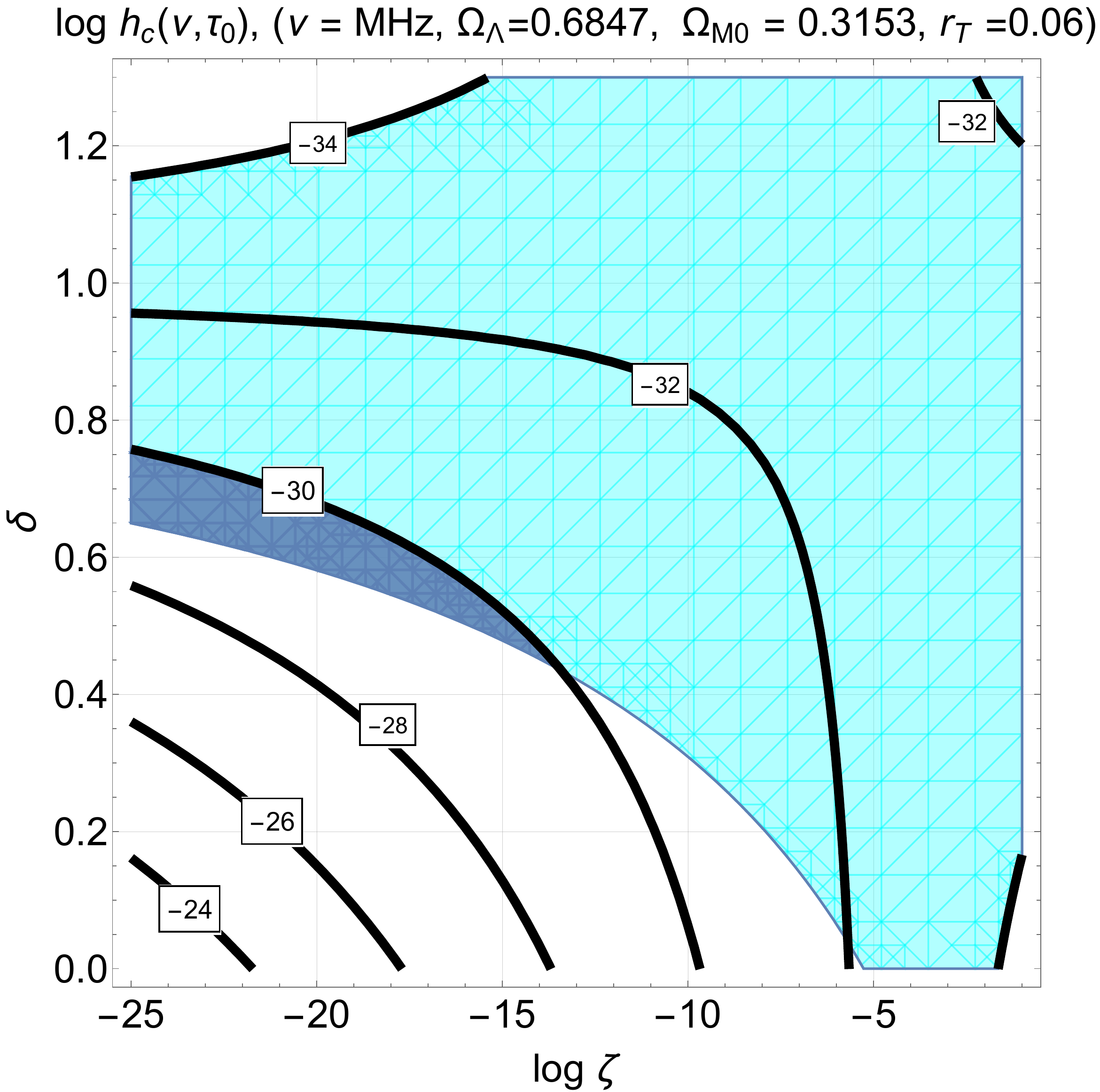}
\includegraphics[height=7.5cm]{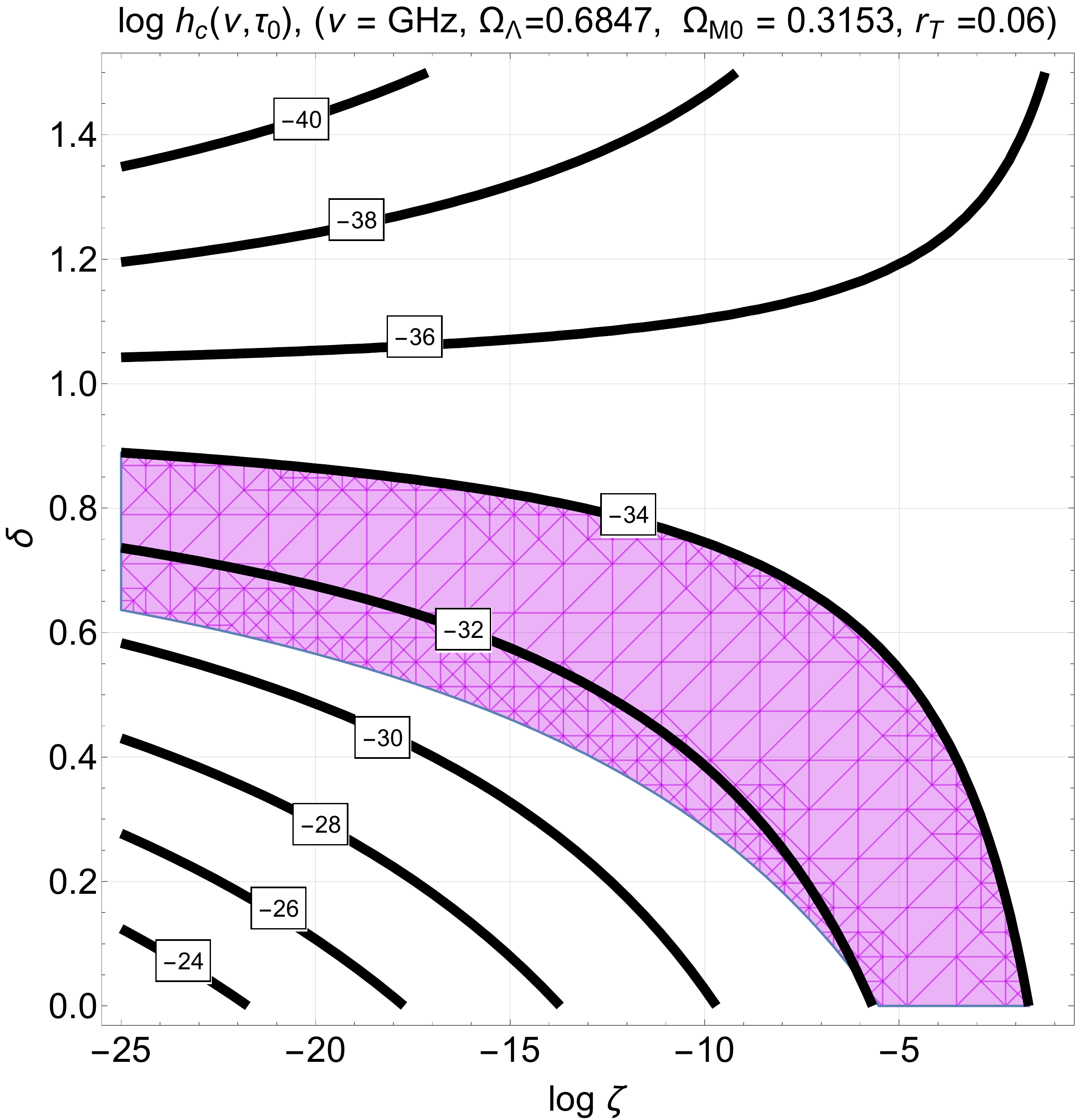}
\caption[a]{The common logarithms of the chirp amplitude are illustrated in both plots 
for different ranges of the comoving frequency.  In the plot at the left 
the typical frequency is ${\mathcal O}(\mathrm{MHz})$ while in the right plot $\nu = {\mathcal O}(\mathrm{GHz})$. The shaded regions define the area where 
$h_{c}(\nu,\tau_{0}) > 10^{-34}$. The darker portion in the left plot corresponds to 
a larger $h_{c}(\nu,\tau_{0})$ (i.e. $h_{c}(\nu,\tau_{0}) > 10^{-30}$) but this region disappears in the GHz range.}
\label{FIGURE10}      
\end{figure}

We are now going to evaluate the chirp and the spectral amplitudes 
in the non-thermal case with the purpose of estimating the minimal detectable $h_{c}(\nu,\tau_{0})$ and $\sqrt{S_{h}(\nu, \tau_{0})}$. In Fig. \ref{FIGURE10} the contours correspond to the common logarithm 
of $h_{c}(\nu,\tau_{0})$ associated with the non-thermal multiplicities illustrated in Figs. \ref{FIGURE5}, \ref{FIGURE7} 
and \ref{FIGURE8}; in the left plot of Fig. \ref{FIGURE10} the typical frequency is ${\mathcal O}(\mathrm{MHz})$ 
while in the plot at the right $\nu = {\mathcal O}(\mathrm{GHz})$.
As usual, in both plots of Fig. \ref{FIGURE10} the common logarithms of the chirp amplitude are 
illustrated by the various labels that are constant along the various contours. The shaded regions correspond to $h_{c}(\nu,\tau_{0}) > 10^{-34}$ while all the other constraints are enforced. The darker area of the left plot  in Fig. \ref{FIGURE10} illustrates the condition $h_{c}(\nu,\tau_{0}) > 10^{-30}$ but this region is actually absent from the right plot since 
the values of the chirp amplitudes are comparatively more suppressed as the 
frequency increases from the MHz to the GHz. This means that while the size of the allowed region shrinks,  in both ranges to detect the non-thermal gravitons $h_{c}(\nu,\tau_{0})$ must be, at least, of the order of $10^{-34}$ (or smaller). The minimal detectable chirp amplitude in the MHz and GHz regions must be therefore $h_{c}^{(min)}< {\mathcal O}(10^{-34})$.
 
\begin{figure}[!ht]
\centering
\includegraphics[height=7.5cm]{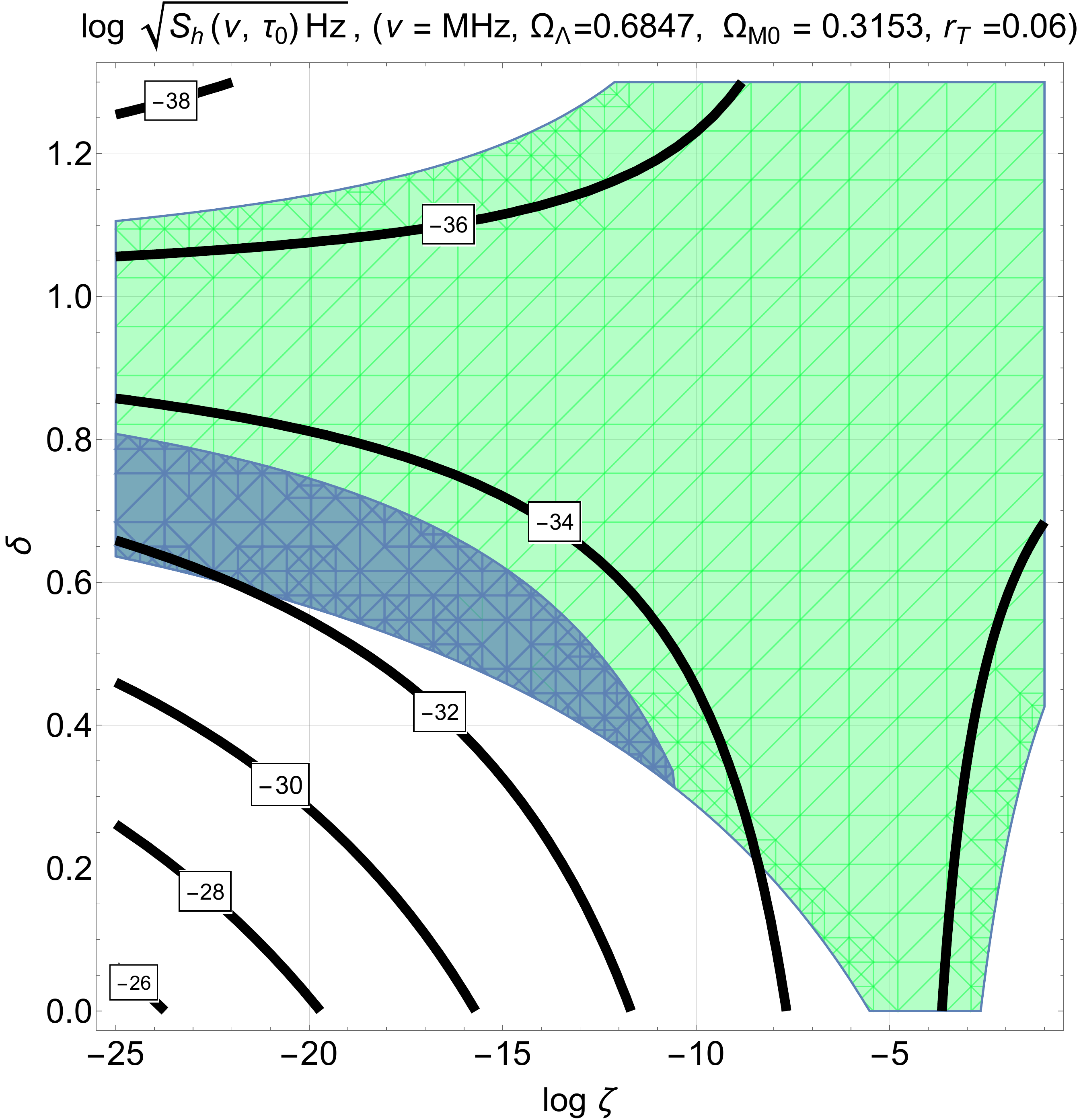}
\includegraphics[height=7.5cm]{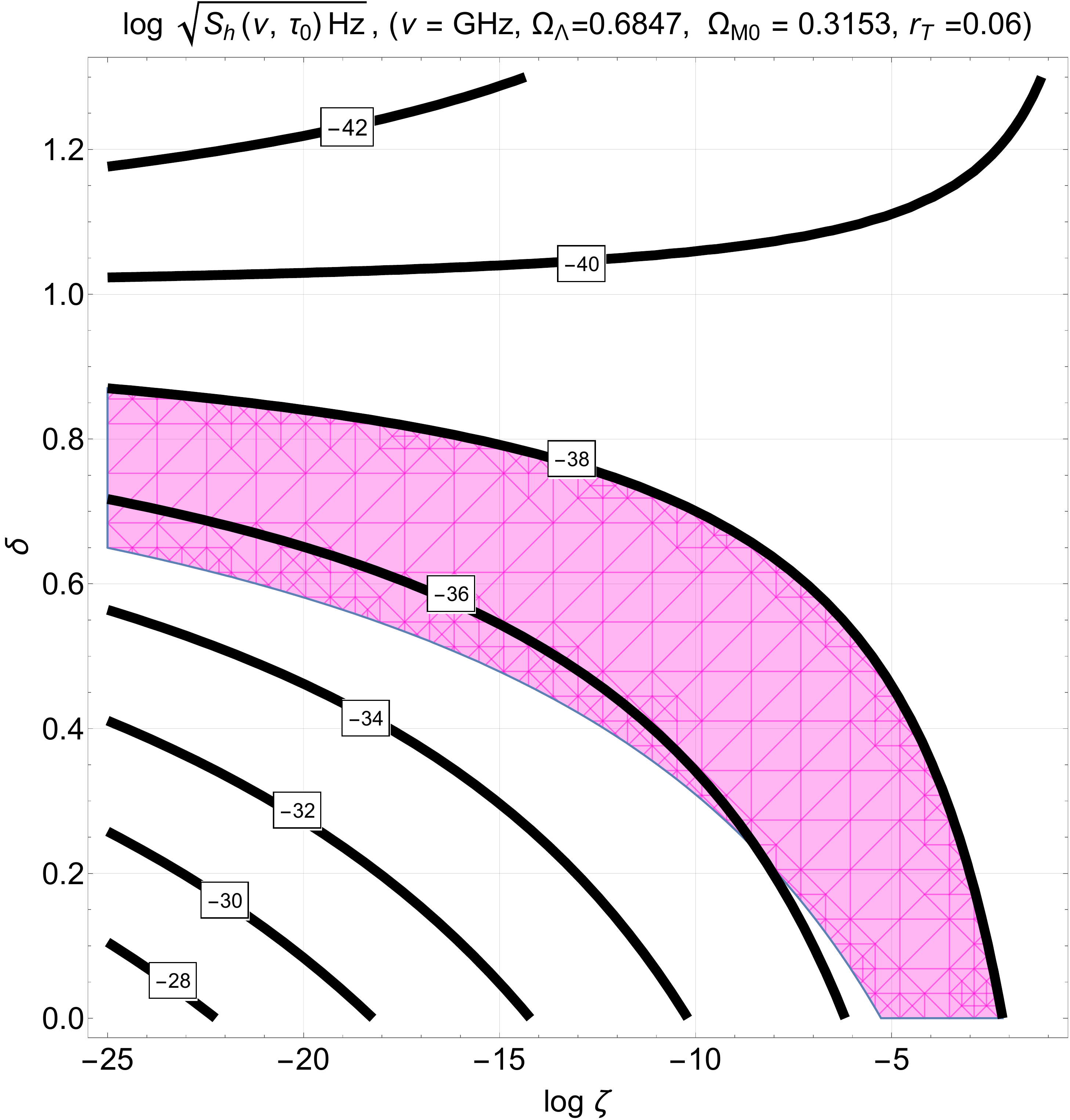}
\caption[a]{We illustrate the common logarithm of $\sqrt{S_{h}(\nu,\tau_{0})}$ in units 
of $\mathrm{Hz}^{-1/2}$. The logic of both plots is the same of Fig. \ref{FIGURE10}; the labels indicate the common logarithm of the square root of the spectral amplitude. In the plot at the left the typical frequency is ${\mathcal O}(\mathrm{MHz})$ while in the right plot $\nu = {\mathcal O}(\mathrm{GHz})$. The shaded regions in both plots correspond to the requirement $\sqrt{S_{h}(\nu,\tau_{0}) } > 10^{-38}\,\,\mathrm{Hz}^{-1/2} $ while the darker 
area in the plot at the left illustrates the condition  $\sqrt{S_{h}(\nu,\tau_{0}) \,} > 10^{-35}\,\,  \mathrm{Hz}^{-1/2}$.}
\label{FIGURE11}      
\end{figure}

 The same analysis of Fig. \ref{FIGURE10} is now translated in terms of the 
 spectral amplitude and  in Fig. \ref{FIGURE11} 
 the labels appearing on the various contours represent in fact the common logarithm 
 of $\sqrt{S_{h}(\nu,\tau_{0})}$ in units of $\mathrm{Hz}^{-1/2}$ (this means that what 
 we effectively illustrate is $\sqrt{S_{h}(\nu,\tau_{0})\, \mathrm{Hz}}$). 
The shapes of the shaded regions in Figs. \ref{FIGURE10} and \ref{FIGURE11} are similar and we purposely set the numerical requirements in order to emphasize this similarity even if the frequency slopes are different in the two cases. In Fig. \ref{FIGURE11} the shaded area of the left plot corresponds to 
$\sqrt{S_{h}(\nu,\tau_{0}) \, } \geq 10^{-38}\, \mathrm{Hz}^{-1/2}$ while the darker region 
corresponds to $\sqrt{S_{h}(\nu,\tau_{0}) } \geq 10^{-35}\, \, \mathrm{Hz}^{-1/2}$. As in the case of Fig. 
\ref{FIGURE9} the area of the allowed region shrinks as the frequency increases
from the MHz to the GHz. With the same logic employed above we can therefore conclude that $\sqrt{S_{h}^{(min)}\, } < {\mathcal O}(10^{-38})\,\, \mathrm{Hz}^{-1/2}$ if we want to cut through the relevant region of the parameter space.  

\begin{figure}[!ht]
\centering
\includegraphics[height=7.5cm]{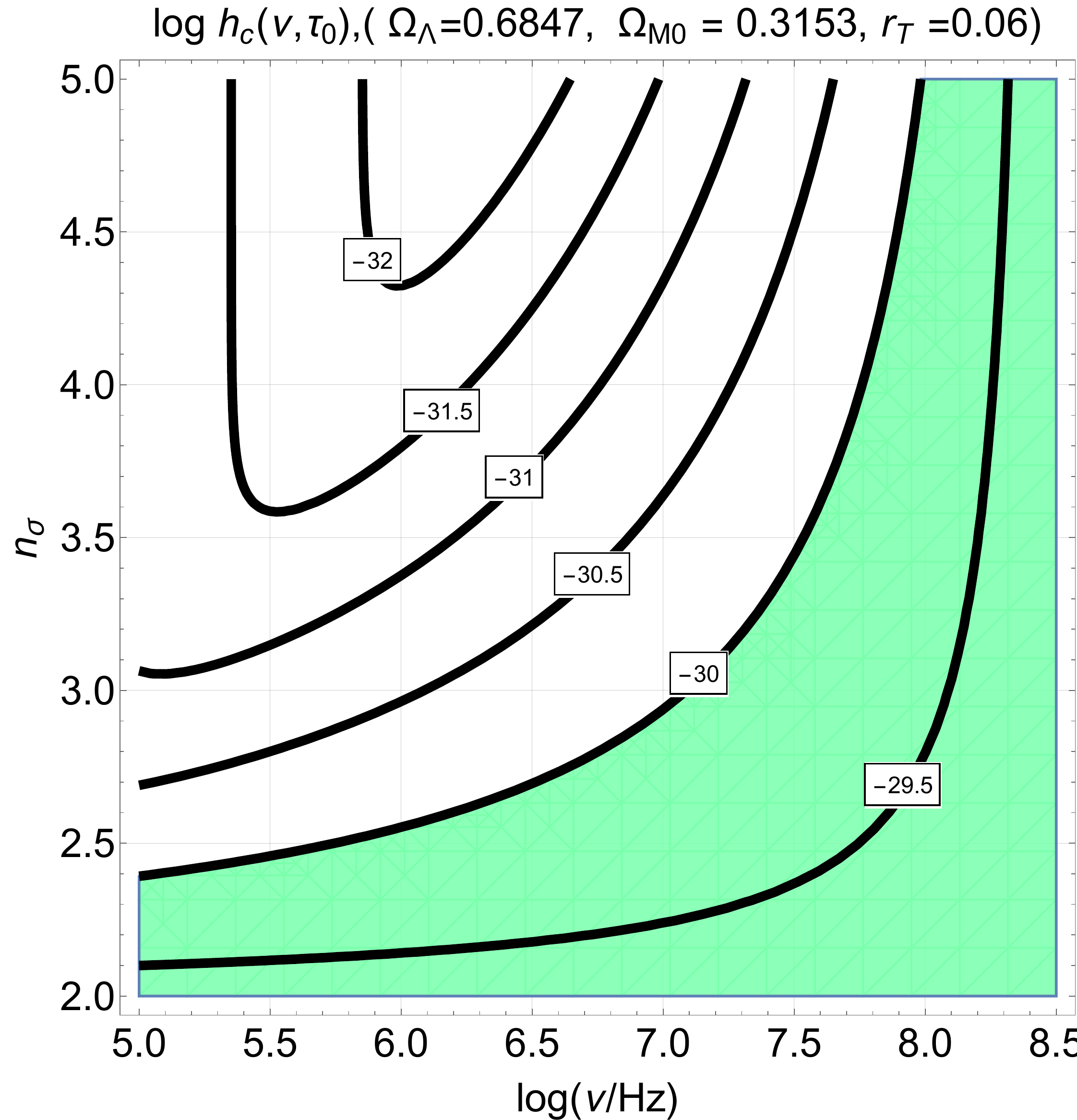}
\includegraphics[height=7.5cm]{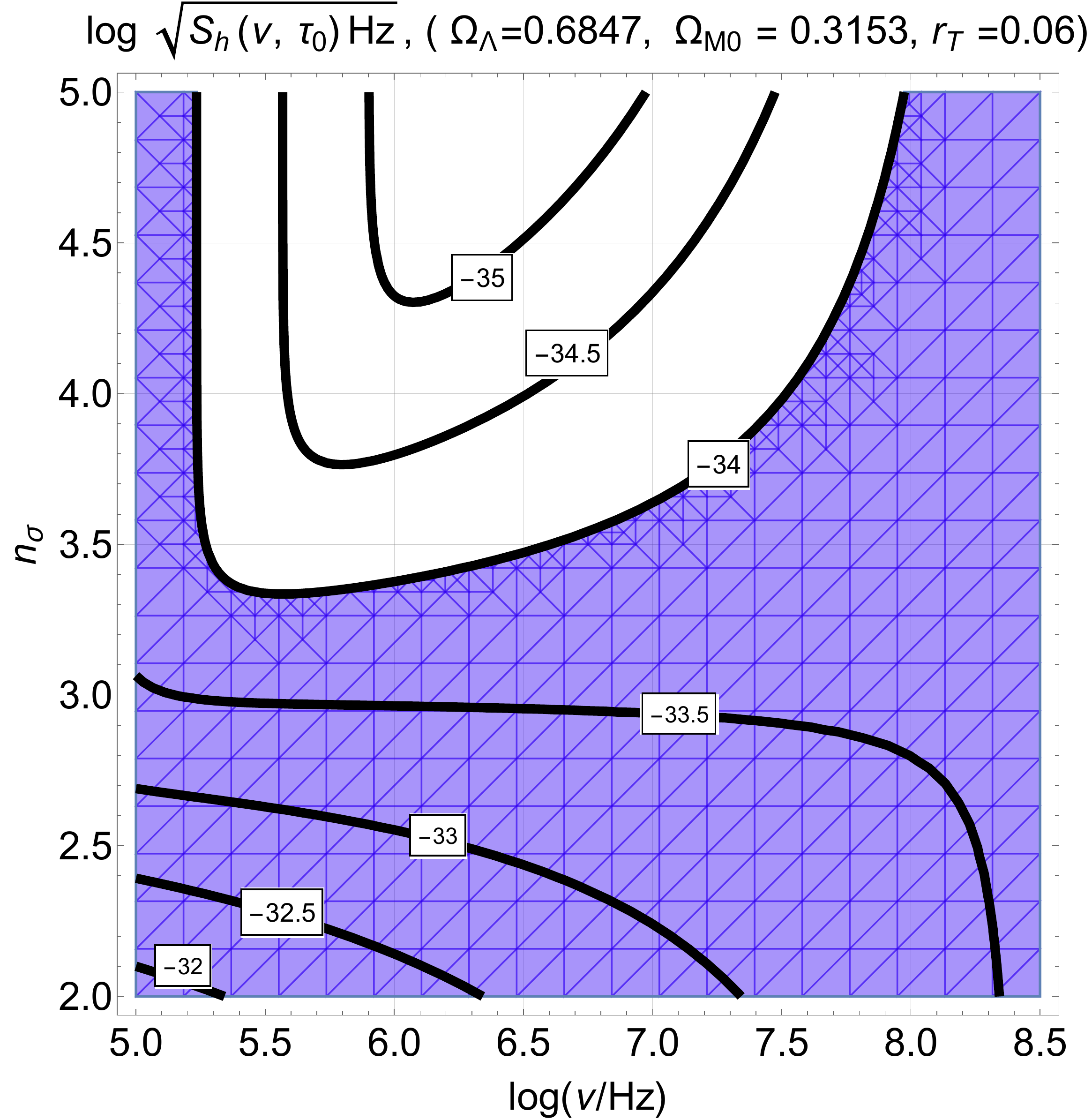}
\caption[a]{We consider the case of a coherent component leading to a sharply increasing 
averaged multiplicity (see also Fig. \ref{FIGURE9}). In the left plot we study the common logarithm of the chirp amplitude and the shaded region corresponds to the largest 
values of the chirp amplitude (i.e. $h_{c}(\nu,\tau_{0}) > 10^{-30}$). 
In the right plot the common logarithm of the spectral amplitude is illustrated and 
within the shaded area $\sqrt{S_{h}(\nu,\tau_{0})} > 10^{-34} \,\, \mathrm{Hz}^{-1/2}$. }
\label{FIGURE12}      
\end{figure}

In the presence of a coherent 
 component (already illustrated in Fig. \ref{FIGURE9}) the situation is slightly different and 
 it is swiftly analyzed in Fig. \ref{FIGURE12} where, in the left plot, we report the common logarithm of the chirp amplitude while in the right plot we illustrate the common logarithm of $\sqrt{S_{h}(\nu,\tau_{0})\, \,\mathrm{Hz}}$.
The interesting physical region suggests that $h_{c}^{min} < {\mathcal O}(10^{-30})$ and $\sqrt{S_{h}^{(min)}\,} < {\mathcal O}(10^{-34})\,\,\mathrm{Hz}^{-1/2}$. In the case of Fig. \ref{FIGURE12} the spectral slopes are larger than in the vacuum case and 
the most favourable region is close to the maximal frequency. All in all, in the case of non-thermal production from the vacuum the minimal chirp and spectral amplitudes probed by a future high-frequency instrument should be  
\begin{equation}
h_{c}^{(min)}\leq {\mathcal O}(10^{-34}), \qquad \sqrt{S_{h}^{(min)}} < {\mathcal O}(10^{-38})\,\,\mathrm{Hz}^{-1/2}, \qquad \nu \geq \mathrm{MHz}.
\label{MIN1}
\end{equation}
In the case of coherent component these requirements are slightly relaxed but in a comparatively 
larger frequency window
\begin{equation}
h_{c}^{(min)}\leq {\mathcal O}(10^{-30}), \qquad \sqrt{S_{h}^{(min)}} < {\mathcal O}(10^{-34})\,\,\mathrm{Hz}^{-1/2}, \qquad \nu \geq \mathrm{GHz}.
\label{MIN2}
\end{equation}
If we now compare the requirements of Eqs. (\ref{MIN1})--(\ref{MIN2}) with the ones of Eqs. (\ref{gravBB11})--(\ref{gravBB12})
and (\ref{gravBB13}) we can preliminarily conclude that the two sets of results are broadly compatible within few orders of magnitude. 
If these sensitivities will be one day achieved the thermal and the non-thermal gravitons could even be distinguished by looking 
at their degree of correlation, as we are going to suggest in the following section.
\renewcommand{\theequation}{5.\arabic{equation}}
\setcounter{equation}{0}
\section{Distinguishing between thermal and non-thermal gravitons}
\label{sec5}
The average multiplicity of the relic gravitons is ${\mathcal O}(\mathrm{few})$ around the 
maximal frequency $\nu_{max}$ so that in Figs. \ref{FIGURE5} and \ref{FIGURE7} {\em the smallest averaged multiplicity} would correspond, by definition, to the production {\em of a single pair of relic gravitons}. If the comoving frequency is large enough, a sufficiently sensitive detector operating for 
$\nu= {\mathcal O}(\nu_{max})$ in the range of the maximal frequency could detect bunches graviton pairs. In this section we are going to argue that, under the conditions of section \ref{sec4}, the statistical properties of the bunches of the gravitons can be directly assessed.
 One might also say   
 that since $h_{0}^2 \Omega_{gw}(\nu, \tau_{0})$ scales as
$\nu^4$  when $\overline{n}(\nu,\tau_{0})$ is (approximately) frequency-independent (as in the coherent case),  single gravitons (or bunches of few gravitons) could be preferentially detected in the high-frequency range. A similar argument is in fact due to Dyson \cite{FA} who suggested that only at high frequencies it will be eventually possible to detect single gravitons. There are however two important differences between the present considerations and the suggestions of Ref. \cite{FA}. Unlike the situation of Ref. \cite{FA}, 
the maximal frequency {\em is not arbitrary}\footnote{In Ref. \cite{FA} the maximal frequency is actually arbitrary and it can 
even go up to $10^{6}$ GHz while in the present case $\nu_{max}$ is fixed in terms of $\delta$ and $\zeta$ (see Fig. \ref{FIGURE6} 
and discussion therein). In the thermal case the maximal frequency is discussed in Eqs. (\ref{gravBB7})--(\ref{gravBB8}).} and, in the non-thermal case, it is ultimately determined by the curvature scale at the end of inflation and by the post-inflationary evolution. The second difference is that, in the present approach, we consider the potential high-frequency signals caused by the relic gravitons and, in some cases, 
these signals may even exceed the contribution of the single graviton line at lower frequencies.

As already discussed above, the  microwave cavities operating in the MHz and GHz regions \cite{EB,cav1a,EC,EC2,ED,EE,EF,EG,MG1,MG2,MG3} can be employed for the detection of gravitons at high-frequencies but other classes of detectors (based on the interaction of gravitons with dynamical electromagnetic fields) have been proposed through the years \cite{EH,EI,EL,EM,EN,EO,EP,EQ}. It is also interesting to consider, in our perspective, the experiments where gravitons are converted into photons in strong magnetic fields, as originally suggested in Ref. \cite{FA} (see also \cite{FA2, FA3,FA4}). 
In most of these studies the potential sources are not mentioned so that, as a consequence, the required chirp amplitudes are optimistically set in the range $h_{c}^{(min)} = {\mathcal O}(10^{-20})$. These goals do not follow from any specific analysis of the potential signals and the present considerations clarified that $h_{c}^{(min)}$ must be at least ${\mathcal O}(10^{-34})$, or smaller. 

While the high-frequency detectors are more challenging than often suggested, they are essential for 
the detection of relic gravitons in MHz and GHz regions and also for a direct scrutiny of their 
statistical properties. In this spirit, the idea conveyed in this section is, in short, the following: if the single graviton line is reached with the sensitivities of actual instruments and prototypes\footnote{Even if this development is not obvious, the results and the implications could be, in our view, even more essential than the ones associated with the current astrophysical observations taking place in the audio band.} it will be possible to scrutinize the statistical properties of the relic gravitons by analyzing their second-order correlation effects as already pointed out in the recent past \cite{HBT1,HBT2,HBT3}.
In this respect the high-frequency domain turns out to be, once more, 
the most promising for the study of the Hanbury-Brown Twiss correlations \cite{QO1} 
associated with the relic gravitons\footnote{The quantum mechanical properties of visible radiation are not apparent in the standard interference experiments conducted within the tenets of Young interferometry where quantum concepts have been considered not directly relevant until the celebrated experiment of R. Hanbury-Brown and R.Twiss in the1950s \cite{HBT0a,HBT0b}. This is the  origin of the terminology employed in this section.}.

\subsection{Single graviton detection}
Equation (\ref{NP1}) can be expressed in the limit $\overline{n}(\nu,\tau_{0}) \to {\mathcal O}(1)$ and since this requirement defines the maximal frequency $\nu_{max}$ we also have that the spectral energy density 
in critical units for $\nu = {\mathcal O}(\nu_{max})$ becomes:
\begin{equation}
h_{0}^2 \Omega_{gw}(\nu_{max}, \tau_{0}) = \frac{128\, \pi^3}{3} \frac{\nu_{max}^4 }{H_{0}^2 \, M_{P}^2} \, \overline{n}(\nu,\tau_{0}) = 3.66\times 10^{-49} \biggl(\frac{\nu_{max}}{\mathrm{Hz}}\biggr)^{4}.
\label{HBTC1}
\end{equation}
We may now recall Eqs. (\ref{BBN1a}) and (\ref{NT24})--(\ref{NT25}) (as well as Figs. \ref{FIGURE6} and \ref{FIGURE7}) and insert the explicit form of $\nu_{max}$ into Eq. (\ref{HBTC1}); the resulting expression is
\begin{equation}
h_{0}^2 \Omega_{gw}(\nu_{max}, \tau_{0})  = 1.92\times 10^{-15}  \zeta_{1}^{\alpha_{1}}\,\, \zeta_{2}^{\alpha_{2}}\,.\,.\,.\,.\,\zeta_{N-1}^{\alpha_{N-1}},
\label{HBTC1a}
\end{equation}
where $\alpha_{i} = 2\,(\delta_{i} -1)/(\delta_{i} +1)$. Equation (\ref{HBTC1a}) estimates the spectral energy density in critical units for $\nu = {\mathcal O}(\nu_{max})$ and it depends on the post-inflationary evolution via the $\zeta_{i}$ and $\delta_{i}$. Since, by definition $\zeta_{i} = H_{i+1}/H_{i}<1$, 
the maximal signal can be expected for $\alpha_{i}<0$ and this happens when 
all the successive stages expand at a rate that is slower than radiation. Furthermore 
it can be argued that the most favourable situation for the maximization of the signal 
occurs when {\em all} the $\delta_{i}$ 
coincide so that, as we already saw, $\zeta= (H_{r}/H_{1})= 
\zeta_{1}\, \zeta_{2} \,.\,.\,.\,.\,\zeta_{N-1}^{\alpha_{N-1}}> 10^{-38}$.
\begin{figure}[!ht]
\centering
\includegraphics[height=7cm]{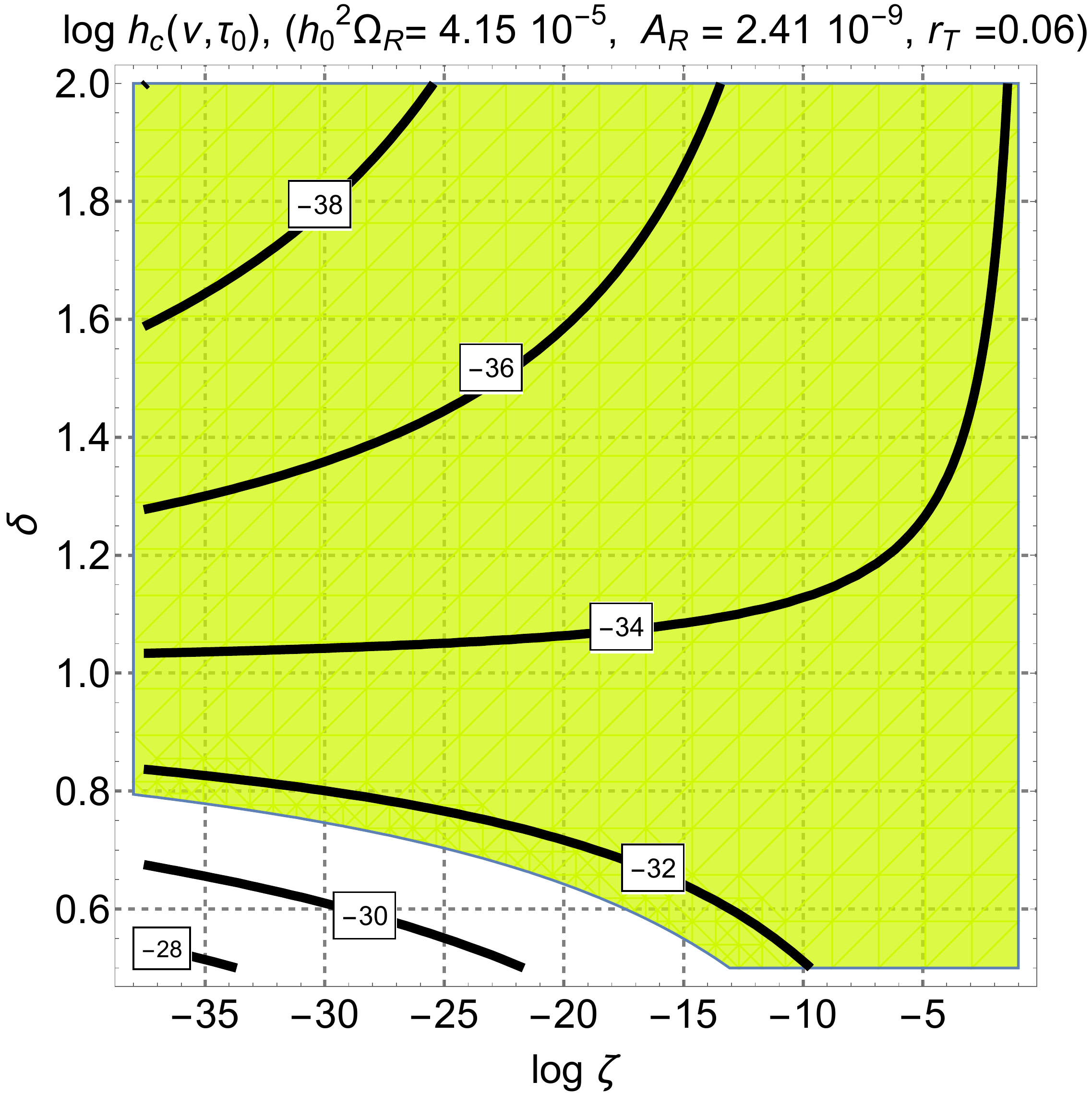}
\includegraphics[height=7cm]{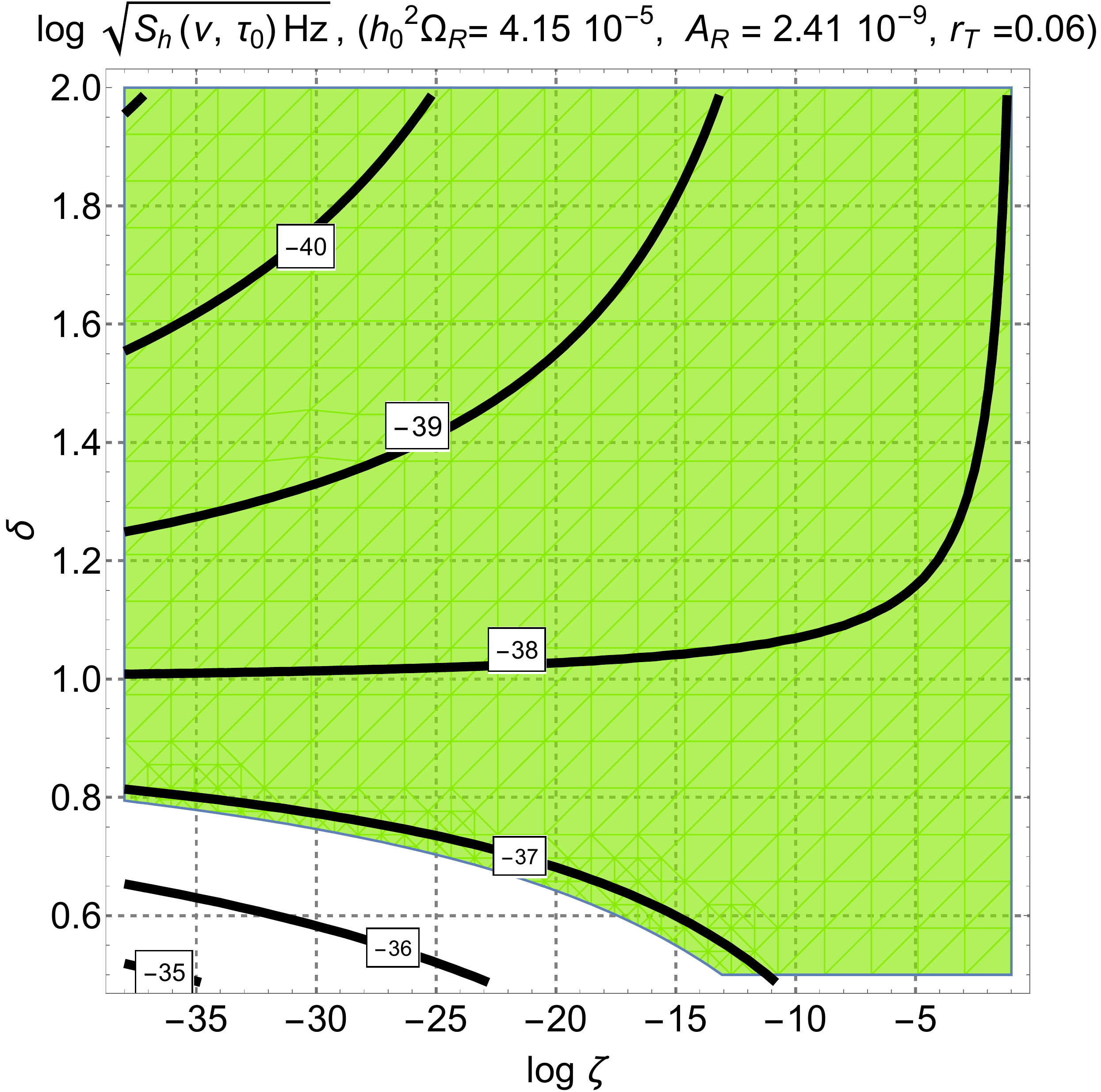}
\caption[a]{The chirp and the spectral amplitudes are computed when the comoving 
frequency coincides with $\nu_{max}$ and in the plane defined by $\log{\zeta}$ and $\delta$. From both plots it appears that the largest $h_{c}(\nu,\tau_{0})$ and $S_{h}(\nu,\tau_{0})$ arise when $\delta<1$ for different values of $\zeta$. This corresponds to the situation 
where the post-inflationary stage is not excessively long and 
its expansion rate is slower than radiation.
 The minimal detectable chirp and spectral amplitudes should 
be comparable with the maximal signals in this portion of the parameter space. }
\label{FIGURE13}      
\end{figure}
This aspect is illustrated in Fig. \ref{FIGURE13} (see also, in this respect, Fig. \ref{FIGURE6}); 
the  shaded areas in both plots of Fig. \ref{FIGURE13} indicate the region of the parameter space where all the phenomenological constraints are concurrently satisfied. Even if the relevant portion of the parameter space is the one where $\delta < 1$,  for completeness  we also reported
the range $\delta >1$. According to the results of Fig. \ref{FIGURE13} the minimal detectable chirp and spectral amplitudes at $\nu_{max}$ should be:
\begin{equation}
h_{c}^{(min)} \leq {\mathcal O}(10^{-38}), \qquad \sqrt{S_{h}^{(min)}} \leq {\mathcal O}(10^{-40}) \, \mathrm{Hz}^{-1/2}, \qquad \mathrm{MHz} < \nu_{max} < \mathrm{THz},
\label{HBTC2}
\end{equation}
where the frequency range can be read-off from the left plot of Fig. \ref{FIGURE6}
by recalling that $\nu_{max}$ {\em grows well above the GHz as long as $\delta<1$}.
The requirements of Eq. (\ref{HBTC2}) are slightly more constraining than the ones 
deduced in Eq. (\ref{MIN1}) however the two are not mutually exclusive. 
On the contrary, as long as we focus on the case $\delta<1$, the conditions 
of Eq. (\ref{MIN1}) are broadly sufficient to reach the single graviton line. 

\subsection{Hanbury-Brown Twiss interferometry}
If the sensitivities compatible with Eq. (\ref{HBTC2}) are 
eventually realized in practice,  the thermal and the non-thermal gravitons lead to markedly different 
second-order correlation effects. Hanbury-Brown and Twiss \cite{HBT0a,HBT0b} first proposed intensity interferometry for stellar measurements and the idea, in a nutshell, is to detect averaged products of intensities rather than averaged products of amplitudes as it happens in the case of Young interferometry\footnote{ In quantum optics the  correlation functions involve vector quantities (e.g. the electric field operators) in the present case we are dealing with tensor fields but the Glauber theory of quantum coherence can be generalized to the tensor case  \cite{HBT1}. }. To introduce the correlation functions in the Glauber form it is 
useful to observe that the operators $\widehat{\mu}_{ij}(x)$ consist of a positive and of a negative frequency part, i.e.  $\widehat{\mu}_{ij}(x) = \widehat{\mu}_{ij}^{(+)}(x) + \hat{\mu}_{ij}^{(-)}(x)$, with 
$\widehat{\mu}_{ij}^{(+)}(x)= \widehat{\mu}_{ij}^{(-)\,\dagger}(x)$. If  $|\mathrm{vac}\rangle$ is the state that minimizes the tensor Hamiltonian when all the modes are inside the effective horizon (for instance at the onset of inflation) the operator $\hat{\mu}_{ij}^{(+)}(x)$ annihilates the vacuum (i.e.  $\hat{\mu}_{ij}^{(+)}(x) |\mathrm{vac} \rangle=0$ and $\langle \mathrm{vac} |\, \hat{\mu}_{ij}^{(-)}(x) =0$). We now recall that the field operators describing the positive and negative frequency parts can be expressed as:
\begin{eqnarray}
\widehat{\mu}_{i\,j}^{(-)}(\vec{x}, \tau) &=& \frac{\sqrt{2} \ell_{P}}{(2\pi)^{3/2}} \sum_{\alpha} \int \frac{d^{3} k}{\sqrt{2 k}} e^{(\alpha)}_{ij} \, \widehat{a}_{- \vec{k},\, \alpha}^{\dagger}(\tau) \,\, e^{- i \vec{k}\cdot\vec{x}},
\label{HBTG7}\\
\widehat{\mu}_{i\,j}^{(+)}(\vec{x}, \tau) &=& \frac{\sqrt{2} \ell_{P}}{(2\pi)^{3/2}} \sum_{\alpha}
\int \frac{d^{3} k}{\sqrt{2 k}} e^{(\alpha)}_{ij} \, 
\widehat{a}_{\vec{k},\, \alpha}(\tau) \,\, e^{- i \vec{k}\cdot\vec{x}},
\label{HBTG8}
\end{eqnarray}
where the creation and annihilation operators enter the Hamiltonian (\ref{NT6}) 
that describes the quantum theory of parametric amplification. In terms of $\widehat{\mu}_{i\,j}^{(\pm)}(\vec{x}, \tau)$ the first-order Glauber correlator in the tensor case is:
\begin{equation}
{\mathcal T}^{(1)}(x_{1},\, x_{2}) =  \langle  \hat{\mu}^{(-)}_{i\,j}(x_{1}) \, \hat{\mu}^{(+)}_{i\,j}(x_{2}) \rangle,
\label{HBTG1}
\end{equation}
where $x_{1}= (\vec{x}_{1}, \tau_{1})$ and $x_{2}=(\vec{x}_{2}, \, \tau_{2})$. While in the discussion 
of section \ref{sec2} the expectation values could also be viewed as ensemble averages defining the properties of the tensor random fields, the quantum mechanical viewpoint is the only one examined in this section. From the first-order Glauber correlation function we can also define the degree of first-order coherence:
\begin{equation}
g^{(1)}(x_{1},\, x_{2}) = \frac{{\mathcal T}^{(1)}(x_1,\, x_{2})}{\sqrt{{\mathcal T}^{(1)}(x_1,x_{1})} \,\, \sqrt{{\mathcal T}^{(1)}(x_2,x_{2})}}.
\label{HBTG2}
\end{equation}
 Equations (\ref{HBTG1})--(\ref{HBTG2}) are probed by Young interferometry in a standard two-slit experiment where the amplitudes of the radiation field coming from the two different pinholes appearing on a 
first screen interfere on a second screen. Since the degree of first-order coherence 
is only sensitive to the averaged multiplicity, $g^{(1)}(x_{1},\, x_{2})$
 does not disambiguate the nature of the quantum state. For this purpose we must instead introduce 
 the second-order Glauber correlation function:
\begin{equation}
 {\mathcal T}^{(2)}(x_{1}, x_{2}) = 
 \langle  \widehat{\mu}^{(-)}_{i\,\,j}(x_{1}) \, \widehat{\mu}^{(-)}_{k\,\,\ell}(x_{2})\widehat{\mu}^{(+)}_{k\,\ell}(x_{2}) \widehat{\mu}^{(+)}_{i\,j}(x_{1}) \rangle.
\label{HBTG3}
\end{equation}
The operator appearing inside 
the expectation value of Eq. (\ref{HBTG3}) is Hermitian; furthermore  for a single tensor 
polarization the second-order (Glauber) correlation function is even simpler, i.e.
${\mathcal S}^{(2)}(x_{1}, x_{2}) = 
 \langle  \widehat{\mu}^{(-)}(x_{1}) \, \widehat{\mu}^{(-)}(x_{2})\widehat{\mu}^{(+)}(x_{2}) \widehat{\mu}^{(+)}(x_{1}) \rangle$ where $\widehat{\mu}^{(\pm)}(x)$ are now defined as in Eqs. (\ref{HBTG7})--(\ref{HBTG8}) but by excluding the sum over the polarizations. From Eq. (\ref{HBTG3}) we can define the normalized degree of second-order coherence:
\begin{equation}
g^{(2)}(x_{1},\, x_{2}) = \frac{{\mathcal T}^{(2)}(x_1,\, x_{2})}{{\mathcal T}^{(1)}(x_1, x_{1}) \,\, {\mathcal T}^{(1)}(x_2, x_{2})}.
\label{HBTG4}
\end{equation}
In the case of a single polarization we can introduce exactly the same quantity but with a slightly different 
notation that allows its distinction from  Eq. (\ref{HBTG4}):
\begin{equation}
\overline{g}^{(2)}(x_{1},\, x_{2}) = \frac{{\mathcal S}^{(2)}(x_1,\, x_{2})}{{\mathcal S}^{(1)}(x_1, x_{1}) \,\, {\mathcal S}^{(1)}(x_2, x_{2})}.
\label{HBTG4a}
\end{equation}
Equations (\ref{HBTG3}) and (\ref{HBTG4})--(\ref{HBTG4a}) are particular cases of the $n$-th order Glauber 
correlator \cite{GG1,GG2,GG3}. More specifically, Eq. 
(\ref{HBTG3}) arises when considering the $2$-fold delayed  coincidence measurement of the tensor field at the space-time points $(x_{1},  x_{2})$. We can imagine to have a high-frequency instrument able to attain the single graviton line and consider the matrix element corresponding to the absorption of gravitons at different times and at different locations of the hypothetical detectors namely $\langle \{a\} \, |  \,\,\widehat{\mu}_{k\,\,\ell}^{(+)}(x_{2}) \widehat{\mu}_{i\,\,j}^{(+)}(x_{1})|\, \{b\}\rangle$ where we 
introduced $|\{\,a\}\rangle$ as the state of the field {\em after} the measurement 
and  $|\{\,b\}\rangle$ the state of the field  {\em before} the measurement. To obtain the rate at which the absorptions occur we must sum over the final states, i.e. 
\begin{eqnarray}
&& \sum_{\{a\}} \langle \{a\} \, | \ \,\,\widehat{\mu}_{k\,\,\ell}^{(+)}(x_{2}) \widehat{\mu}_{i\,\,j}^{(+)}(x_{1})|\, \{b\}\rangle\biggr|^2 =
 \nonumber\\
&& \equiv \sum_{\{a\}} \langle \{ b\}| \widehat{\mu}_{i\,\,j}^{(-)}(x_{1})\,\,\widehat{\mu}_{k\,\,\ell}^{(-)}(x_{2})| \{a\}\rangle \langle \{ a \}| 
 \widehat{\mu}_{k\,\,\ell}^{(+)}(x_{2})\,.\,.\,.\, \widehat{\mu}_{i\,\,j}^{(+)}(x_{1}) |\{ b \}\rangle,
\label{HBTG5}
\end{eqnarray}
that coincides, thanks to the completeness relation, with the expectation value 
\begin{equation}
\langle \{\, b\}|\widehat{\mu}_{i\,\,j}^{(-)}(x_{1})\,\,\widehat{\mu}_{k\,\,\ell}^{(-)}(x_{2})  \widehat{\mu}_{k\,\,\ell}^{(+)}(x_{2})\widehat{\mu}_{i\,\,j}^{(+)}(x_{1}) |\{ b\} \rangle,
\label{HBTG6}
\end{equation}
which is exactly the same quantity appearing in Eq. (\ref{HBTG3}).  All in all the logic of the HBT interferometry is  rooted in the quantum mechanical analysis of intensity correlations as stressed for the first time by Glauber and Sudarshan \cite{GG1,GG2}. In the language of the quantum theory of optical coherence the current observations of gravitational waves by wide-band interferometers  between few Hz and $10$ kHz are only sensitive to the average multiplicity of the gravitons and hence to their degree of first-order coherence. This is true even assuming, rather optimistically, that the current interferometers will be one day sufficiently sensitive to detect a background of relic gravitons. This is why it is relevant to analyze this possibility at higher frequencies 
where the signal is also potentially larger.

\subsection{Inclusive and exclusive approaches}
Let us first consider the explicit form of the degrees of first- and second-order
given in Eqs. (\ref{HBTG1}) and (\ref{HBTG3}):
\begin{eqnarray}
{\mathcal T}^{(1)}(x_{1},\, x_{2}) &=& \frac{1}{(2\pi)^3} \int \frac{d^{3} k_{1}}{\sqrt{2 k_{1}}} \int \frac{d^{3} k_{2}}{\sqrt{2 k_{2}}} \, \sum_{\alpha_{1}} \sum_{\alpha_{2}} \, e^{(\alpha_{1})}_{ij}(\widehat{k}_{1}) \, e^{(\alpha_{2})}_{i\,j}(-\widehat{k}_{2}) \, e^{ - i (\vec{k}_{1}\cdot\vec{x}_{1} + \vec{k}_{2}\cdot\vec{x}_2)}
\nonumber\\
&\times& \langle \widehat{a}_{\vec{k}_{1},\,\alpha}^{\dagger}(\tau_{1}) \,  \widehat{a}_{-\vec{k}_{2},\,\beta}(\tau_{2})\rangle,
\label{HBTG9}\\
{\mathcal T}^{(2)}(x_{1}, x_{2}) &=& \frac{1}{(2\pi)^6} \int \frac{d^{3} k_{1}}{\sqrt{2 k_{1}}} 
  \int \frac{d^{3} k_{2}}{\sqrt{2 k_{2}}}  \int \frac{d^{3} k_{3}}{\sqrt{2 k_{3}}}  \int \frac{d^{3} k_{4}}{\sqrt{2 k_{4}}} 
 e^{- i (\vec{k}_{1} + \vec{k}_{4})\cdot\vec{x}_{1}}  e^{- i (\vec{k}_{2} + \vec{k}_{3})\cdot\vec{x}_{2}} 
\nonumber\\
&\times& \sum_{\alpha_{1}} \, \sum_{\alpha_{2}}  \, \sum_{\alpha_{3}}  \,\sum_{\alpha_{4}}  \,\,e_{ij}^{(\alpha_{1})}(\widehat{k}_{1}) \,\, e_{kl}^{(\alpha_{2})}(\widehat{k}_{2}) \,\, e_{kl}^{(\alpha_{3})}(\widehat{k}_{3}) \,\,
e_{ij}^{(\alpha_{4})}(\widehat{k}_{4}) 
\nonumber\\
&\times& \langle \widehat{a}^{\dagger}_{- \vec{k}_1, \, \alpha_{1}}(\tau_{1}) \,\, \widehat{a}^{\dagger}_{- \vec{k}_2, \, \alpha_{2}}(\tau_{2}) \,\,\widehat{a}_{ \vec{k}_3, \, \alpha_{3}}(\tau_{2})  \,\,\widehat{a}_{ \vec{k}_4, \, \alpha_{4}}(\tau_{1})  \rangle.
\label{HBTG10}
\end{eqnarray}
If the momenta and the polarizations are neglected in 
Eqs. (\ref{HBTG9})--(\ref{HBTG10}) we obtain, in practice, the {\em single-mode approximation}. This strategy is often employed in Mach-Zender and Hanbury Brown-Twiss interferometry \cite{QO1}.  Since many experiments use plane parallel light beams whose transverse intensity profiles are not important for the measured quantities, it is often sufficient in interpreting the data to consider the light beams as exciting a single mode of the field. In this sense the  quantum optical perspective is {\em exclusive} since a particular mode of the field is selected. With 
these caveats the degrees of first- and second-order coherence in the single-mode approximation are given by:
\begin{eqnarray}
\overline{g}^{(1)}(\tau_{1}, \tau_{2}) &=& \frac{\langle \widehat{a}^{\dagger}(\tau_{1}) \, \widehat{a}(\tau_{2})\rangle}{\sqrt{\langle \widehat{a}^{\dagger}(\tau_{1}) \, \widehat{a}(\tau_{1})\rangle}
\, \sqrt{\langle \widehat{a}^{\dagger}(\tau_{2}) \, \widehat{a}(\tau_{2})\rangle}},
\label{HBTG11}\\
\overline{g}^{(2)}(\tau_{1}, \tau_{2}) &=& \frac{\langle \widehat{a}^{\dagger}(\tau_{1})  \widehat{a}^{\dagger}(\tau_{2}) \, \widehat{a}(\tau_{2})\, \widehat{a}(\tau_{1})\rangle}{\langle \widehat{a}(\tau_{1}) \, \widehat{a}(\tau_{1})\rangle \langle \widehat{a}^{\dagger}(\tau_{2}) \, \widehat{a}(\tau_{2})\rangle}.
\label{HBTG12}
\end{eqnarray}
Equations (\ref{HBTG11})--(\ref{HBTG12}) define the degrees of  (temporal) coherence;  different quantum states lead to the same degree of first-order coherence but the corresponding intensity correlations of Eq. (\ref{HBTG12}) are sensitive to the distinct statistical properties of the corresponding states. 
To give an example that seems relevant for the present ends we then consider the case of a thermal mixture described by the following density operator:
\begin{equation}
\widehat{\rho} = \sum_{n=0}^{\infty} \, p_{n} \, |\,n \,\rangle\, \langle \, n\,|, \qquad\qquad p_{n} = \overline{n}^{\,n}/(\overline{n} +1)^{n +1},
\label{HBTG13}
\end{equation}
where $\overline{n}$ may coincide with the Bose-Einstein occupation number\footnote{ Equation (\ref{HBTG13}) may arise also in situations far from the local thermal equilibrium. This is what happens for chaotic (i.e. white) light 
where photons are distributed as in Eq. (\ref{HBTG13}) for each mode of the radiation field but they
are produced by sources in which atoms are kept at an excitation level higher than that in thermal equilibrium. In this case $\overline{n}$ 
will not have the standard Bose-Einstein form. What matters for the degrees of quantum coherence are not the explicit 
forms of the occupation numbers but the statistical properties of the states.}.
If we now insert Eq. (\ref{HBTG13}) inside Eqs. (\ref{HBTG11})--(\ref{HBTG12}) the calculation of the first-order degree of quantum coherence is immediate. For the second-order correlations the numerator of Eq. (\ref{HBTG12}) 
can be evaluated as $\mathrm{Tr}[\widehat{\rho}\,\hat{a}^{\dagger}(\tau_{1})  \hat{a}^{\dagger}(\tau_{2}) \, \hat{a}(\tau_{2})\, \hat{a}(\tau_{1})]$ and the results are:
\begin{equation}
\lim_{|\tau_{1}- \tau_{2}|\to 0} \overline{g}^{(1)}(\tau_{1},\, \tau_{2})  =\, 1,\qquad \lim_{|\tau_{1}- \tau_{2}|\to 0} \, \overline{g}^{(2)}(\tau_{1},\, \tau_{2})  = 2.
\label{HBTG14}
\end{equation}
Equation (\ref{HBTG14}) shows that $g^{(2)}(\tau_{1},\tau_{2})$ 
evaluated in the case of thermal mixture always exceeds the result of a coherent state 
which is, by definition, an eigenstate of the annihilation operator (i.e. $\hat{a} |\alpha \rangle = \alpha  |\alpha \rangle$). In the zero time-delay limit 
(i.e. $|\tau_{1}- \tau_{2}|\to 0$) Eqs. (\ref{HBTG11}) and (\ref{HBTG12}) imply $\overline{g}^{(1)} = \overline{g}^{(2)} =1$ and, according  to Glauber theory, this property holds to all orders; this means, in practice,  
that $\overline{g}^{(1)} = \overline{g}^{(2)} =\,.\,.\,.\, = \overline{g}^{(n-1)} = \overline{g}^{(n)} =1$.  This result is often dubbed by saying that the
chaotic (i.e. white) light is {\em bunched} and it exhibits super-Poissonian statistics. In the case of a single Fock state we have instead $\overline{g}^{(2)} = (1 - 1/n )< 1$ showing that Fock states always lead to sub-Poissonian behaviour.  While chaotic light is an example of bunched quantum state (i.e. $\overline{g}^{(2)} > 1$ implying more degree of second-order coherence than in the case of a coherent state), Fock states are instead antibunched (i.e. $\overline{g}^{(2)} <1$) indicating a degree of second-order coherence smaller than in the case of a coherent state.

Let us finally come to the case of the non-thermal state characterising the relic gravitons. Neglecting the polarizations and the momentum dependence we have that, in the single-mode approximation, Eq. (\ref{NT9}) becomes $\widehat{a} = u \, \widehat{b} - v \widehat{b}^{\dagger}$ so that the degree of second-order coherence is: 
\begin{equation}
\langle \widehat{a}^{\dagger}\,\widehat{a}^{\dagger}\, \widehat{a}\, \widehat{a}^{\dagger}\rangle= 
2 |\,v\,|^4 + |\,v\,|^2 \, |\,u\,|^2.
 \label{HBTG15}
\end{equation}
But this means that, in the zero-delay limit, 
\begin{equation}
\lim_{|\tau_{1}- \tau_{2}|\to 0} \overline{g}^{(1)}(\tau_{1},\, \tau_{2})  =\, 1,\qquad \lim_{|\tau_{1}- \tau_{2}|\to 0} \overline{g}^{(2)}(\tau_{1},\, \tau_{2})  = 3 + \frac{1}{|\,v\,|^2},
\label{HBTG16}
\end{equation}
where, by definition, $\langle \widehat{a}^{\dagger}\, \widehat{a} \rangle = |\, v\,|^2$. This means 
that, in the limit of large averaged multiplicities the thermal and non-thermal sates
have degrees of second-order coherence that are clearly distinguishable. 

The single-mode approximation must be extended to incorporate the momentum and the polarization 
dependence. In short the conclusion stemming from the previous analyses \cite{HBT1,HBT2,HBT3} is that 
the results derived in the single-mode approximations remain valid if we 
take into account the momentum and the polarization dependence. Neglecting, for simplicity, the 
polarizations but keeping the various momenta the density matrix of a thermal state can be written, in the Fock basis, as: 
\begin{equation}
\hat{\rho} = \sum_{\{n\}} \, P_{\{n\}} \, | \{n\} \rangle\, \langle \{n\}|,\qquad \sum_{\{n\}} \, P_{\{n\}} =1.
\label{AMB1}
\end{equation}
The multimode probability distribution appearing in Eq. (\ref{AMB1}) is given by:
\begin{equation}
P_{\{n\}} = \prod_{\vec{k}} \frac{\overline{n}_{k}^{n_{\vec{k}}}}{( 1 + \overline{n}_{k})^{n_{\vec{k}} + 1 }},
\label{AMB2}
\end{equation}
where $\overline{n}_{k} = \mathrm{Tr}[ \hat{\rho} \, \hat{d}_{\vec{k}}^{\dagger}\, \hat{d}_{\vec{k}}]$ is the average occupation number of each Fourier mode; following the usual habit we also employed the notation $ |\{n \}\rangle = |n_{\vec{k}_{1}} \rangle \, | n_{\vec{k}_{2}} \rangle \, | n_{\vec{k}_{3}} \rangle...$ where the ellipses stand for all the occupied modes of the field. For the degree of second-order 
coherence we need to analyze either ${\mathcal T}^{(2)}(x_{1},\,x_{2})$ or ${\mathcal S}^{(2)}(x_{1},\,x_{2})$(see Eq. (\ref{HBTG3}) and discussion thereafter); the relevant step, in this respect, is to compute $\langle \hat{a}^{\dagger}_{- \vec{k}_1}(\tau_{1}) \,\, \hat{a}^{\dagger}_{- \vec{k}_2}(\tau_{2}) \,\,\hat{a}_{ \vec{k}_3}(\tau_{2})  \,\,\hat{a}_{ \vec{k}_4}(\tau_{1})  \rangle$. Considering for simplicity the zero-delay limit we have that the preceding term becomes:
\begin{equation}
\sum_{\{n\}} \, P_{\{n\}} \,  \langle \{n\} |\hat{a}^{\dagger}_{- \vec{k}_1} \,\, \hat{a}^{\dagger}_{- \vec{k}_2} \,\,\hat{a}_{ \vec{k}_3}  \,\,\hat{a}_{ \vec{k}_4}  | \{n\} \rangle.
\label{AMB2a}
\end{equation}
The expectation value appearing Eq. (\ref{AMB2a}) can then be expressed as:
\begin{eqnarray}
&& \langle \hat{d}_{i}^{\dagger} \,  \hat{d}_{j}^{\dagger} \, \hat{d}_{k}\,  \hat{d}_{\ell} \rangle = 
\langle \hat{d}_{i}^{\dagger} \,  \hat{d}_{i}^{\dagger} \, \hat{d}_{i}\,  \hat{d}_{i}\rangle \delta_{i\,j} \,
\delta_{j\,k}\, \delta_{\ell\,k} +
\nonumber\\
&& \langle \hat{d}_{i}^{\dagger} \,  \hat{d}_{j}^{\dagger} \, \hat{d}_{i}\,  \hat{d}_{j} \rangle  \, \delta_{i\, k} \,
 \delta_{j\, \ell} [ 1 - \delta_{ij}] +  \langle \hat{d}_{i}^{\dagger} \,  \hat{d}_{j}^{\dagger} \, \hat{d}_{j}\,  \hat{d}_{i} \rangle  \, \delta_{i\, \ell} \,
 \delta_{j\, k} [ 1 - \delta_{ij}], 
\label{AMB4}
\end{eqnarray}
where $\hat{d}_{i}$ and $\hat{d}_{j}^{\dagger}$ denote the annihilation and creation operators related two generic momenta, 
i.e. for instance $\hat{d}_{\vec{q}}$ and $\hat{d}^{\dagger}_{\vec{p}}$; furthermore, following the same shorthand 
notation, $\delta_{i\, j}$ denotes the delta functions over the three-momenta, i.e. 
$ \delta^{(3)}(\vec{q}- \vec{p})$.  All the momenta are equal  in the first line of Eq. (\ref{AMB4}); in the 
second line of Eq. (\ref{AMB4}) the momenta are paired two by two is such a way that double counting is avoided. The normalized degree of second-order coherence finally becomes
\begin{equation}
g^{(2)}(\vec{r}, \tau_{1}, \tau_{2}) = \frac{\int d^{3} k_{1} \overline{n}_{k_{1}}(\tau_{1})/k_{1} \, \int d^{3} k_{2} \,\overline{n}_{k_{2}}/k_{2}\, \biggl[ 1 + e^{- i (\vec{k}_{1} + \vec{k}_{2})\cdot \vec{r}}\biggr]}{\int d^{3} k_{1} \overline{n}_{k_{1}}(\tau_{1})/k_{1} \, \int d^{3} k_{2} \,\overline{n}_{k_{2}}(\tau_{2})/k_{2}}.
\label{AMB3}
\end{equation}
If we integrate over the angular coordinates and take the zero-delay limit we obtain 
\begin{equation}
g^{(2)}(\vec{r}, \tau) = 1 + \frac{\int k_{1} d k_{1} \overline{n}_{k_{1}}(\tau)\,j_{0}(k_{1} r)  \int k_{2} d k_{2} \,\overline{n}_{k_{2}}(\tau)j_{0}(k_{2} r)}{\int k_{1} d k_{1} \overline{n}_{k_{1}}(\tau)\, \int k_{2} d k_{2} \,\overline{n}_{k_{2}}(\tau)},
\label{AMB5a}
\end{equation}
where there is a residual time-dependence coming from the averaged multiplicities.
In the large-scale limit Eq. (\ref{AMB5a}) then implies that $g^{(2)}(\vec{r},\tau) \to 2$.

In the case of the non-thermal gravitons Eqs. (\ref{NT9})--(\ref{NT10}) must be inserted 
into the expectation value of Eq. (\ref{HBTG10}) so that the relevant term 
for the degree of the second-order coherence is given by:
\begin{eqnarray}
&& \langle \hat{a}^{\dagger}_{- \vec{k}_1, \, \alpha_{1}}(\tau_{1}) \,\, \hat{a}^{\dagger}_{- \vec{k}_2, \, \alpha_{2}}(\tau_{2}) \,\,\hat{a}_{ \vec{k}_3, \, \alpha_{3}}(\tau_{2})  \,\,\hat{a}_{ \vec{k}_4, \, \alpha_{4}}(\tau_{1})  \rangle
\nonumber\\
&&= v_{k_{1},\,\alpha_{1}}^{*}(\tau_{1}) v_{k_{2},\,\alpha_{2}}^{*}(\tau_{2}) v_{k_{3},\,\alpha_{3}}(\tau_{2}) v_{k_{4},\,\alpha_{4}}(\tau_{1})
\langle \hat{b}_{\vec{k}_{1},\,\alpha_{1}} \hat{b}_{\vec{k}_{2}, \alpha_{2}}
 \hat{b}_{-\vec{k}_{3},\,\alpha_{3}}^{\dagger} \hat{b}_{-\vec{k}_{4},\,\alpha_{4}}^{\dagger}\rangle
 \nonumber\\
 &&+  v_{k_{1},\,\alpha_{1}}^{*}(\tau_{1}) u_{k_{2},\,\alpha_{2}}^{*}(\tau_{2}) u_{k_{3},\,\alpha_{3}}(\tau_{2}) v_{k_{4},\,\alpha_{4}}(\tau_{1})
\langle \hat{b}_{\vec{k}_{1},\,\alpha_{1}} \hat{b}_{\vec{k}_{2}, \alpha_{2}}^{\dagger}
 \hat{b}_{-\vec{k}_{3}, \alpha_{3}} \hat{b}_{-\vec{k}_{4}, \alpha_{4}}^{\dagger}\rangle.
 \label{degB2}
 \end{eqnarray}
Since the relic graviton background  is not polarized\footnote{This means that
$v_{k,\,\alpha}(\tau)$ and $u_{k,\,\alpha}(\tau)$ are the 
same for the two polarizations, i.e. 
$v_{k,\,\oplus}(\tau) = v_{k,\,\otimes}(\tau) = v_{k}(\tau)$ 
and $u_{k,\,\oplus}(\tau) = u_{k,\,\otimes}(\tau) = u_{k}(\tau)$.} 
thanks to Eq. (\ref{degB2}),  the explicit form of the HBT correlations becomes: 
 \begin{eqnarray}
 {\mathcal T}^{(2)}(x_{1}, x_{2}) &=& \frac{1}{(2\pi)^6} \int \frac{d^{3} k}{k} \, \int \frac{d^{3} p}{p}\, \biggl\{
 4 |v_{k}(\tau_{1})|^2 \,\, |v_{p}(\tau_{2})|^2 
 \nonumber\\
 &+& \frac{1}{4} [ 1 + (\hat{k}\cdot\hat{p})^2] [ 1 + 3(\hat{k}\cdot\hat{p})^2] \biggl[ v_{k}^{*}(\tau_{1}) 
 v_{p}^{*}(\tau_{2}) v_{k}(\tau_{2})v_{p}(\tau_{1})  
 \nonumber\\
 &+& v_{k}^{*}(\tau_{1}) u_{k}^{*}(\tau_{2}) u_{p}(\tau_{2})v_{p}(\tau_{1})\biggr] 
 e^{- i (\vec{k} - \vec{p})\cdot\vec{r}}
 \biggr\}.
 \label{degB3}
 \end{eqnarray}
Because of the sum over the polarizations, Eq. (\ref{degB3}) differs a bit from the single-polarization approximation which is given by:
 \begin{eqnarray}
 {\mathcal S}^{(2)}(x_{1}, x_{2}) &=& \frac{1}{4(2\pi)^6} \int \frac{d^{3} k}{k} \, \int \frac{d^{3} p}{p} 
\nonumber\\
&\times& \biggl\{ |v_{k}(\tau_{1})|^2 \,\, |v_{p}(\tau_{2})|^2 
 +\biggl[ v_{k}^{*}(\tau_{1}) 
 v_{p}^{*}(\tau_{2}) v_{k}(\tau_{2})v_{p}(\tau_{1}) 
 \nonumber\\
 &+& v_{k}^{*}(\tau_{1}) 
 u_{k}^{*}(\tau_{2}) u_{p}(\tau_{2})v_{p}(\tau_{1})\biggr] 
 e^{- i (\vec{k} - \vec{p})\cdot\vec{r}}\biggr\}.
 \label{degB4}
\end{eqnarray}
The degrees of 
second-order coherence will receive the dominant contribution for $k r \sim p r \sim {\mathcal O}(1)$ so that the final result can be written as:
\begin{eqnarray}
&& g^{(2)}( \tau_{1}, \tau_{2}) \simeq \frac{41}{30} \frac{\int k\,d k\,\,
|v(k)|^2 \int p\,d p \,|v(p)|^2 e^{- i (k -p)\Delta\tau}}{\int k \,d k  
|v(k)|^2 \int p \,d p \,|v(p)|^2  },
\label{degB18}\\
&& \overline{g}^{(2)}(\tau_{1}, \tau_{2}) \simeq 3 \frac{\int k \, d k  
|v(k)|^2 \int p \,d p \,|v(p)|^2 e^{- i (k -p)\Delta\tau}}{\int k \,d k\,
|v(k)|^2 \int p \, d p \,|v(p)|^2 },
\label{degB19}
\end{eqnarray}
where, recalling Eqs. (\ref{NT14})--(\ref{NT15}), we used the shorthand notation $v_{k}(\tau) = - v(k) e^{-i k\tau}$; note also that $\Delta\tau = \tau_{1}- \tau_{2}$. When $\tau_{1} \neq \tau_{2}$ it can be demonstrated that $|g^{(2)}(\tau) |< g^{(2)}(0)$ and  $|\overline{g}^{(2)}(\tau)| < \overline{g}^{(2)}(0)$ which implies, in a quantum optical language, that the degree of second-order coherence is not only super-Poissonian but also bunched. 
Bearing in mind the results for the first-order correlations, it turns out that 
the intensity correlations are factorized as follows:
\begin{equation} 
{\mathcal T}^{(2)}(r,\tau) \simeq \frac{41}{30}\, {\mathcal T}^{(1)}(\tau) {\mathcal T}^{(1)}(\tau),\qquad
{\mathcal S}^{(2)}(r,\tau_{1}, \tau_{2}) \simeq 3 \,{\mathcal S}^{(1)}(\tau_{1}) {\mathcal S}^{(1)}(\tau_{2}).
\label{degB12}
\end{equation}
This means that, in the zero-delay limit, the degrees of second-order coherence are
$g^{(2)}(r, \tau) \to 41/30$ and $\overline{g}^{(2)}(r, \tau) \to 3$. The second result in Eq. (\ref{degB12}) 
reproduces the single-mode approximation discussed in Eq. (\ref{HBTG16}) whereas the sum over 
the polarization partially reduce the estimate obtained within the single-mode approximation.
In spite of that  the degree of second-order coherence is always super-Poissonian since both $g^{(2)}$ and $\overline{g}^{(2)}$ are larger than $1$. 

\newpage
\renewcommand{\theequation}{6.\arabic{equation}}
\setcounter{equation}{0}
\section{Concluding remarks}
\label{sec6}
The frequency window of wide-band detectors notoriously ranges between few Hz and $10$ kHz
where successful astrophysical observations 
are ongoing. However, as far as diffuse backgrounds are concerned, the current limits imply that the sensitivity of correlated interferometers for the detection of a flat spectral energy density of relic gravitons is approximately $h_{c}^{(min)}= {\mathcal O}(10^{-24})$ for typical frequencies in the audio band. Sharp deviations from scale-invariance lead to similar orders of magnitude and while these figures may improve in the years to come, the frequency domain of ground-based interferometers will remain the same. For this reason it is important to promote new instruments operating in a much higher frequency domain where the potential signals coming from the past history of the plasma are dominant. 

The first suggestions that microwave cavities (operating between the MHz and the GHz regions) 
could be used for the detection of relic gravitons associated with post-inflationary phases stiffer than radiation are almost twenty years old. While in the 1980s the typical sensitivities of these instruments were $h_{c}^{(min)} = {\mathcal O}(10^{-17})$ they improved later on and reached $h_{c}^{(min)} = {\mathcal O}(10^{-20})$. Similar prototypes aimed at the detection of dark matter could be used as high-frequency detectors of gravitational waves and the target sensitivities of these instruments are often set by requiring in the MHz (or even GHz regions) the same sensitivities reached today in the audio band by interferometers. These requirements 
are in fact not guided by the signals of the available sources in the corresponding frequency domain and are therefore arbitrary. Both thermal and non-thermal gravitons lead to a large 
cosmic signal in the MHz--GHz domain. If we collect all the 
current phenomenological bounds together with the basic features of the potential signals we are led to consider chirp amplitudes that are much smaller than ${\mathcal O}(10^{-20})$.
The origin of thermal spectra may follow from graviton decoupling but similar 
signals may also have a geometric origin especially below the maximal frequency of the spectrum To detect directly thermal gravitons with high-frequency instruments operating between the MHz
and the GHz the minimal detectable chirp amplitude should be $h_{c}^{(min)} = {\mathcal O}(10^{-28})$ (or smaller)
while $\sqrt{S_{h}^{(min)}} \leq {\mathcal O}(10^{-32})\, \mathrm{Hz}^{-1/2}$. 

A complementary class of high-frequency signals involves non-thermal graviton spectra that are generated from the amplification of the zero-point fluctuations, for instance during a quasi-de Sitter stage of expansion.  The averaged multiplicity of the produced gravitons is typically suppressed as $\nu^{-4}$ below the Hz but at higher frequencies this result is not compelling since the wavelengths of the gravitons reenter the Hubble radius when the plasma is not yet dominated by radiation. This happens, in particular, when a long stiff phase precedes the current dominance of dark energy. While in the concordance paradigm the maximal frequency of the spectrum is ${\mathcal O}(250)$ MHz, $\nu_{max}$ decreases if the post-inflationary expansion rate is faster than radiation but it increases (even beyond the GHz) 
when the post-inflationary expansion rate is slower than radiation. 
In the non-thermal case in the MHz--GHz domain $h_{c}^{(min)}$ and $\sqrt{S_{h}^{(min)}}$ are grossly comparable with the ones determined for thermal gravitons but they are slightly smaller (i.e. ${\mathcal O}(10^{-32})$ and ${\mathcal O}(10^{-36})\, \mathrm{Hz}^{-1/2}$ respectively).

In a quantum mechanical perspective the maximal frequency of the spectrum corresponds to the 
production of a single pair of gravitons with opposite (comoving) three-momenta. If high-frequency instruments will ever be operating around $\nu_{max}$ with the sensitivities summarized in the 
previous paragraph they will also be able to detect bunches of gravitons. For this reason 
it is natural to argue that detectors operating in the MHz and GHz regions are particularly suitable for the analysis of second-order interference effects. 
As in the case of optical photons, the interferometric techniques 
pioneered by Hanbury-Brown and Twiss in the 1950s could allow, in this context to 
distinguish the statistical properties of thermal and non-thermal gravitons. 

All in all a sensitivity ${\mathcal O}(10^{-20})$ or even ${\mathcal O}(10^{-24})$ in the chirp amplitude for frequencies in the MHz or GHz regions 
is a technological achievement but it is not a reasonable physical goal as long as it deliberately ignores the potential sources in the high-frequency and ultra-high-frequency domains. The considerations developed here necessarily lead to requirements that are far more severe both in the chirp and in the spectral amplitudes. Bearing in mind these caveats, it is important to stress 
that high-frequency detectors could be the only instruments able to access the single graviton lines. The detection of graviton bunches and of their statistical 
properties is, in this perspective,  crucial for a direct scrutiny of the quantum aspects of gravitational interactions. While the interferometers operating in the audio band (i.e. between few Hz and $10$ kHz) are and will be relevant for astrophysical applications, MHz and GHz detectors are equally essential to probe the relic gravitons and their quantumness. High-frequency detectors may also provide a unique information on the post-inflationary expansion rates and, more generally, on the early stages of the evolution of the primeval plasma. 

\section*{Acknowledgements}
The author acknowledges countless hours of discussions with the 
late E. Picasso on high-frequency detectors and on relic gravitons.
It is also a pleasure to thank T. Basaglia, A. Gentil-Beccot, 
S. Rohr and J. Vigen of the CERN Scientific Information 
Service for their kind help.

\end{document}